\documentclass{aa}     

\usepackage{amssymb}
\usepackage{graphicx}
\usepackage{txfonts}

\newcommand\vv{{\mathrm v}  }
\begin{document}

   \title{Induced differential rotation and mixing in asynchronous binary stars}


   \author{G. Koenigsberger          
           \inst{1}
        \and   
            E. Moreno
          \inst{2}
        \and
            N. Langer 
        \inst{3}
          }

   \institute{Instituto de Ciencias F\'{\i}sicas, Universidad Nacional Aut\'onoma de M\'exico,
              Ave. Universidad S/N, Chamilpa, Cuernavaca, M\'exico \\
              \email{gloria@astro.unam.mx}
         \and
             Instituto de Astronom\'{\i}a, Universidad Nacional Aut\'onoma de M\'exico, Apdo. Postal 70-264,
             Ciudad de M\'exico, 04510 M\'exico \\
             \email{edmundo@astro.unam.mx}
         \and
             Argelander-Institut f\"ur Astronomie, Universit\"at Bonn, Auf dem H\"ugel 71, 53121 Bonn, Germany,
             \email{nlanger@astro.uni-bonn.de}    
             }

   \date{Version 2021 April 19 }


 
\abstract
{Rotation contributes to internal mixing processes and observed variability in massive stars. 
A significant number of binary stars are not in strict synchronous rotation, including all
eccentric systems. This leads to a tidally induced and time-variable differential rotation structure.
       }
{We present a method for exploring the rotation structure of asynchronously rotating binary stars.
       }
{The method consists of solving the equations of motion of a 3D grid of volume elements located above the 
rigidly rotating core of a binary star in the presence of gravitational, centrifugal, Coriolis, gas 
pressure and viscous forces to obtain the angular velocity as a function of the three spatial
coordinates and time. The method is illustrated for a short period massive binary in a circular
orbit and in an eccentric orbit.
        }
{We find that for a fixed set of stellar and orbital parameters, the induced rotation structure and 
its temporal variability depend on the degree of departure from synchronicity. In eccentric systems, the structure
changes over the orbital cycle with maximum amplitudes occurring potentially at orbital phases other than 
periastron passage.   We discuss the possible role of the time-dependent tidal flows in enhancing the mixing 
efficiency and speculate that, in this context, slowly rotating asynchronous binaries could have more 
efficient mixing than the analogous more rapidly rotating but tidally locked systems. We find that
some observed nitrogen abundances depend on the orbital inclination, which, if real, 
would imply an inhomogeneous chemical distribution over the stellar surface or that tidally induced 
spectral line variability, which is strongest near the equator, affects the abundance determinations.  
Our models predict that, neglecting other angular momentum transfer mechanisms, a pronounced initial 
differential rotation structure converges toward average uniform rotation on the viscous timescale.
        }
{A broader perspective of binary star structure, evolution and variability can be gleaned by taking into
account the processes that are triggered by asynchronous rotation.
}


   \maketitle
%

\section{Introduction }

Rotation in massive stars plays a crucial role in transporting nuclear-processed chemical 
elements  toward surface layers and dragging fresh fuel into the nuclear region. Faster
rotation is associated with more efficient mixing  
\citep{1987A&A...178..159M,1991A&A...243..155L,1992A&A...265..115Z,1993SSRv...66..285Z, 1997IAUS..189..343L, 
2000ApJ...528..368H, 2000A&A...361..101M, 2011A&A...530A.115B,Ekstrom_etal2012, 2012ARA&A..50..107L}. 
A star in which the mixing efficiency is high lives longer on the main sequence, becomes brighter, 
and grows a larger convective core than one with a low mixing efficiency. This increases the probability 
that the end product will be a black hole instead of a neutron star.  Thus, considerable effort is
invested in analyzing the mechanisms that intervene in the mixing processes.

There are numerous processes that can contribute to mixing in massive stars 
\citep{2000ApJ...528..368H, 2000ARA&A..38..143M, 2016ApJ...821...49G, 2019ARA&A..57...35A},
but the dominant ones are most likely meridional circulation and the shear instability
\citep{1992A&A...265..115Z, 2000A&A...361..101M, 2000ApJ...528..368H}. Shear arises 
in contiguous layers in a fluid moving at different velocities and turbulence arises when the
relative motions are large enough to trigger instabilities.
In massive stars, differential rotation is considered to be the prime
source of shear instabilities  \citep{2000ARA&A..38..143M}, and
it is modeled in terms of the functional dependence on radius of rotation 
angular velocity, often referred to as the Omega-gradient or $\omega$ profile.

Differential rotation arises as a consequence of angular momentum removal 
by stellar winds from the surface  \citep{1992A&A...265..115Z, 1997A&A...322..209T,1998A&A...329..551L, 
2000ARA&A..38..143M, 2000A&A...361..101M, 2016CeMDA.126..249L}. It also appears during late stages of 
main sequence evolution when the convective core contracts and speeds up as the envelope expands 
and slows down (see, for example, \citealt{2013A&A...556A.100S}  and references therein) and due to 
the long term action of  meridional circulation currents driven by baroclinicity.
Because massive stars are in general assumed to have
spherical symmetry, the angular momentum transport and removal by winds impacts only the radial 
differential rotation structure which allows the analysis to be simplified to a 1D calculation.  However,
this assumption is invalid for very rapidly rotating stars and close binary stars for which 2D and 3D 
calculations are needed \citep{2019A&A...625A..89G, 2017PASA...34....1D, 2020FrASS...6...77L}.

Stars in binaries are generally assumed to be in uniform rotation \citep{1982ApJ...261..265T, 
2009A&A...497..243D}, a condition that follows from  the assumption that they are in the 
equilibrium state, often referred to as being tidally locked.  The equilibrium state is 
attained when the stellar equator is coplanar with the orbital plane, the orbit is strictly 
circular and the spin and the orbital angular velocities are equal (synchronous rotation). 
They are generally modeled as single stars in uniform rotation. However, the equilibrium 
conditions do not hold in eccentric-orbit systems, nor in systems in which evolutionary changes 
cause the core to contract and the envelope to expand, nor in pre-main-sequence systems if they 
were born in asynchronous rotation \citep{1975A&A....41..329Z}.

It is well known that a departure as small as 1\% from synchronous rotation induces a 3D,
time-dependent perturbation \citep{1981ApJ...246..292S, 1987ApJ...322..856T, Dolginov:1992up, 
2009ApJ...704..813H}. The presence of these effects led \citet{2013EAS....64..339K} to speculate 
that the induced shearing flows might provide an additional mixing mechanism in binary stars 
that is absent in single stars.   In this paper we take a next step in following up on this 
hypothesis by introducing a method that allows the evaluation of the amplitude of the tidal flow 
velocities of multiple stellar layers in a binary star with arbitrary rotation velocity and 
orbital eccentricity. 

The method is described in Sect. 2. In Sect. 3 we describe the dependence of the velocity 
and its radial gradient on azimuth for the case in which the average
rotation velocity is nearly uniform as well as for the case in which it is has a steep radial gradient.   
In Sect. 4 we illustrate the example of a system with an eccentric orbit. A discussion of the results
and implications for internal mixing and observational characteristics is presented in Sect. 5,
and the results are summarized in Sect. 6 together with our conclusions.
Complementary and supporting material are included in the appendices.

\section{Method}

Tidal perturbations have historically been analyzed in the context of the long-term evolution of the orbital 
elements and rotation rates in binary stars (see, for example, \citet{2014ARA&A..52..171O} and references therein).  
More recently, focus has shifted toward the shorter-term  interaction of a star's internal oscillation modes 
with the external gravitational potential of the companion thanks to the discovery in {\it Kepler} satellite 
data of photometric variability on orbital timescales in eccentric binaries, and which is associated with 
the predicted oscillation modes \citep{1995ApJ...449..294K, 2020ApJ...888...95G, 2011ApJS..197....4W}.  
Clearly, the short and long timescale effects are not independent of each other, as the former are responsible 
for the energy and angular momentum transfer that produce the latter.

Our approach has been to focus on the short-timescale spectroscopic variability, aiming to  understand the
tidally induced photospheric line-profile variability and its impact on the determination of the 
stellar and orbital parameters. The method that is described below was introduced in \citet{1999RMxAA..35..157M},
where we examined the time-dependent behavior of the equatorial latitude in the eccentric system $\iota$ Orionis,
predicting an increase in the tidal bulge around periastron passage and the ensuing  appearance of short-timescale
oscillations. A brightening at periastron and the presence of short-timescale oscillations have now been 
observed \citep{2017MNRAS.467.2494P}.  The next steps in the development of the method involved modeling the entire
stellar surface, not only the equatorial latitude, and incorporating the calculation of photospheric absorption
line profiles using the projected surface velocities along the line-of-sight to the observer \citep{Moreno:2005cq}.
The predicted line-profile variations for the eccentric binary systems  $\epsilon$ Persei (P=14d, e=0.6)
and $\alpha$ Virginis (P=4d, e=0.1) were found to be very similar to those that are observed 
\citep{Moreno:2005cq,  2009ApJ...704..813H,  2016A&A...590A..54H}.
The application of the same model to the optical counterpart of the Vela X-1
pulsar showed that the distortion in the radial velocity curve is significant enough to affect the
determination of the neutron star companion mass  \citep{2012A&A...539A..84K}.  In \citet{2016A&A...590A..54H},
we showed that in a binary system that undergoes orbital precession, the line profile variations 
can result in (fictitious) epoch-dependent determinations of the orbital eccentricity.  We have also
examined the possibility that tidally induced shear energy dissipation contributed to the onset of
the red nova event in V1309 Sco \citep{2016RMxAA..52..113K} and the eruption of the Wolf-Rayet system
HD\,5980 which was similar to that observed in luminous blue variables  \citep{2007ASPC..367..437T}. 

The mathematical method, which in its more recent version is fully described in \citet{Moreno:2011jq}, is
an {\it ab initio} dynamical calculation of the response of the perturbed star's surface to the gravitational,
inertial, gas pressure and viscous forces to which it is subjected as a function of time.   In this
paper we report the results of upgrading the model to be able to model a sufficient number of stellar layers
so as to explore the amplitude of tidal perturbations closer to the stellar core, a region
that is relevant for answering the question of the relevance of tidal perturbations in mixing processes.

\subsection{TIDES code}

The algorithm that is used for the numerical simulation is an upgraded version of the 
{\it Tidal Interactions with Dissipation of Energy due to Shear (TIDES)} code\footnote{The TIDES 
code is available upon request and is easily implemented in any operating system running a Fortran
or GNU Fortran compiler. The particular versions used to produce the results in this paper are named 
lx.osc.disip.2.2.f (one layer), lx.capas.vr0.3.1.f (n layers).},  and  which is fully documented 
in \citet{1999RMxAA..35..157M}, \citet{2007A&A...461.1057T} and \citet{Moreno:2011jq}. It employs 
a time-marching  quasi-hydrodynamic Lagrangian scheme to solve simultaneously the equations of motion 
of a grid of 3D volume elements covering the inner, rigidly rotating region  of a tidally perturbed star.  
The equations are solved in the non-inertial reference frame with origin in the center of the perturbed 
star ($m_1$) and that rotates with angular velocity \mbox{\boldmath ${\Omega}$}, which corresponds
to the orbital motion of the companion.  Summarizing from \citet{Moreno:2011jq}, 
the total acceleration {\boldmath $a$}$'$ of a volume element  measured in this non-inertial frame is:

\begin{eqnarray} \mbox{\boldmath $a$}'& =& \mbox{\boldmath $a$}_{\star}-
 \frac{Gm_1({|\mbox{\boldmath $r$}'|}) \mbox
{\boldmath $r$}'}{|\mbox{\boldmath $r$}'|^3}-Gm_2 \left [
\frac{\mbox{\boldmath $r$}'- \mbox{\boldmath $r$}_{21}}
{|\mbox{\boldmath $r$}'-
\mbox{\boldmath $r$}_{21}|^3}+ \frac{\mbox{\boldmath $r$}_{21}}
{|\mbox{\boldmath $r$}_{21}|^3}
\right ]- \nonumber \\
&& - \mbox{\boldmath ${\Omega}$} \times \left ( \mbox{\boldmath
 ${\Omega}$} \times
\mbox{\boldmath $r$}' \right )
-2 \mbox{\boldmath ${\Omega}$} \times \mbox{\boldmath $\vv$}'-
\frac{d \mbox{\boldmath ${\Omega}$}}{dt} \times \mbox{\boldmath $r$}',
\label{ecmov} \end{eqnarray}

\noindent where {\boldmath $r$}$'$ and {\boldmath $\vv$}$'$ are the position and velocity of the element in the
non-inertial frame, {\boldmath $r_{21}$} is the instantaneous orbital separation, $m_1({|\mbox{\boldmath $r$}'|})$
is the mass of $m_1$ that is contained inside the radius ${|\mbox{\boldmath $r$}'|}$, and  
{\boldmath $a$}$_{\star}$ is the acceleration of a volume element produced by 
gas pressure and viscous forces exerted by the stellar material surrounding the element. 
The companion, $m_2$ is treated as a point mass and the orbital plane is coplanar with $m_1$'s equator.

With $h$ the orbital angular momentum per unit mass and $u_r$ the radial orbital speed of $m_2$,
the orbital motion is found by solving the equations:

\begin{eqnarray}
\dot{r}_{21} &  =  & u_r                                                         \nonumber \\ 
\dot{u}_r    &  =  & \frac{h^2}{r^3_{21}} - \frac{G(m_1 m_2)}{r^2_{21}}          
\label{eq_orbital_motion}
\end{eqnarray}

Equation~\ref{ecmov} is solved simultaneously for all volume elements, together with the equation of 
orbital motion of $m_2$ around $m_1$ (Equation~\ref{eq_orbital_motion}), using  a seventh order Runge-Kutta 
integrator. The values of the orbital separation $\boldmath $r$_{21}$, the orbital 
velocity $\boldmath \Omega$, and its time derivative $\boldmath \dot{\Omega}$ are obtained from the 
solution of the orbital motion. A polytropic state equation is used to compute the pressure inside 
the elements as they expand and contract  thus providing  a representation of the hydrodynamical 
response of the fluid to the perturbing forces.  The motions of individual elements are coupled to 
those of neighboring elements and the core through a viscosity parameter, which is given as an input 
parameter and which, as discussed below, corresponds to a turbulent viscosity.  

The output includes values of the radius and the velocity of each grid element at user-specified time 
intervals. The time-dependent solution of the coupled equations  captures the nonlinear dynamical 
evolution of the system for arbitrary stellar rotation, orbital period and orbital eccentricity. 
However, the detailed microphysical processes, heat diffusion and buoyancy are neglected.

The upgrade that we use in this paper performs the calculation for multiple interacting layers, 
instead of only one layer as in the original version of the code.
Adjoining layers are coupled to each other in the same manner as with the core; that is, 
via the viscous coupling. The only other modification with respect to the one-layer calculation, 
is that the motion in the radial direction is now suppressed, a simplification that is justified 
by the larger azimuthal motions (by close to a factor of ten) compared to those in the polar 
and radial directions \citep{1981ApJ...246..292S,  2009ApJ...704..813H}. This simplification has 
a measurable impact on the quantitative results, as illustrated by the comparison  between the one-layer
and $n$-layer outputs shown in the appendix Figs.~\ref{fig_radius1D} and \ref{fig_compare_TIDES1and2}, but 
this difference does not affect the conclusions to which we arrive regarding the general behavior of the 
angular velocities and their dependence on radius.

\subsection{Definitions and reference frames}

The perturbed binary star, which we  call the primary, is assumed to consist of a 
rigidly rotating inner region, called the core (which does not necessarily coincide with the nuclear 
burning convective core in massive stars) and a number of discrete layers above it, extending 
to the surface. The core rotates at a constant rate $\omega_{0}$, measured in an inertial 
reference frame $S$, and is parametrized in terms of a synchronicity parameter $\beta_0^0$ that is 
defined as: $\beta_0^0$=$\omega_{0}/\Omega_0$ where $\Omega_0$ is the orbital angular velocity at 
periastron for eccentric orbits or the constant orbital angular velocity for circular orbits, 
$\Omega_0$=2$\pi/P$, with $P$ the orbital period. The allowed values for 
$\beta_0^0\in \{0, ..., \beta_{0,max}  \}$, with the upper bound  set by the maximum velocity of the 
outer layers, which must remain below the critical rotation velocity. 

The thickness of each layer $\Delta R$ is given in terms of  $R_1$, the outer radius of the star, 
with the input parameter being $\Delta R/R_1$. The layers are coupled to each other and to the core 
by a kinematic viscosity $\nu$, which is assumed to be isotropic and constant. The viscosity
also couples the motion of neighboring volume elements.

As already noted, the equations of motion are solved in the non-inertial reference frame with origin at the center 
of the primary star $m_1$ and with the $x'$ axis along the line joining the two stellar centers. This 
is the $S'$ reference frame  and it rotates at a rate $\Omega_0$, for the circular orbits considered 
in this paper. For eccentric orbits, the $S'$ reference frame rotates at a varying speed, dictated by
the orbital motion of $m_2$ around $m_1$. 

We define a third reference frame, $S''$,  which is also centered on $m_1$  and rotates at a constant 
rate $\omega_0$.  This is the rest frame of the rotating primary star.  The direction of the $S''$ rotation 
is the same as that of $S'$.  

The model is based on the assumption that the primary's equator coincides with the orbital plane, so
$\theta'$=$\theta''$. Also, the radial distance from the center of the primary is 
independent of the reference frame, so  $r'$=$r''$.  For simplicity, in this paper we drop the 
prime notation for $\theta$ and $r$.    For the circular orbits considered here, 
$\varphi''$=$\varphi'+(\Omega_0 - \omega_0)t$,  
where $t$ is time after an initial time when 
$\varphi'$= $\varphi''$=0.  We retain $\varphi'$ as our azimuthal coordinate (instead of $\varphi''$) because 
at any given time, the sub-binary longitude provides a unique definition for the point of origin for
this coordinate.  

\begin{table}
\caption{Description of the input parameters and nominal values. }
\label{table1}
\centering
\begin{tabular}{c l c }     
\hline\hline
                                                  
Param         & Description               & Value        \\      
\hline
$P_{orb}$     &Orbital period (d)                              &    4.0145        \\
$e$           &Orbital eccentricity                            & 0                     \\
$m_1$         &Perturbed star mass (${\rm M_\odot}$) &10.25             \\
$m_2$         &Companion star mass (${\rm M_\odot}$) & 6.97             \\
$R_1$         &Primary equilibrium radius (${\rm R_\odot}$)    & 6.84          \\
$\beta_0^0$   & Core synchronicity parameter                   & 1.8                  \\
$\nu$         &Kinematical (turbulent) viscosity (${\rm R_\odot^2\, d^{-1}}$) &0.1              \\
$n$           &Polytropic index                                &3.0      \\
$\Delta$R/R$_1$  &Layer thickness                              &0.06              \\
$N_r$         &Number of layers                                & 10    \\
$N^{eq}_\varphi$&Num partitions in longitude at equator &200                      \\
$N_\theta$    & Num partitions in latitude               &20                      \\
              & (one hemisphere)                               &          \\
$N_{cy}$      & Number of orbital cycles of the run            &$>$40              \\
$Tol$         & Tolerance for the Runge-Kutta integration      &10$^{-9}$        \\ 
\hline
\hline
\end{tabular}

\end{table}

\begin{table}
\caption{Case numbers and properties of models}
\label{table_models_nominal}
\centering
\begin{tabular}{ c c  r r  c  }
\hline\hline
Case &$\beta_0^0$   & N$_{r}$     &  Cy  & Notes  \\
\hline
\hline
30  & 1.01       &5            &  100 &  u     \\
31  &  1.8       &10           &   47 &  u     \\ 
34  &  1.8       &10           &   59 &  b     \\ 
35  &  1.8       &10           &   89 & u,e    \\ 
41  & 2.2       &10           &   78 &  u,*   \\
\hline
\hline
\end{tabular}
\tablefoot{The models listed in this table were run with $\nu$=0.1 R$_\odot^2~d^{-1}$. Column 5 indicates whether
the initial rotation structure is uniform (u) or differential (b) as given in  
Table~\ref{table_case34_non-uniform}; Case 35 is an eccentric orbit with $e$=0.1; 
*Case 41 is the model run with parameters from the Bonn Evolutionary Code (BEC):
$m_1$=10 M$_\odot$, R$_1$=5.48\,R$_\odot$, $n$=3.5.
}
\end{table}

\begin{table}
\caption{Initial differential rotation structure and structure after 196 days (Case 34)}
\label{table_case34_non-uniform}
\centering
\begin{tabular}{ r l l r ll}
\hline\hline
$k$ &$r_{mid}$ &$\beta_0^k$ & $\omega''_0$ & $\left <\omega''\right >$ &$\Delta\omega''$  \\
....    & $R_\odot$&....        & \multicolumn{3}{c}{|--------- rad\,d$^{-1}$ --------|}     \\
\hline
  1  &   2.94   & 2.0    &  0.31   &  -0.03  &   0.005                     \\
  2  &   3.35   & 1.9    &  0.17   &  -0.06  &   0.007                     \\
  3  &   3.76   & 1.8    &  0.00   &  -0.09  &   0.011                     \\
  4  &   4.17   & 1.7    & -0.16   &  -0.11  &   0.018                     \\
  5  &   4.58   & 1.6    & -0.31   &  -0.13  &   0.028                     \\
  6  &   4.99   & 1.4    & -0.63   &  -0.14  &   0.043                     \\
  7  &   5.40   & 1.25   & -0.86   &  -0.15  &   0.072                     \\
  8  &   5.81   & 1.14   & -1.03   &  -0.16  &   0.126                     \\
  9  &   6.22   & 0.92   & -1.38   &  -0.17  &   0.272                     \\
 10  &   6.63   & 0.87   & -1.46   &  -0.24  &   1.054                     \\
\hline
\hline
\end{tabular}
\tablefoot{$\omega''_0$ is the initial angular velocity of each layer in the $S''$ reference 
frame; 
$\left <\omega''\right >$  is the average (over azimuth) angular velocity at the equator after
the run has advanced for 196 days.
$\Delta\omega''$=$|\omega''_{max}-\omega''_{min}|$ is the peak to peak amplitude of the angular velocity.
The radius listed in column 2 corresponds to the midpoint of each layer.}
\end{table}

\subsection{Input parameters}

The nominal input parameters are described in Table~\ref{table1}. The values for orbital period, 
masses, and primary star radius are the same as those that we used for the analysis 
of the $\alpha$ Virginis (Spica) system \citep{2009ApJ...704..813H,2016A&A...590A..54H, 2013A&A...556A..49P}.  
The nominal orbital eccentricity, however, is here adopted to be $e$=0 whereas the Spica system has
$e \sim$0.1.  Cases with this eccentricity are also presented below.  We chose for the nominal
synchronicity parameter $\beta_0^0$=1.8, which is somewhat smaller than the value $\beta_0^0$=2.07 that 
we previously employed for the primary star in Spica, but lies within the values that are possible 
for the system, given the uncertainties in the rotation velocity and radius that are derived from 
observations. The value $\beta_0^0$=1.8 corresponds to a surface equatorial velocity of 155\,km~s$^{-1}$.
Examples of results for $\beta_0^0$=0, 1.6, and 2.1 are also briefly discussed in the appendix. 

For the nominal case, we also adopt the value of kinematical viscosity $\nu$ used in 
our earlier investigations.  This parameter, however, merits special mention as it is the 
single input parameter with the largest uncertainty. As we have discussed previously
\citep{2016RMxAA..52..113K}, there is currently no clear criterion in TIDES for the choice of 
its value except for the fact that for ``small'' values the numerical integration is halted 
due to overlap of volume elements. This  condition  occurs primarily in the surface layer, since 
it suffers the largest amplitude perturbations.  For the binary system analyzed in this paper, 
``small''  means $\nu \leq$10$^{13}$ cm$^2$\,s$^{-1}$, and the value used in our nominal calculations 
is  1.6\,10$^{15}$ cm$^2$ s$^{-1}$. 

In general, $\nu$ depends on the amplitude of the shearing motions \citep{1999A&A...347..734R, 
2009ApJ...705..285P, 2018A&A...620A..22M}.  Because the perturbation amplitude due to tidal
effects increases toward the stellar surface, the value of $\nu$  should vary with radius
as modeled by \citet{2016CeMDA.126..249L}. For the same reason, \citet{1975ApJ...202L.135P} suggested 
that its value ought to depend on the stellar and orbital parameters. We performed a preliminary 
test of this suggestion in our analysis of the pre-eruption decline in the orbital period of the 
red nova V1309 Sco, where we assumed that tidal shear energy dissipation caused envelope expansion
removing energy from the orbit \citep{2016RMxAA..52..113K}. The results led to estimated viscosity values 
10$^{12} \lesssim \nu \lesssim$ 10$^{16}$ cm$^2$\,s$^{-1}$, the largest value corresponding to the 
shortest orbital period of the system prior to the outburst. 
It is also important to note that, in general, $\nu$ is  non-isotropic 
\citep{1992A&A...265..115Z, 2004A&A...425..243M}.
Hence, for the purposes of this paper in which we examine only the horizontal flows, our viscosity values 
will refer to the corresponding component.

The structure of the star is modeled as a polytrope, and the same equation of state describes the 
gas pressure exerted by a volume element on its neighbors. In our nominal case, we used $n$=3.
For addressing the shear instabilities (Sects. 5.1 and 5.2), we performed the TIDES calculation using
$n$=3.5 which gives a similar density structure to that of a rotating  main-sequence star in the
grid of \citet{2011A&A...530A.115B}. We also performed numerous experiments with $n$=1.5 driven 
by the suggestion that,  because the turbulent eddy turnover timescale is much shorter than the Kelvin 
thermal timescale for stars that depart even slightly from synchronization, turbulent processes are 
likely to prevail in the outer stellar layers which should thus be roughly an $n$=3/2 polytrope rather 
than the $n$=3 polytrope usually used for main sequence stars \citep{1975ApJ...202L.135P}.   However, 
this hypothesis has not been verified \citep{2014ARA&A..52..171O}. Sample results of models with this 
smaller polytropic index are illustrated in the appendix.

Each grid element in the TIDES computation may be viewed as if it were a parcel of stellar 
material that is enclosed by a 3D (fictitious) membrane of initial linear dimensions 
($\ell_r$, $\ell_\varphi$, $\ell_\theta$) and which contains an amount of mass that remains constant 
throughout the calculation.  After the start of the calculation, these linear dimensions change 
dynamically in response to the forces to which the element is subjected.  The nominal computational 
grid size is ($N_r$, $N^{eq}_\varphi$, $N_\theta$) =(5, 200, 20), where $N_r$ and $N_\theta$ are, respectively, 
the number of grid elements in the radial and polar direction, and   N$^{eq}_\varphi$ is the number of grid 
elements in the azimuthal direction at the equator.  Once N$^{eq}_\varphi$ is specified, the number of elements at 
other colatitudes, N$_\varphi$($\theta$) is determined based on the initial condition that all elements 
possess the same initial values of ($\ell_r$, $\ell_\varphi$, $\ell_\theta$).  This means that  
N$_\varphi$($\theta$) decreases with colatitude.   The list of colatitudes and corresponding number of grid 
elements may be found in Table \ref{table_parsnum} for the nominal case in which $N^{eq}_\theta$=20.  
For the nominal case, the initial values of ($\ell_r$, $\ell_\varphi$, $\ell_\theta$)=(0.41, 0.21, 1.08),
listed here in units of \,R$_\odot$.  The number of elements in radial shells below the surface is the same
at each colatitude as at the surface.

The input grid size was chosen so that $\ell_r$, $\ell_\varphi$, and $\ell_\theta$ do not
differ significantly from one another and also to properly sample the time-dependent behavior 
of the azimuthal motions. In the radial direction, there is a very slow and monotonic decline 
in the amplitude of  perturbations with decreasing radius, except very near the surface, making 
a large radial grid unnecessary.  Clearly, a larger radial grid size is possible but is 
computationally expensive. In the latitudinal direction, the computation is performed only for the 
hemisphere contained in colatitudes 0$^\circ$ to 85$^\circ$, under the assumption that there is 
north-south symmetry, and the decreasing amplitude of perturbations toward the poles justifies
a relatively small grid size.  This is not the case for the azimuthal grid. While it is true that 
the topology of the equilibrium tide can be described by a low-order  Fourier mode, we have found
that the dynamical perturbations have significantly higher modes and a large grid in the azimuthal 
direction is needed to resolve these oscillations.  

The other computational parameters are  the thickness of each layer ($\Delta R/R_1$), the number of 
orbital cycles over which the computation is performed ($Cy$)  and the tolerance for the Runge-Kutta 
integration.  The value of the layer thickness is constrained to lie within the interval
0.02$<\Delta R/R_1<$0.1 to guarantee relatively small density and pressure gradients in the grid
elements.  Our nominal value is $\Delta R/R_1$=0.06, which is approximately two times larger than
the extension of the tidal bulge with respect to the unperturbed radius.  The nominal number of
orbital cycles over which the time-marching algorithm operates was chosen to be 100, which is 
sufficient time for the calculations with the nominal input parameters to attain the stationary
state.  However, cases with larger viscosity values may require shorter times.  

We list in Table \ref{table_models_nominal} the input parameters of the model runs that are discussed
in the main section of this paper.  The case number in Col.~1 will be used to refer to individual models.  
Col.~2 indicates the value of the core asynchronicity parameter $\beta_0^0$, Col.~3 the number of 
layers, and  Col.~4 the number of cycles over which the model was run.  Col.~5 indicates the initial 
rotation condition; for example, initiated with a uniform rotation or a differential rotation structure.
The initial differential rotation structure is described in Cols.~1-4 of Table~\ref{table_case34_non-uniform}. 
The full set of models is listed in Table \ref{table_models}.  We make occasional reference below 
to some of these. 

The deformation in the radial direction of the surface layer can be estimated with a one-layer TIDES computation.
For the $\beta_0^0$=1.8 nominal case, this yields a maximum value $\sim$0.03R$_\odot$ (0.4\%) at the equator. 
We find that the neglect of the radial deformation in the $n$-layer TIDES calculation for the same nominal case
leads to an under-estimate of the angular velocity around the sub-binary longitude and at 
180$^\circ$, as illustrated in  Appendix~\ref{app_compare_1D} (Fig.~\ref{fig_compare_TIDES1and2}). 
Furthermore, this deformation is equivalent to the size of the $\Delta$R/R$_1$=0.03 layer thickness that is
chosen for some of our computations, and thus, in these cases, the angular velocity behavior near the surface layer 
should only be taken as approximate.
 
\subsection{Output variables}

TIDES is a time-marching algorithm that outputs the instantaneous angular velocity,
angular momentum, and tidal shear energy dissipation rate.
The variable of interest in this paper is the angular velocity measured in the 
frame of reference that is  rotating with the underlying core, $S''$.  This variable is
denoted $\omega''$ throughout this paper.

Angular velocity in single stars with a differential rotation structure is generally dependent only 
on radius and polar angle.\footnote{It is also dependent on time but on significantly longer timescales 
than those considered in this paper.}  In the asynchronous binaries analyzed in this paper, however, it 
is a function of the three spatial coordinates and time, $\omega''$=$\omega''(r,\theta,\varphi',t)$. 
We express it as:

\begin{equation}
\omega''=\left<\omega'' \right>+\delta\omega''
\label{eq_omdp}
\end{equation}

\noindent where 

\begin{equation}
\left<\omega''\right>=\left<\omega''(r,\theta,t)\right>=\frac{1}{2\pi}\int_0^{2\pi}\omega''(r,\theta,\varphi',t)~d\varphi.
\label{eq_average}
\end{equation}

\noindent is the azimuthally averaged angular velocity  and $\delta\omega''$  represents 
the tidal velocity in the azimuthal direction. Average differential rotation occurs 
when the radial gradient of $\left<\omega''\right>$ is nonzero.   

In this paper, we are interested primarily in the dependence of $\omega''$ on radius at a fixed time
$t_i$ and at fixed colatitude $\theta_i$. The function $\omega''(r, \theta_1,\varphi', t_1)$  will 
be referred to as the $\omega''$ profile.  When also analyzing its dependence on azimuth angle,
we use the term meridional $\omega''$ profile to describe it.

For the calculations that are performed with the smallest grid size ($N_r, N_{\varphi'}, N_{\theta}$)=(5, 200, 20),
the output consists of $\sim$10$^4$ data points at each timestep. The typical timestep is $\sim$1~minute and the 
computation is performed over a time span of up to several thousand days. To obtain an overview of the
temporal evolution of this relatively large data set, we define

\begin{equation}
\omega_{max}''(r,\theta,t)=\pm {\rm max}\left(|\omega''(r,\theta,\varphi',t)|\right)
\label{eq_minmax}
\end{equation}

\noindent where $\omega''$($r$,$\theta$,$\varphi'$,$t$) represents the set of angular velocities 
at a fixed radius $r$, colatitude $\theta$, and time. 
The evolution  over time of $\omega_{max}''$ provides information on the maximum value of angular 
velocity and also provides information on the duration of the initial transitory state in the 
calculation.

The sign of $\omega''_{max}$ can be positive or negative, with negative values corresponding 
to motion opposite to the positive direction in the $S''$ reference frame. However, because 
this variable is used primarily to determine when a particular calculation has attained the 
stationary state, its abolute value generally suffices.

\subsection{Initial conditions and transitory state}

The nominal initial condition is one in which the primary is unperturbed and in uniform
rotation at a rate $\omega_0$.  When it is subjected to the companion's gravitational force,
it undergoes a transition as it adjusts to the new condition and during which large amplitude
motions are excited.  These are damped down over time until the system attains the stationary state.
In this state, the maximum amplitude remains constrained to within a small range of angular velocities.
The duration of the transitory state depends on the particular set of input parameters, and may 
last from ten to several hundred orbital cycles.  Examples illustrating the evolution from the
transitory to the stationary state may be found in the appendix.

The TIDES computation can also be started with an arbitrary initial rotation structure, instead
of uniform rotation as the initial condition. In this  case, each layer is assigned an initial value of 
the synchronicity parameter $\beta_0^k$,  where $k$=0, 1, ... is a number that identifies the $k$-th 
layer, with $k$=0 corresponding to the core and $k$=1 corresponding to the layer that interfaces 
with the core.  Such a computation is useful in assessing  the temporal evolution of an initial 
differential rotation structure, an example of which is described in Appendix \ref{sect_diffrotation}.

\subsection{Stability of synchronous rotation}

Synchronous rotation in a circular orbit in which the rotation and orbital planes coincide 
corresponds to the equilibrium state.  In this state, $\beta_0^k$=1;  that is, all layers are 
in uniform rotation and synchronous with the binary orbit.  This is  a well-known classical
result \citep{1973Ap&SS..23..459A, 1981ApJ...246..292S}.  As illustrated in Appendix \ref{appendix_method} 
the TIDES calculation reproduces to great precision this result, showing that the numerical 
integration scheme used in the TIDES calculations is stable and produces results that are 
consistent with theory.  This is particularly significant because the surface rotation velocity 
of our synchronously rotating star is $\sim$85\,km\,s$^{-1}$, a velocity at which Coriolis effects
are non-negligible, and yet the numerical simulation converges to the equilibrium configuration.

\section{The $\omega''$ profile in asynchronous circular orbits}

The internal rotation structure of all stars, except for our Sun, is unknown.  Thus, the rotation in 
single stars is generally analyzed in terms of two general types: uniform rotation, in which the angular 
velocity is constant, and differential rotation, in which it depends on radius and, potentially, also on 
latitude.  The usual assumption for binary stars is that they are in uniform rotation, based on the assumption 
that they are tidally locked.  If a binary star is not tidally locked, the tidal interaction produces a 
velocity field in which the azimuthal velocity component generally dominates \citep{1981ApJ...246..292S, Dolginov:1992up,  
2009ApJ...704..813H}.  Thus, the actual rotation structure includes a contribution from this tidal velocity field.

In this section we analyze the behavior of the rotation velocity in a short-period asynchronously rotating binary 
system in a circular orbit. We first present the results of our calculations for a circular orbit, a very small 
departure from synchronicity and an initial uniform rotation structure in order to compare them with the analytical 
solution that has been derived under those conditions.  We then address the case of a large departure from 
synchronicity and an initial uniform rotation (Sect.~\ref{sec_uniform}) and finally that of an initial differential
rotation structure (Sect.~\ref{sec_differential}).  The different models are referred to by Case number as listed
in Col.~1 of Table~\ref{table_models_nominal} where the input parameters are also specified.

As a general comment, we note that because the effects due to the tidal interaction are strongest at and around the 
stellar equator, we mainly present in what follows the results for this latitude although the computations cover 
up to within 5$^\circ$ of the pole.  Also, unless noted otherwise, radial distances from the stellar center always 
refer to the midpoint of the layers that are modeled.

\begin{figure}
\centering
\includegraphics[width=0.48\columnwidth]{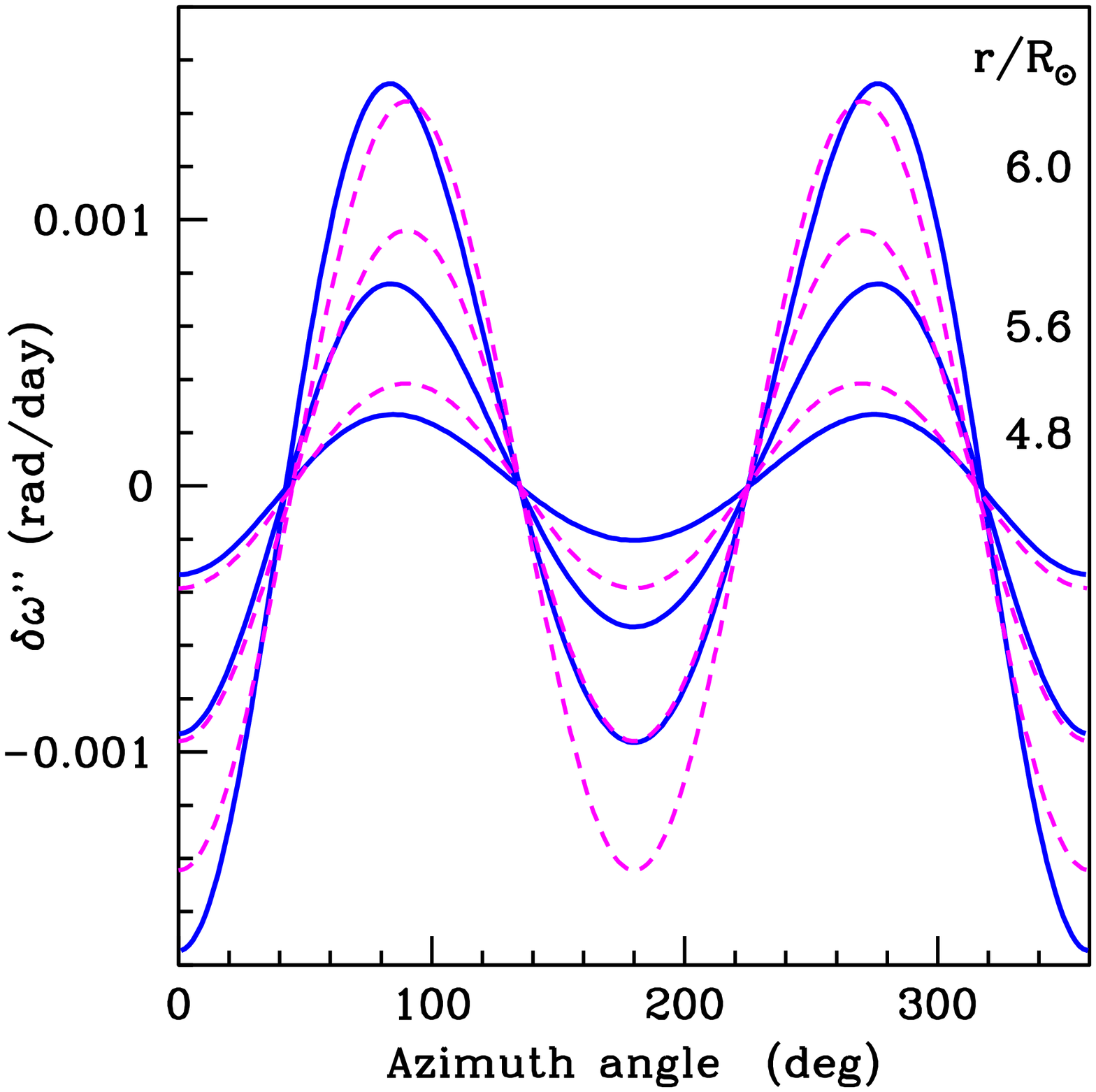}
\includegraphics[width=0.48\columnwidth]{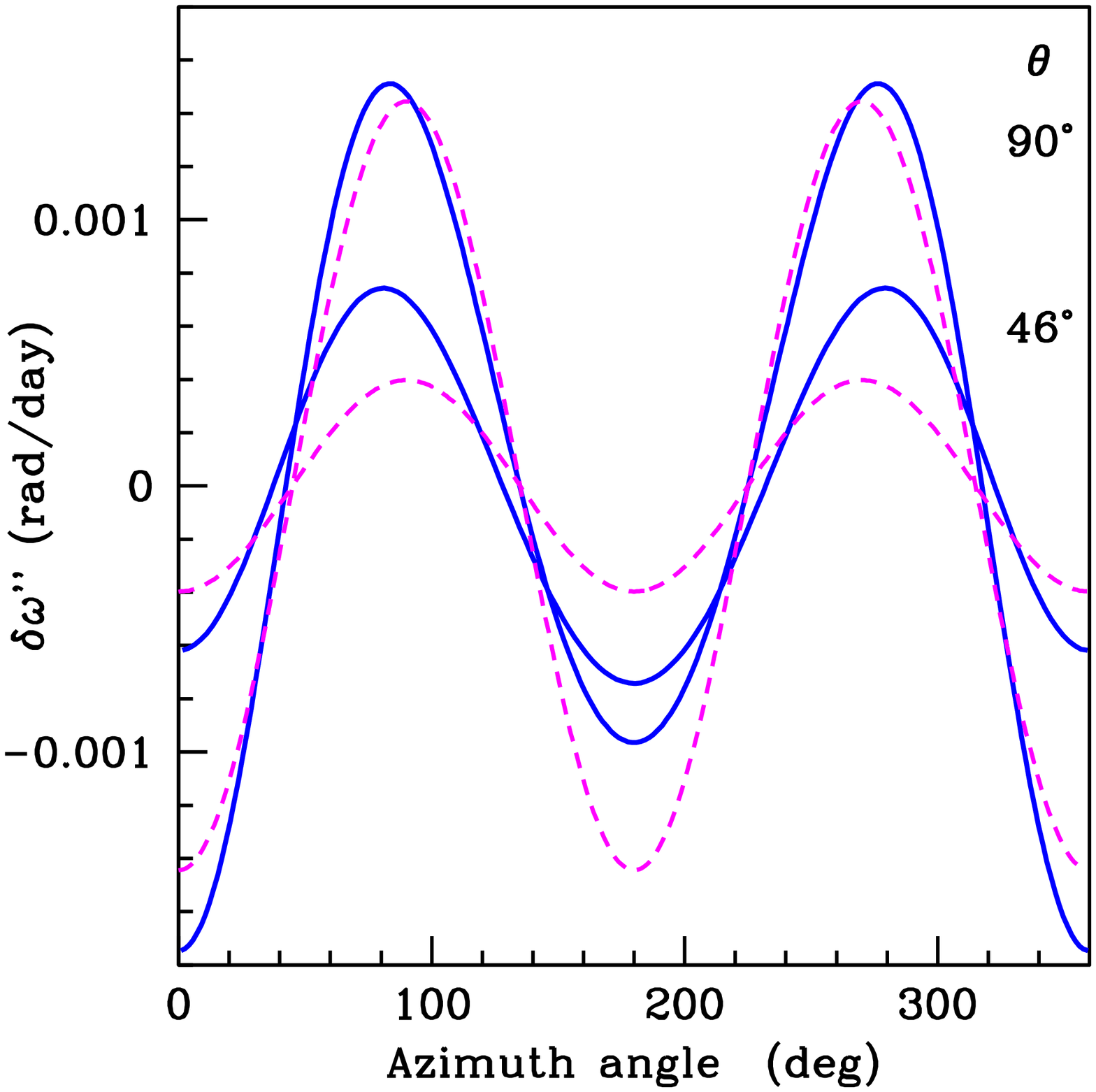}
\caption{Tidal velocity in the azimuthal direction $\delta\omega''$ computed with the $n$-layer TIDES numerical
simulation (Case 30 in Table~\ref{table_models_nominal}; continuous lines) compared to the analytical expression
(Equation~\ref{eq_scharl}). The abscissa is the azimuth angle measured from the sub-binary longitude.
{\it Left:}  Equator ($\theta$=90$^\circ$) at three radii as indicated in the legend. {\it Right:}
Layer at radius 6 R$_\odot$  at colatitudes $\theta$=90$^\circ$ and 46$^\circ$.
\label{fig_propaz_beta1.01}}
\end{figure}

\subsection{Comparison with the analytical solution \label{sec_scharl1}}

\citet{1981ApJ...246..292S} derived an analytical expression for the tidal velocity field 
of a star in a circular orbit with a very small departure from the synchronous rate. 
Using the same rotating reference frame as our $S'$ system, in his Eq.~8 he defines the total velocity  
\mbox{\boldmath $U$} =\mbox{\boldmath $V_{rot}$}  + \mbox{\boldmath $V$} +\mbox{\boldmath $W$},
where \mbox{\boldmath $V_{rot}$} is the rotation velocity, \mbox{\boldmath $V$} is the
tidal velocity, and \mbox{\boldmath $W$} the velocity field due to meridional circulation.  
Neglecting meridional circulation, we can write \mbox{\boldmath $U$}=\mbox{\boldmath $\vv'$}, 
where \mbox{\boldmath $\vv'$} is the velocity field defined in our Equation~\ref{ecmov}.  Hence, 
we can write Scharlemann's tidal velocity field as 
\mbox{\boldmath $V$}=\mbox{\boldmath $\vv'$}$-$\mbox{\boldmath $V_{rot}$}. It is important to
note that the \mbox{\boldmath $V_{rot}$} is a function of radius, thus allowing for differential 
rotation.  In our notation, \mbox{\boldmath $V_{rot}$}=$r \sin(theta) <\omega'>$,  where $<\omega'>$ is the 
average rotation angular velocity measured in the $S'$ reference frame.  It is then 
straightforward\footnote{We can write 
$V_\varphi$=$\vv_\varphi'$-$V_{rot}$= r$\sin(\theta)$$(\omega'-<\omega'>)$. For a circular orbit, 
$\omega'$=$\omega'' + \omega_0 - \Omega_0$. Using this relation (which also applies to $\omega_0'$)  
and the definition in Equation~\ref{eq_omdp},
$V_\varphi$=r$\sin(\theta)$$(\omega''-<\omega''>)$=r$\sin(\theta)$$\delta\omega''$,  since $\omega_0''$=0 by definition.}
to show that the azimuthal component of \mbox{\boldmath $V$} is $V_\varphi$ =$r \sin(\theta)\delta\omega''$,
which is given by Scharlemann in his Equation. 40 as:

\begin{eqnarray}
V_{\varphi}&=&-15 \Omega_0 R_1 f_* \omega_{0*} \left(\frac{r}{R_1}\right)^4 \sin^4\theta \cos (2\varphi')
\label{eq_scharl}
\end{eqnarray}

\noindent where $\theta$ is the colatitude and $\varphi'$ is the longitude measured in
the direction of rotation from the line between the stellar centers, $\Omega_0=2\pi/P$ is 
the angular velocity of the binary and R$_1$ is the stellar radius. The parameter 
$\omega_{0*}$ is a measure of the departure from synchronicity, in the reference frame that 
is rotating with angular velocity $\Omega_0$.  In our notation, $\omega_{0*} \simeq \beta_0-1$.  
The parameter $f_*$ is a measure of the tidal amplitude at the stellar surface and is typically 
given by $f_*\simeq\frac{m2}{m1}\left(\frac{r}{a}\right)^3$ \citep{2008EAS....29...67Z} with $a$ 
the orbital separation.  We use this expression in Eq.~\ref{eq_scharl}, but it is worth noting that 
Scharlemann (1981) adopts a value four times smaller for this tidal deformation parameter, while 
others use a value that is $\sim$1.5  times larger \citep{2019A&A...629A.142V}.
Eq.~\ref{eq_scharl} can be compared directly with the values of $\delta\omega''$ that 
are obtained from our numerical simulations with TIDES. As an example, the results of Case 30 
(see Table~\ref{table_models_nominal}) are plotted in Fig.~\ref{fig_propaz_beta1.01} and compared with 
$V_\varphi (r)/r \sin(\theta)$ from Equation~\ref{eq_scharl}. 

There are three main differences between our numerical calculation and Eq.~\ref{eq_scharl}.
First is the degree of symmetry with respect to the line connecting the stellar centers ($\varphi'$=0).  
Equation~\ref{eq_scharl} predicts equal tidal bulge amplitudes while TIDES predicts an asymmetry.  We
note that the tidal distortion is symmetric only if R$_1$/$a<<$1, where $a$ is the orbital separation, a 
condition that is generally used for calculations in the ``weak tides'' regime. Our test binary system does
not satisfy this condition.  Thus, the tidal perturbation at the sub-binary longitude is stronger than at 
the anti-binary longitude ($\varphi''$=180$^\circ$).  

The second difference is the  phase shift between the maxima of the curves shown in
Fig.~\ref{fig_propaz_beta1.01}.  This is a likely consequence of the large viscosity in the TIDES
calculation.  It is interesting to note that the effect of viscosity is mainly to introduce
a phase shift while not significantly affecting the tidal amplitude, a result that has also been shown to
hold for tidal flows  in  subsurface oceans of Solar System  moons \citep{2013PhDT.......256C}. 
The third difference is more evident in the deeper layers, for which Equation~\ref{eq_scharl}
overpredicts the amplitude.  This is likely related to the assumed stellar structure, since the
departures from the analytical result depend on the polytropic index used to perform the numerical 
simulation.  However, the amplitudes predicted by Equation~\ref{eq_scharl} are seen here to differ only 
by factors $\sim$1.5-2 (excluding the external layer) which is within the range of uncertainties 
implied by the range in adopted tidal deformation values $f_*$  mentioned above.

Finally, it is important to note that the radiation transfer effects on the tidal velocity field
are generally neglected in our numerical approach and the analytical method of Scharlemann (1981). 
This simplification may lead to an excessively large response of the surface layer to the tidally forced 
oscillations. The excess amplitude was shown by \citet{1975A&A....41..329Z} to be damped down when radiative 
dissipation is taken into account. This process  could be incorporated in future versions of our model.

\begin{figure}
\centering
\includegraphics[width=0.48\columnwidth]{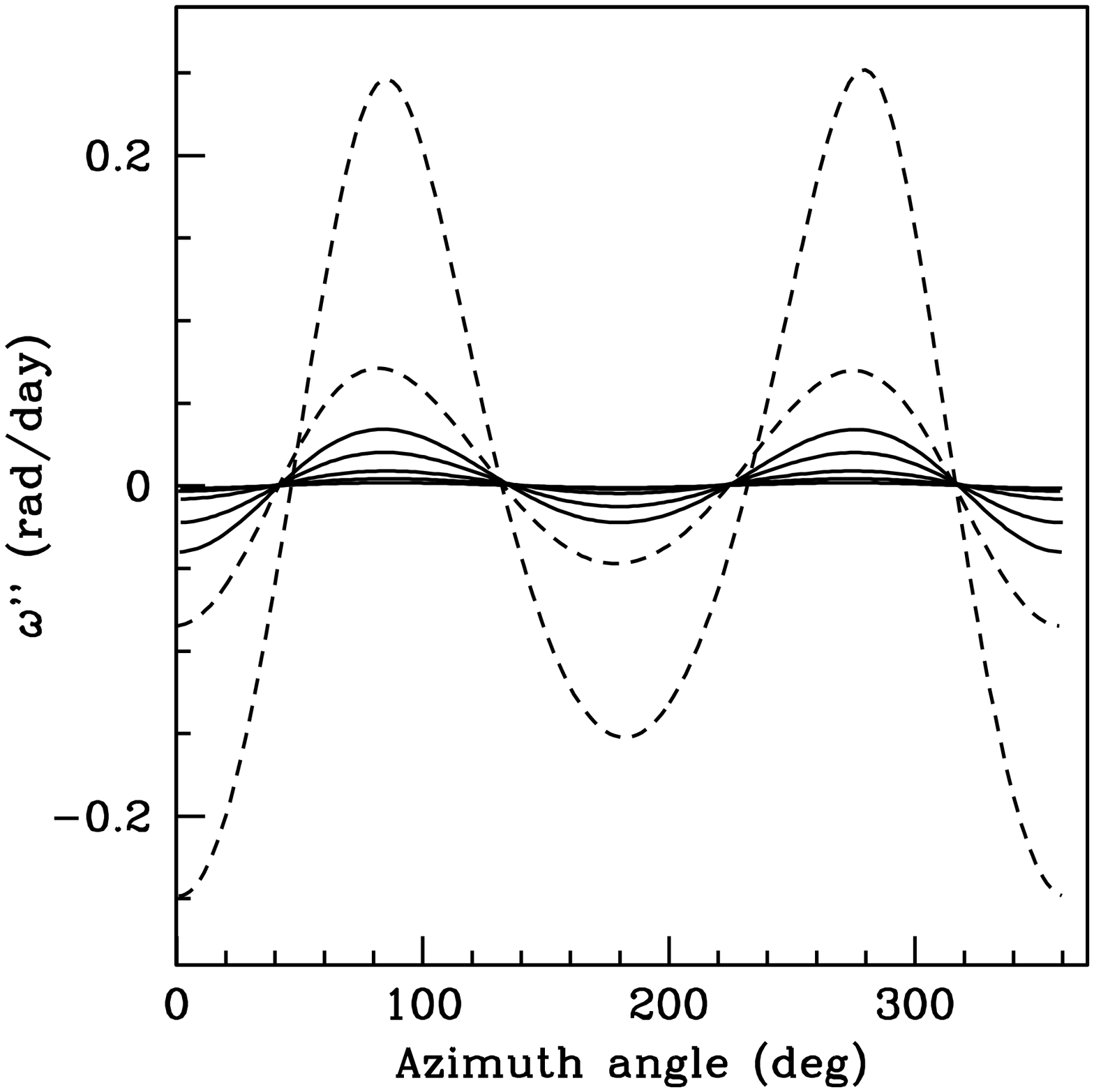}   
\includegraphics[width=0.48\columnwidth]{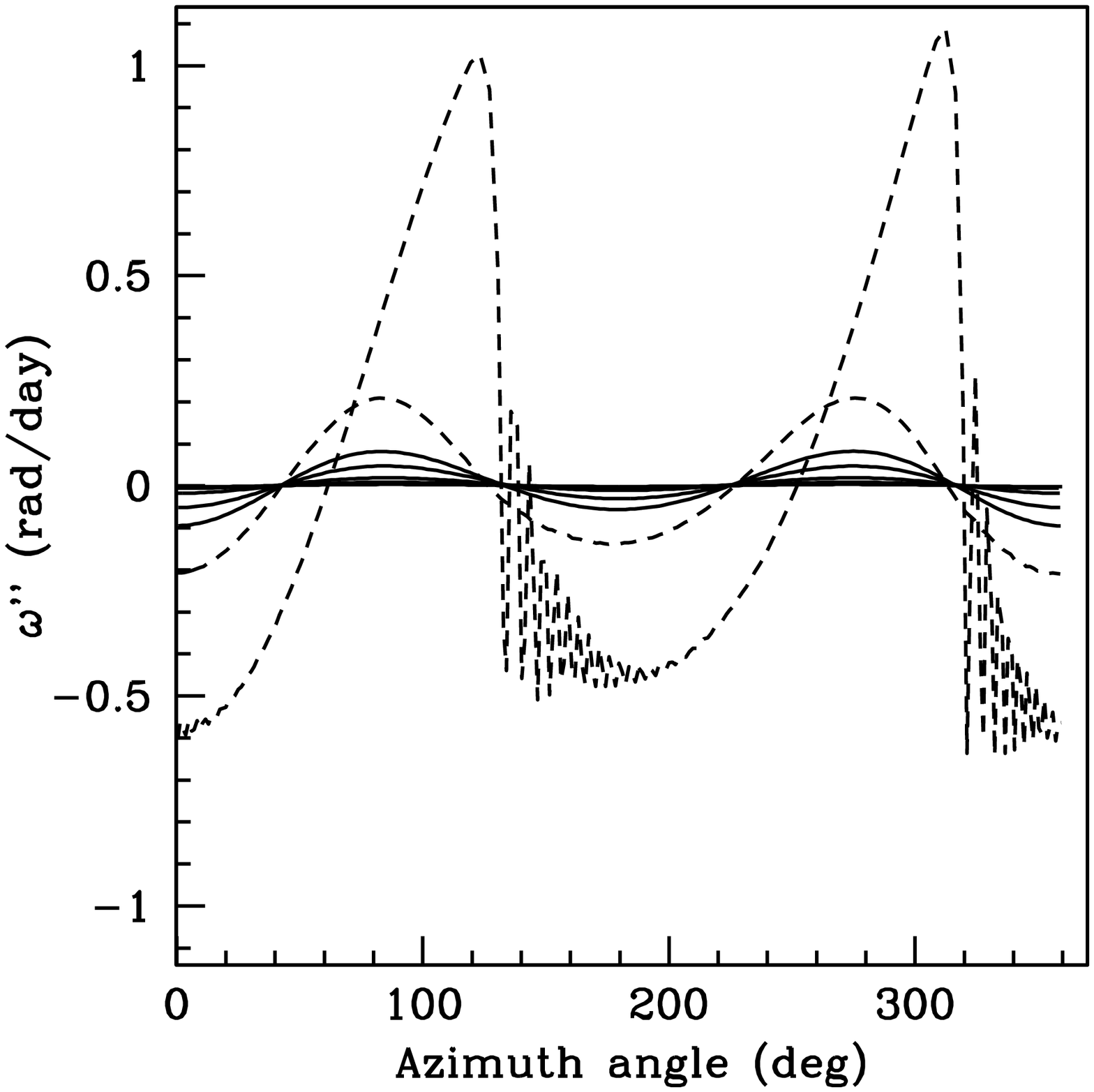} 
\includegraphics[width=0.48\columnwidth]{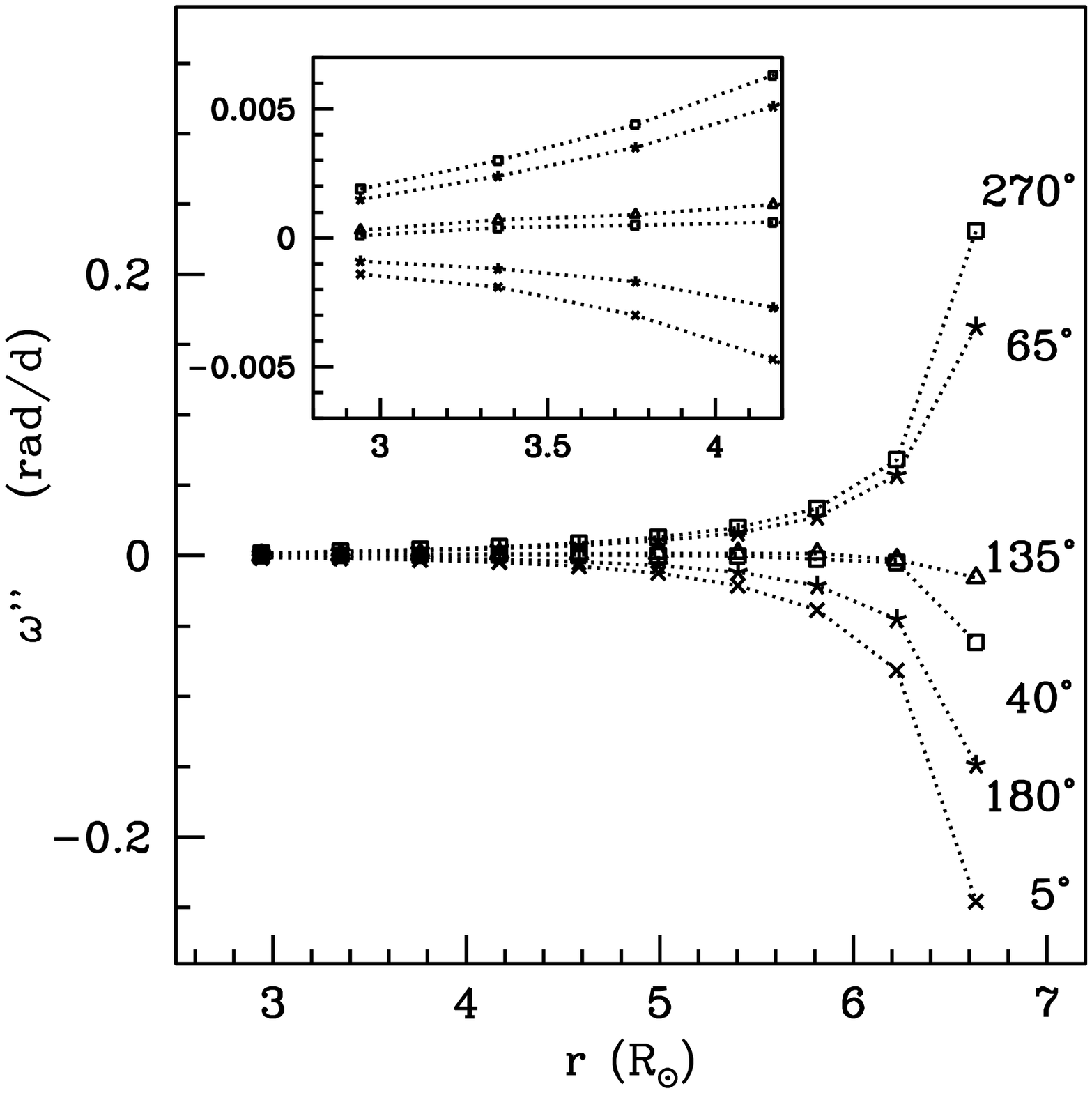}    
\includegraphics[width=0.48\columnwidth]{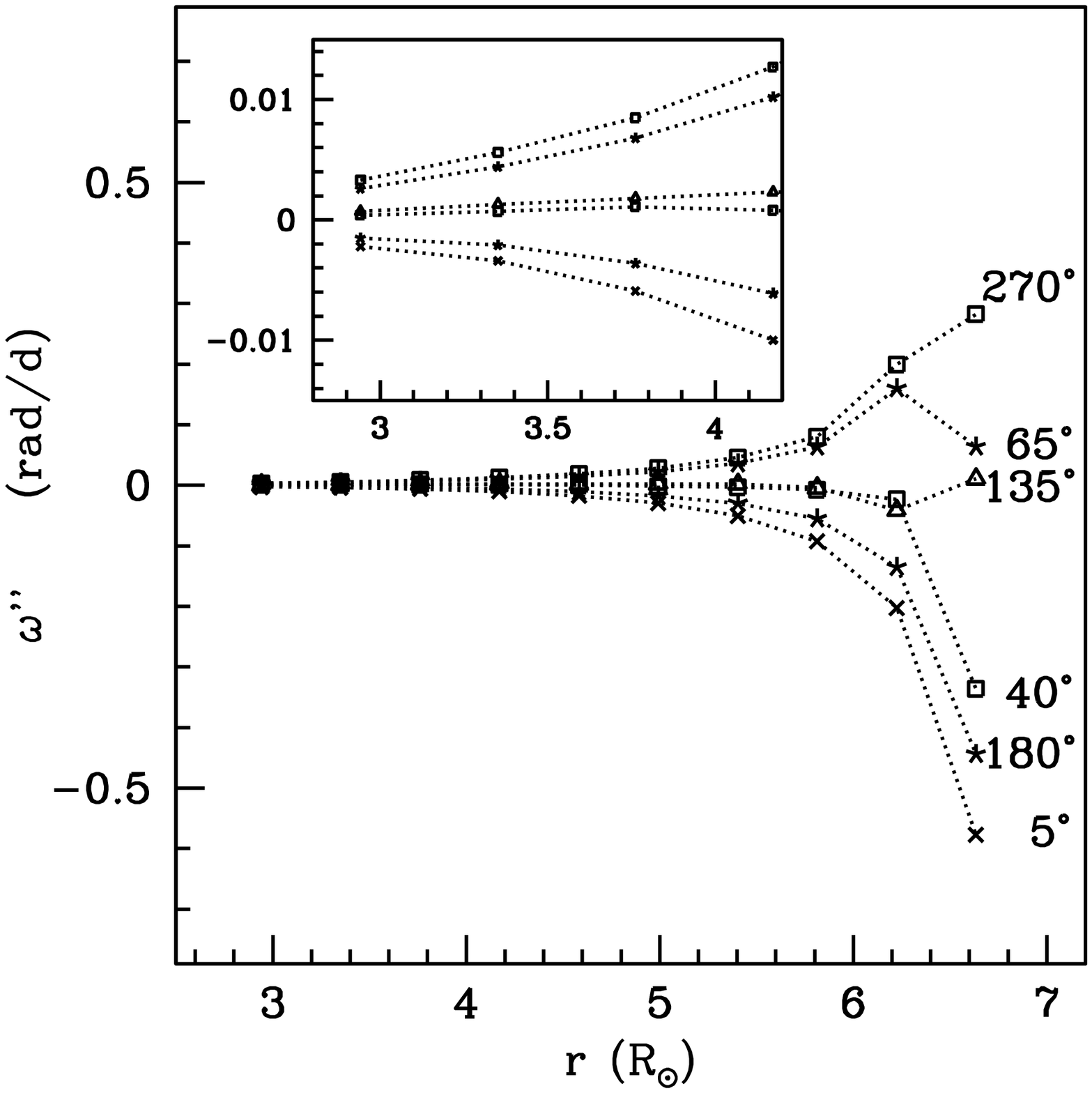}
\caption{Results for supersynchronous rotation in a circular orbit for polytropic indices  $n$=1.5 
(left, Case 3) and $n$=3 (right, Case 31). {Top:} Angular velocity ($<\omega''>$+$\delta\omega''$) at the 
equator in the ten layers used for this calculation.  Dashes correspond to layers closest to the surface, 
$r/R_\odot$=6.65 and 6.22. The abscissa is azimuth angle measured in degrees with $\varphi'$=0 
corresponding to the sub-binary longitude.
{\it Bottom:} Meridional $\omega''$ profiles at the equator for a selection of meridians $\varphi'$ as 
indicated by the labels. The abscissa is the radius at the midpoint of each  layer.
The behavior for $r/R_\odot$<5 is shown on an amplified scale in the inset.
}
\label{fig_om_and_gradient}
\end{figure}

\begin{figure}
\centering
\includegraphics[width=0.48\columnwidth]{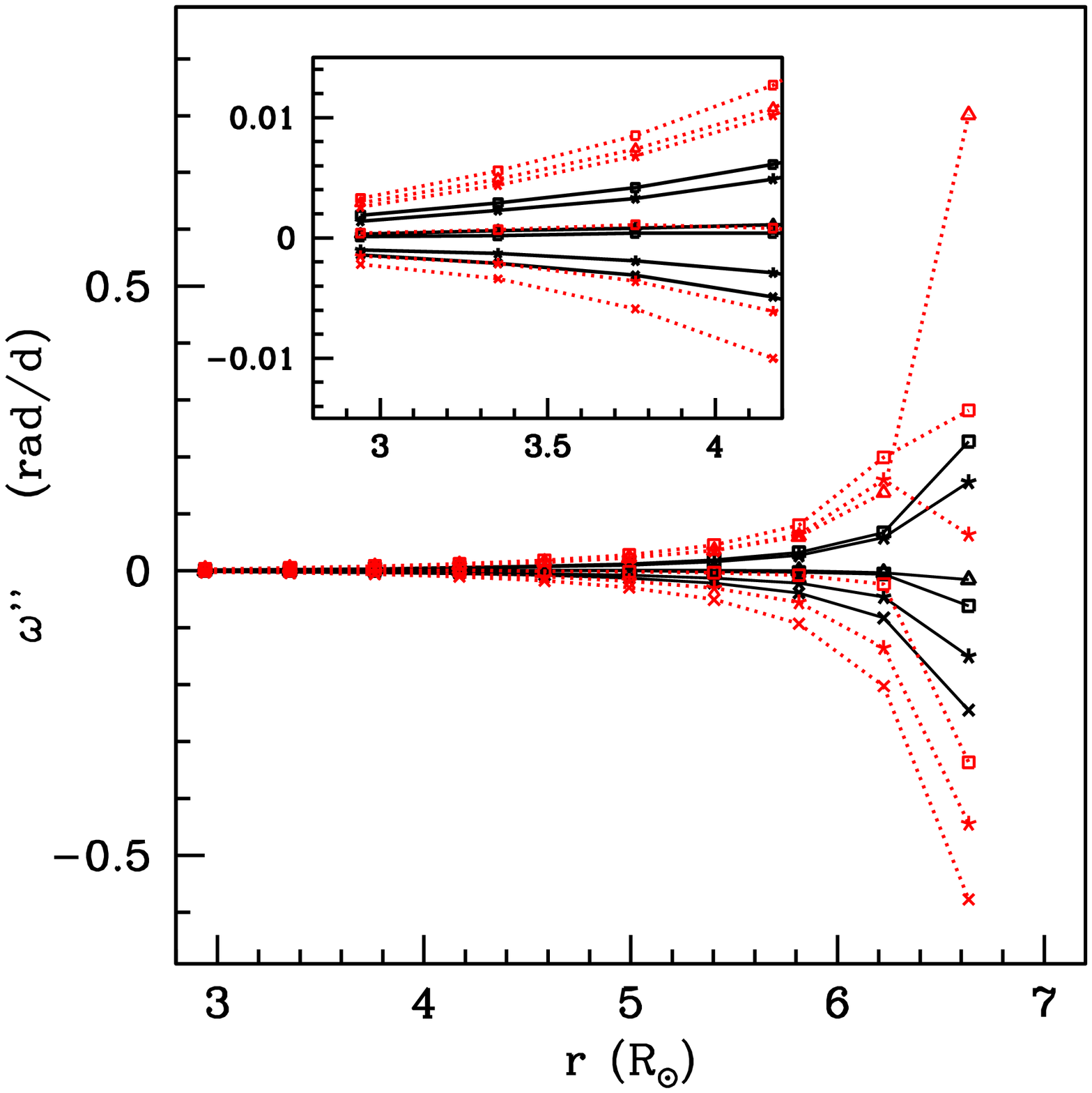}
\includegraphics[width=0.48\columnwidth]{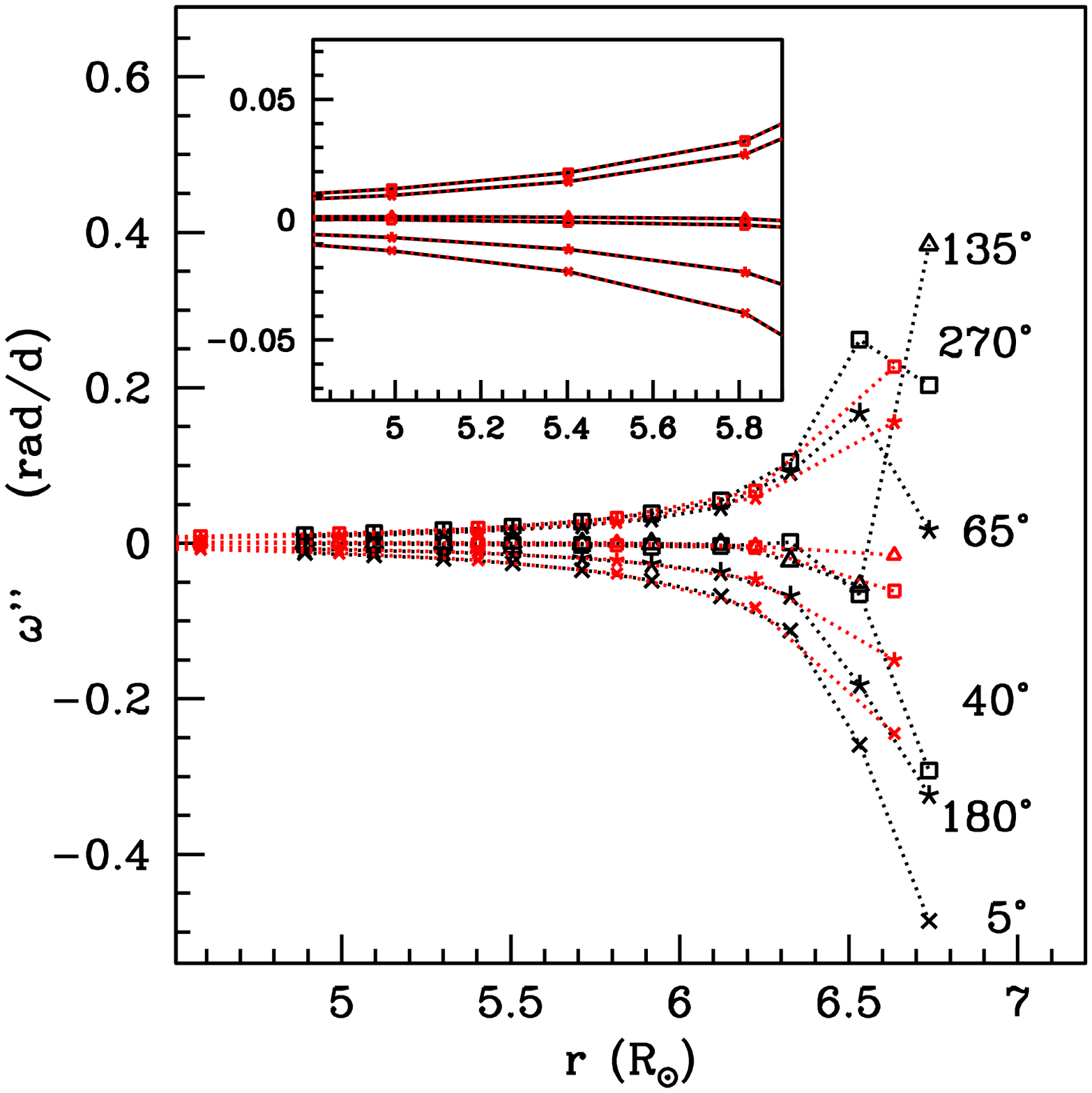}
\caption{Comparison of the meridional $\omega''$ profiles: {\it Left:} the polytropic indices $n$=1.5 (Case 3)
and $n$=3 (dots, Case 31); {\it Right:}  For layer thicknesses $dR/R_1$=0.03 (black, Case 11) and 0.06 (red, Case 3).
 The angles that are labeled correspond to Case 11.
}
\label{om_gradient_compareIND}
\end{figure}

\begin{figure}
\centering
\includegraphics[width=0.48\columnwidth]{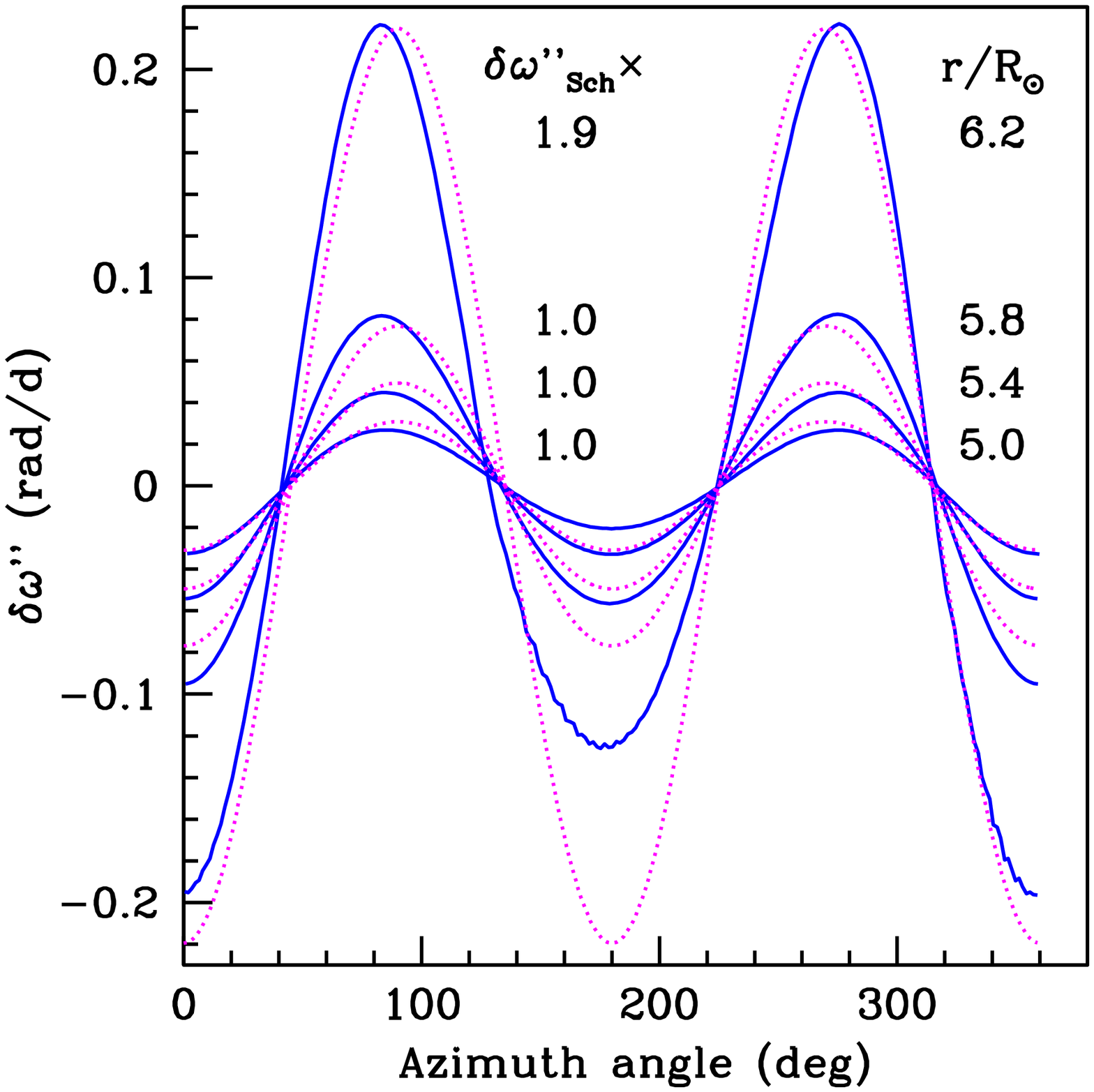}
\includegraphics[width=0.48\columnwidth]{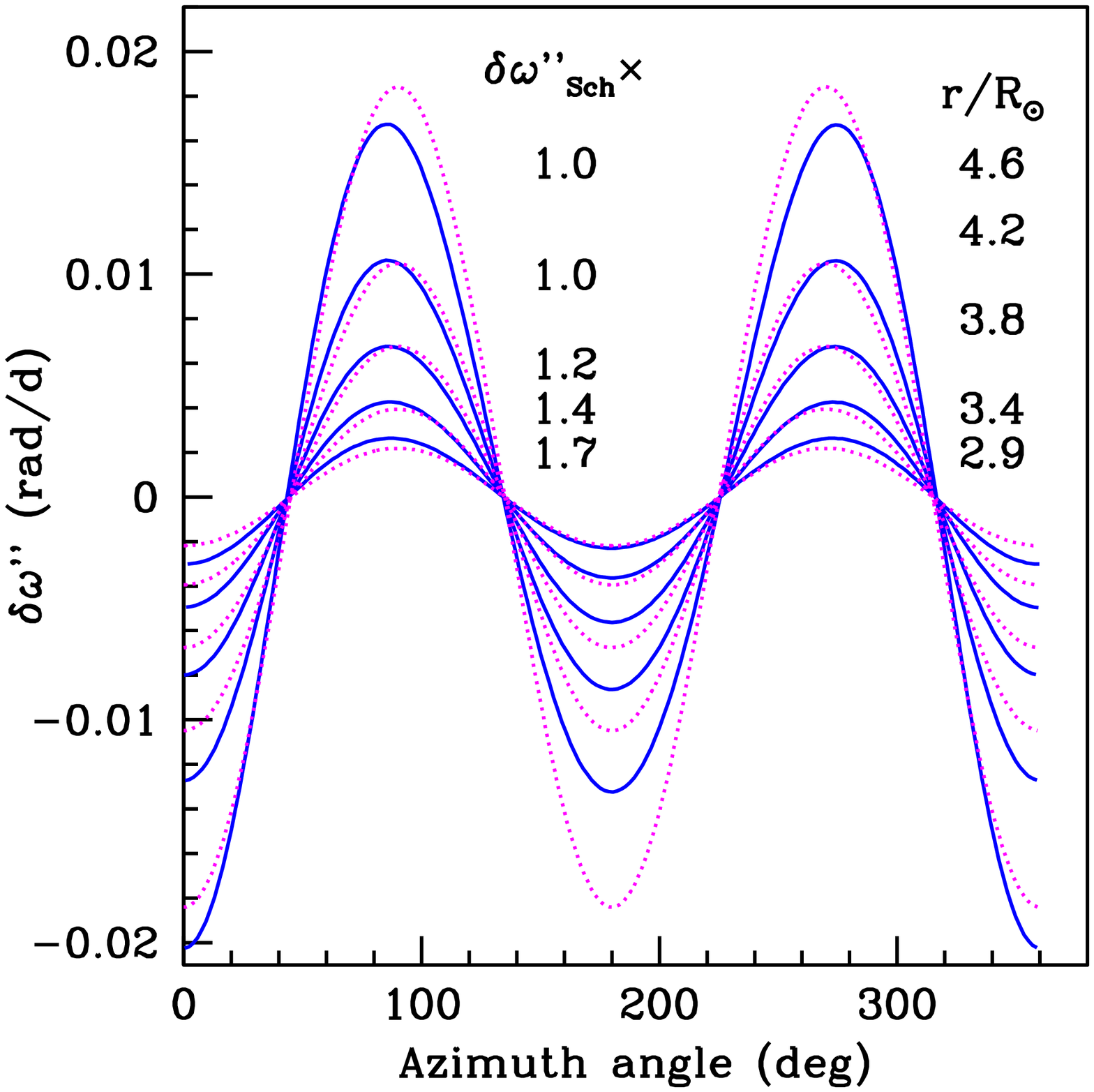}
\caption{Tidal angular velocity $\delta\omega''$ computed with the $n$-layer TIDES numerical
simulation (Case 31 in Table~\ref{table_models}; continuous lines) compared to the scaled analytical expression
(Equation~\ref{eq_scharl}). The scaling for each radius is listed in the center and the radius on
the right side of each panel. The abscissa is the azimuth angle measured from the sub-binary longitude.
{\it Left:}  Layers closest to the surface. {\it Right:} Layers closest to the core.}
\label{fig_propaz_beta1.8_scaled}
\end{figure}
 
\subsection{Initial uniform rotation \label{sec_uniform}}

The general characteristic of the angular velocity of a perturbed star in a circular orbit whose core 
rotates with a velocity that is 1.8 times that of the orbital angular velocity is illustrated in the 
top panels of Fig.~\ref{fig_om_and_gradient}.  The rotation angular velocity  as measured in the 
star's rest frame, $\omega''$, is plotted as a function of azimuth angle $\varphi'$, where  
$\varphi'$=0 corresponds to the sub-binary longitude.  
This plot shows a similar sinusoidal morphology as that illustrated in Fig.~\ref{fig_propaz_beta1.01}, but
with a significantly larger amplitude.  The left panels correspond to Case 3 in Table~\ref{table_models}, which 
is computed with a polytropic index $n$=1.5 and the right panels to Case 31, which is computed with $n$=3 and 
show that the response to the tidal force is influenced by the internal stellar structure.
The reason for this is that the models with larger polytropic indices are more centrally condensed,
which means that the density declines more rapidly with radius. Hence, the restoring action of
gas pressure is weaker than in the $n$=1.5 models allowing the volume elements to attain faster
velocities.

The faster velocities that are evident in the surface layer of the $n$=3 model cause two additional
effects.  The first is a significantly larger phase lag in the outer layers, compared to a smaller
polytropic index model.  The second is the appearance of high-frequency oscillations, most evident
in the surface, which results from the faster tidal velocities, leading to nonlinearities
in the system.

As it is evident that the shape of the $\omega''$ profile changes from one meridian to the next,
we use the term meridional $\omega''$ profile to describe the dependence of $\omega''$ over radius at a 
fixed azimuthal angle $\varphi'$.  This is illustrated in the bottom panels of Fig.~\ref{fig_om_and_gradient}
where we plot a selection of the meridional $\omega''$ profiles.  This representation is useful, for example, 
for comparing the velocity structure of different models, as in Fig.~\ref{om_gradient_compareIND} (left)
which shows that the larger polytropic index allows for steeper meridional $\omega''$ profiles.  The right panel of 
this figure shows the manner in which the results are affected by the choice of the layer thickness.  For the 
internal layers, the angular velocity is relatively unaffected by this parameter (Fig.~\ref{om_gradient_compareIND},
right).  However, the velocity of the surface layer (and on occasion, the layer below it) does depend on
this parameter.  This is because the radius of the surface grid point (always chosen as the midpoint of the layer) 
is larger for a thinner layer, and is thus subjected to a larger external force than the midpoint of a thicker 
layer.  A second (nonphysical) factor is a consequence of the computational limitations.  Specifically, the 
time that is required for the outer layer to attain the stationary state is significantly longer than that
of the inner layers.  Thus, unless the objective is to model the behavior of the surface (for example,
for the calculation of spectral line profile variability), it is not cost-effective to run the
simulation beyond the time for which the inner layers have attained the stationary state.             
The dependence of $\omega''$ on other input parameters is analyzed in the appendix.

Given the convenience of an analytical representation for the tidal velocity field, the natural question
that arises concerns the limits of applicability of Equation~\ref{eq_scharl}.  This is explored in 
Fig.~\ref{fig_propaz_beta1.8_scaled}, which again shows the result of our Case 31 numerical simulation
compared to $\delta\omega''$ obtained from Equation~\ref{eq_scharl}, but scaled so as to compare more favorably
with the numerical simulation. We find that the required scaling is less than a factor of $\sim$2 for all layers
except the stellar surface.  This suggests that Equation~\ref{eq_scharl}  may be used as a rough first 
approximation to the tidal velocity field for circular orbits in which the average velocity structure is
close to uniform.

\subsection{Initial differential rotation  \label{sec_differential}}

\begin{figure}
\centering
\includegraphics[width=0.49\columnwidth]{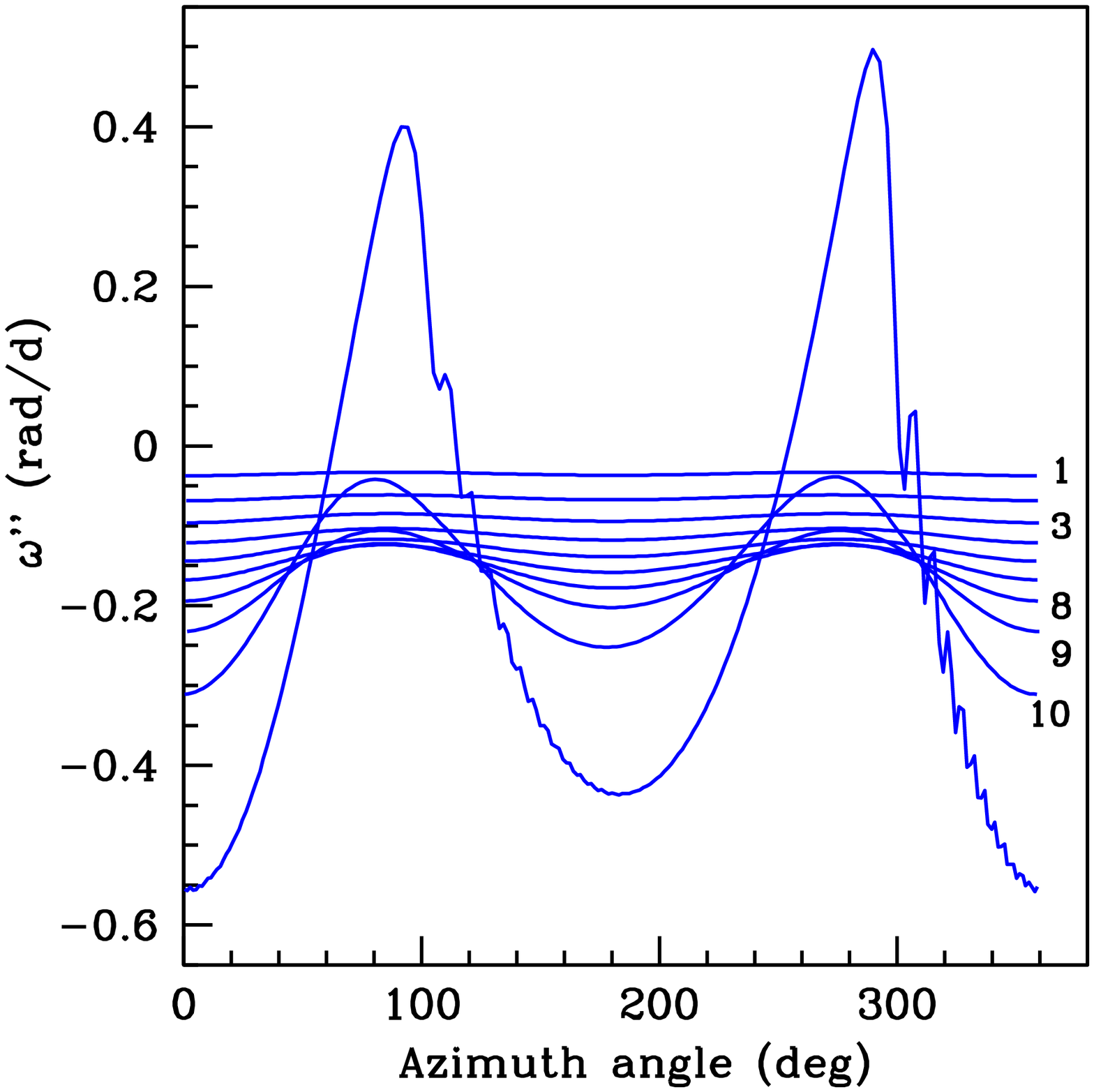}          
\includegraphics[width=0.49\columnwidth]{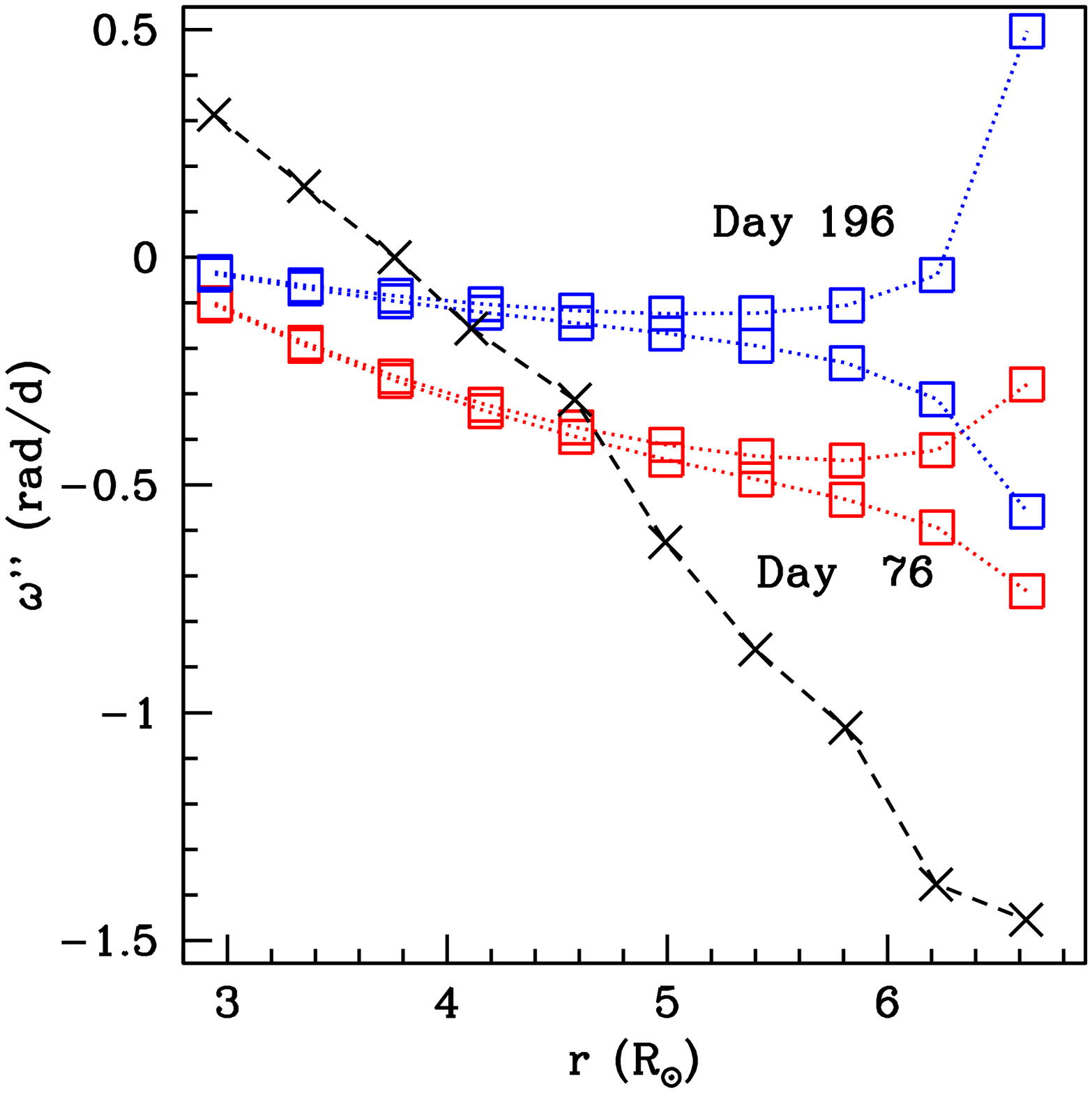}
\caption{Results for a star with an initial steep differential rotation structure (Case 34). 
{\it Left:} Angular velocity at the equator at Day 196. Each curve corresponds to 
one of the layers, as labeled on the right side of the curves, where 1 corresponds to the deepest layer.  
{\it Right:}  $\omega''$ profiles at Day 76 (square) and Day 196 (pentagon) for azimuth angles at 
which the velocity gradient is greatest.   The crosses connected by a dash line indicate the initial $omega''$ 
profile at the start of the computation.
}
\label{omdp_VarBeta_Case34}
\end{figure}

An example of the results that are obtained from a model in which the computation is initiated 
with differential rotation is shown in  Fig.~\ref{omdp_VarBeta_Case34}.  The initial angular velocity 
distribution is listed in Table~\ref{table_case34_non-uniform} and is displayed in the right panel of 
Fig.~\ref{omdp_VarBeta_Case34} with crosses connected by a dash line.  Its very steep shape 
was inspired by the \citet{2000A&A...361..101M} evolutionary models near the end of the core 
hydrogen burning phase, but is otherwise arbitrary.  As discussed in Appendix~\ref{sect_diffrotation}, 
the manner in which the temporal evolution proceeds leads to results that are insensitive to the 
precise shape of the initial $\omega''$ profile. 

Recalling that $\omega''$=$\left<\omega'' \right>+\delta\omega''$ (Equation~\ref{eq_omdp}), we now 
define the term $\left<\omega''\right>$ profile to be the variation over radius of the average
angular velocity and our results show that the $\left<\omega''\right>$ profile flattens over
time.  Thus, at any time during the computation, the tidal velocity $\delta\omega''$ is
superposed on what may be regarded as the background differential rotation structure defined
by $\left<\omega'' \right>$.  When $\left<\omega'' \right> \rightarrow$ 0, one may consider the
star to be in a state of average uniform rotation upon which a tidal component is
superposed.  It is only when $\delta\omega''\rightarrow$ 0 that one may speak of a true uniform 
rotation structure.

Fig.~\ref{omdp_VarBeta_Case34} (right) presents the $\omega''$ profiles for azimuth angles 
at which  $\delta\omega''$ has the largest amplitude. There are two sets of curves for each of the
times chosen for this illustration, 76\,d and 196\,d after the start of the calculation.  Each set 
of curves shows the extrema of angular velocities and encloses the azimuthally averaged angular velocity 
$\left<\omega''\right>$ (not shown in the figure). By Day 76, the $\left<\omega''\right>$ profile 
has flattened significantly, with only a weak average differential rotation structure still present. 
We find that the flattening first occurs very rapidly but then systematically slows down as 
average uniform rotation is approached. We plot $\omega''$  versus $\varphi'$ in the left panel of 
Fig.~\ref{omdp_VarBeta_Case34}, showing that a differential rotation structure at Day 196 still persists 
but is flatter than the one at which the computation was started.

An additional feature that we observe is that the tidal amplitude $\delta\omega''$ of each layer depends 
on the degree of asynchronicity of the particular layer. This results from the dependence of $\delta\omega''$ 
on the instantaneous value of the average synchronicity parameter 
$\left<\beta(r)\right>$ =$\left<\omega(r)\right>/\Omega_0$. Specifically, the tidal amplitude is larger 
for larger departures from synchronicity, as is also evident from Equation~\ref{eq_scharl}. 

We note that since it is the value of $\left<\beta(r)\right>$ that determines the tidal amplitude 
of a particular layer, and  not $\beta_0$ (the core synchronicity parameter), this means that,
from an observational perspective,  an evolving value of $\left<\beta(R_*)\right>$ could lead to
patterns that change on timescales longer than the orbital timescale. Here, $R_*$ represents 
the layers from which the dominant photospheric absorption lines arise. For example, if these 
layers approach a state in which $\left<\beta(R_*)\right>\simeq$1, the tidal flow amplitudes 
would diminish considerably as would any spectral variability associated with these flows.   

\section{The $\omega''$ profile in eccentric orbits}

In the previous section, we showed that the azimuthal tidal velocities in circular-orbit binary stars 
retain a general sinusoidal-like shape as a function of azimuth angle similar to that predicted by 
the analytical expression in Equation \ref{eq_scharl}, albeit with different amplitudes, and that this
shape remains relatively stable as a function of orbital phase in the reference frame that rotates 
with the companion ($S'$).  In this section, we examine the manner in which this behavior changes in 
an eccentric binary system. This question has already been addressed for the surface layer in the primary 
star of the $\alpha$ Virginis (Spica) system by \citet{2009ApJ...704..813H} where  the azimuthal 
dependence of the tidal velocity was found to have very pronounced, small frequency oscillations superposed
on a general sinusoidal shape and was found to be orbital-phase dependent.  Here, we now extend the 
analysis to layers below the surface. 

\begin{figure}
\centering
\includegraphics[width=0.48\columnwidth]{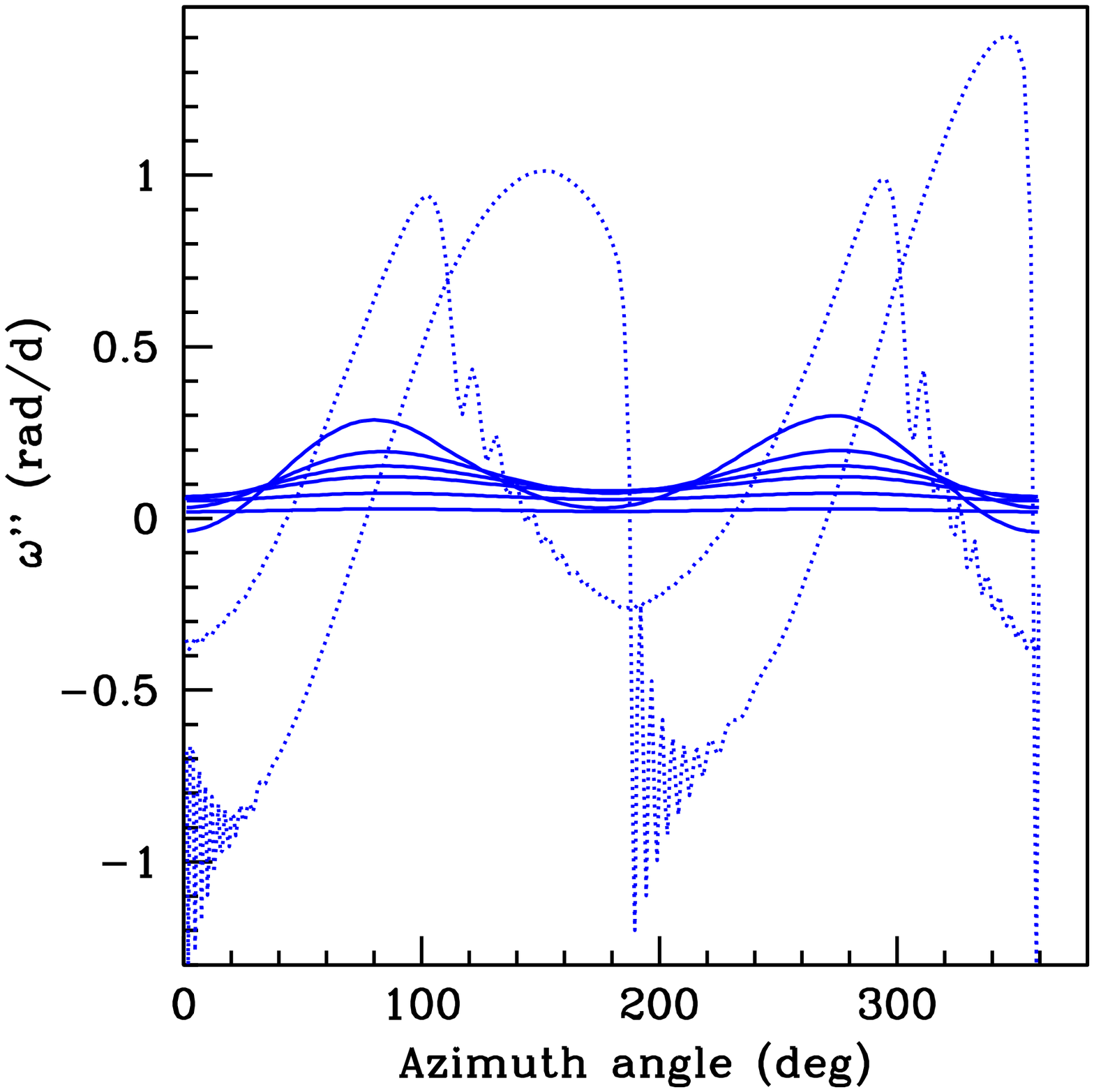}
\includegraphics[width=0.48\columnwidth]{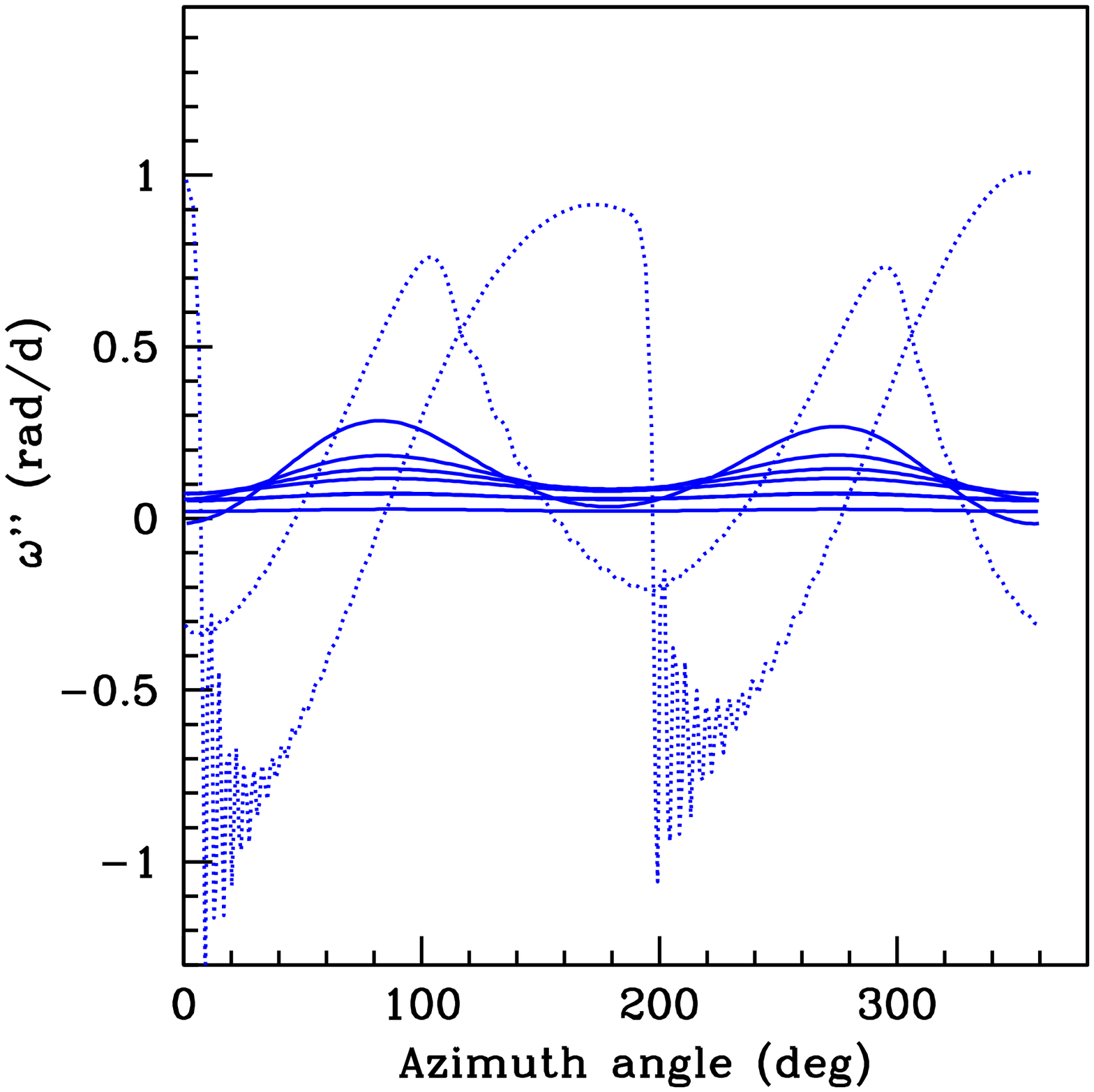}
\includegraphics[width=0.48\columnwidth]{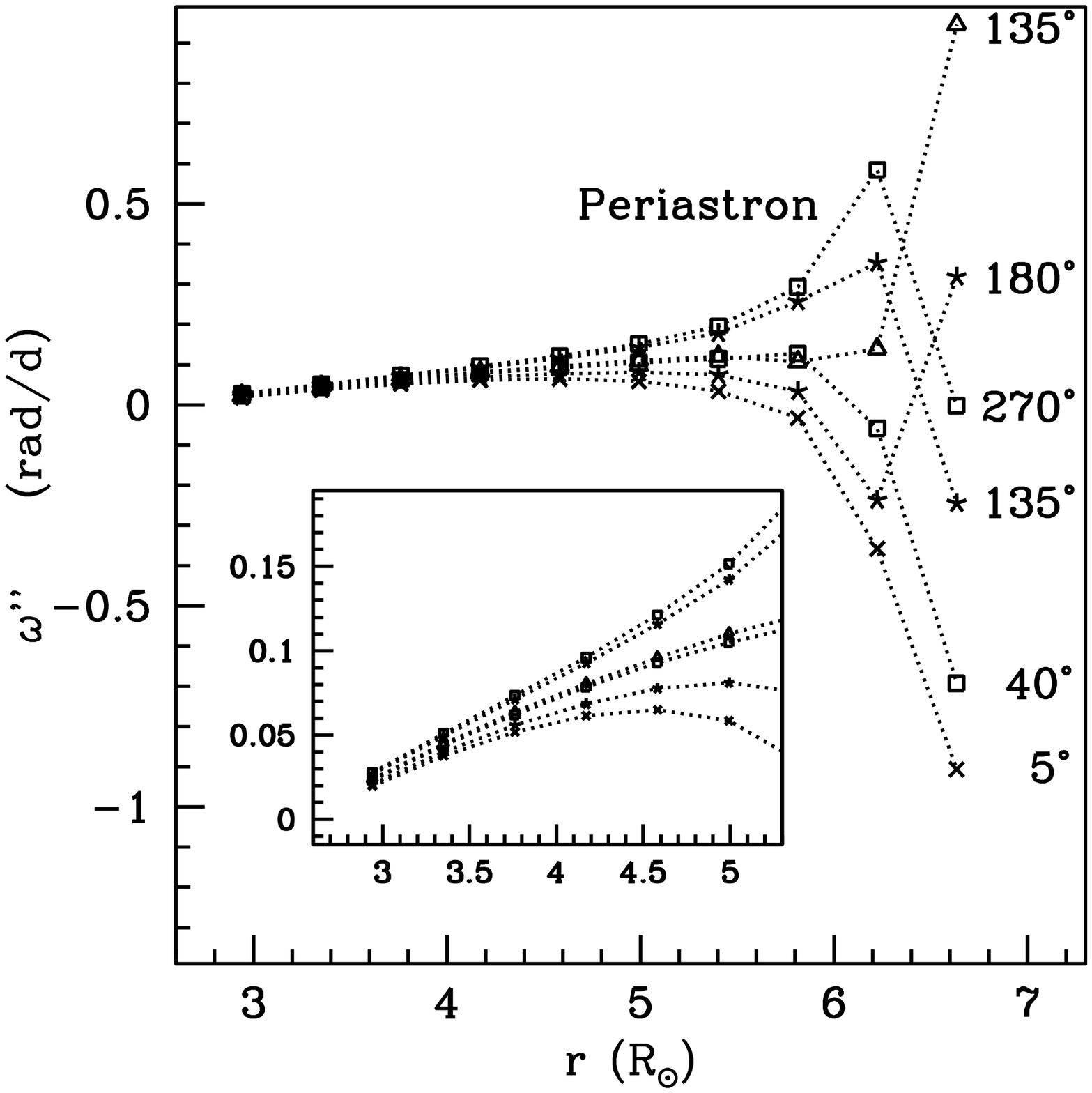}
\includegraphics[width=0.48\columnwidth]{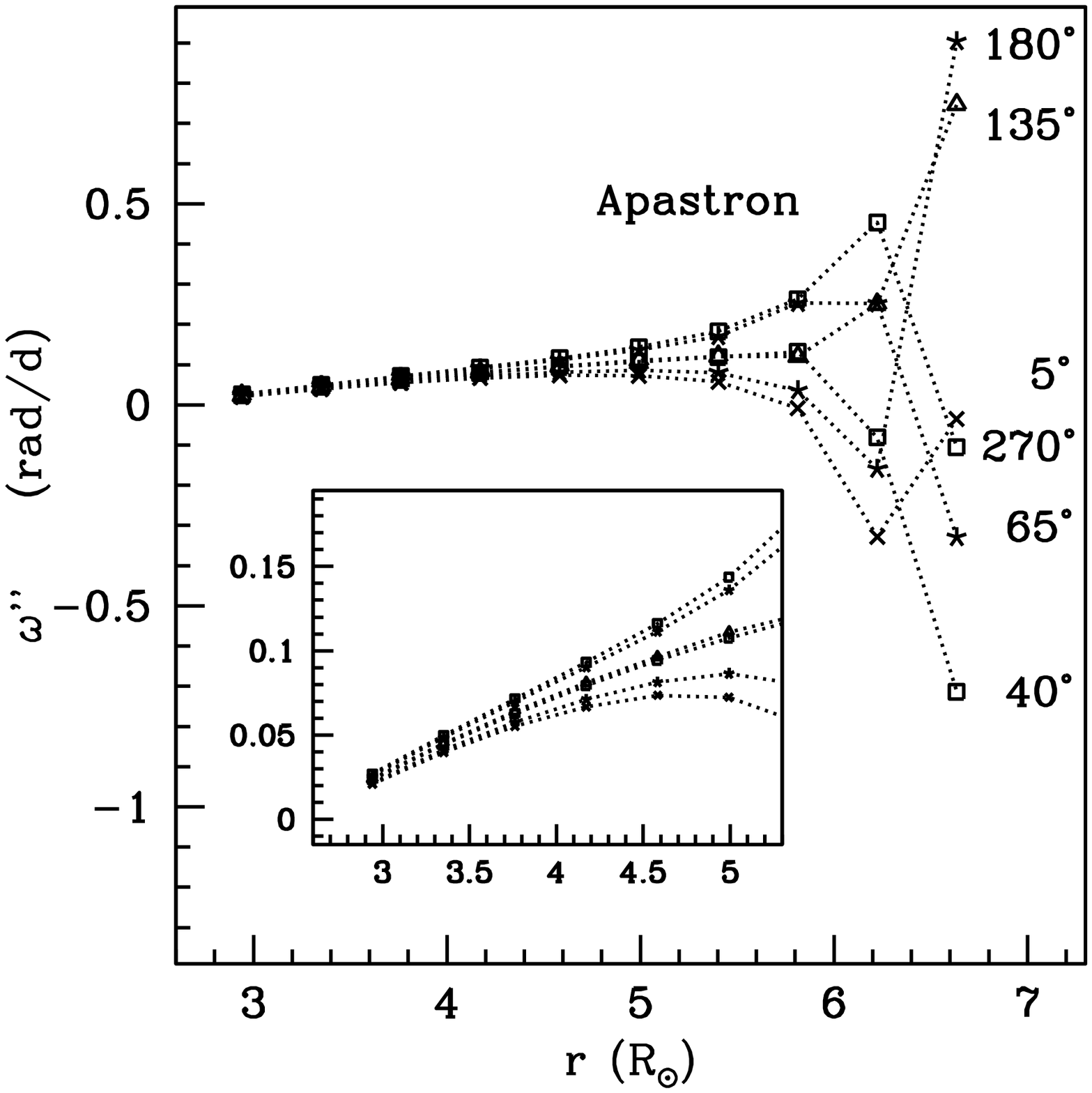}
\caption{Angular velocity and  meridional $\omega''$ profiles at the equator for the
eccentric binary Case 35. The inset shows the  $r/R_\odot$<6 layers, which have developed an average 
differential rotation structure. The amplitudes and $\omega''$ profiles change between periastron and apastron. 
}
\label{omdp_eccentric_IND3}
\end{figure}

\begin{figure}
\centering
\includegraphics[width=0.98\columnwidth]{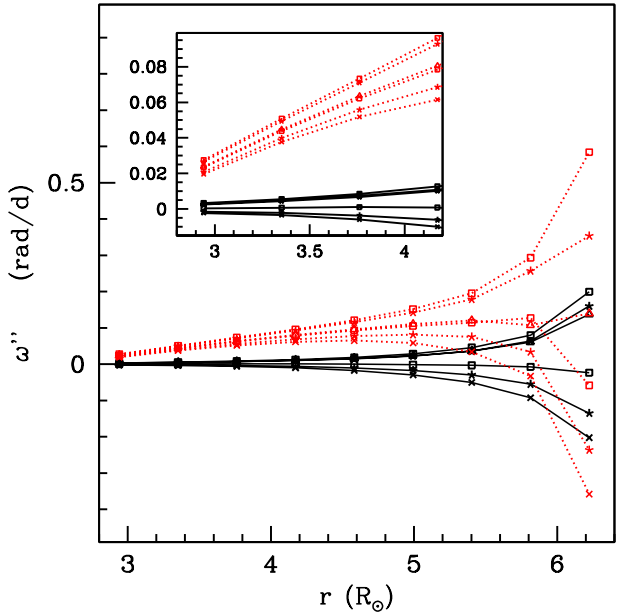}
\caption{Meridional $\omega''$ profiles at the equator for the circular orbit Case 31 (black) 
and the eccentric Case 35 (red); both models have polytropic index $n$=3. 
}
\label{omdp_compare_circ_ecc_IND3}
\end{figure}

The azimuth-dependent shape of $\omega''$  and the $\omega''$ profiles in the eccentric  $e$=0.1 Case 35
model are shown in Fig.~\ref{omdp_eccentric_IND3}. 
The most striking differences with respect to the analogous circular orbit case is the larger tidal 
amplitude, even at apastron and high-frequency oscillations, the latter similar to those reported 
by Harrington et al. (2009) based on the one-layer model.  Additionally, the inner layers develop a 
differential rotation in which the gradient of the $\left<\omega''\right>$ profile increases over time.

Some of the differences with respect to the $e$=0 analogous cases can be understood by recalling that 
the periastron distance is smaller than the semimajor axis of the equivalent circular orbit.
Also, we recall that $\beta_0^0$ describes the asynchronicity at periastron and that 
the instantaneous value, $\beta(r,t)$ increases over orbital phase between periastron and apastron.  
As mentioned in Appendix \ref{sect_diffrotation}, the tidal amplitude $\delta\omega''$ 
increases with increasing $\beta(r,t)$.  Hence, there is an interplay between these two processes.  
In addition, it is important to note that the ``equilibrium tide" configuration (as described, for example,
by the Scharlemann (1981) equations) holds as long as the perturbations caused by the tidal
force are small enough so that the fluid has time to adjust  to the altered hydrostatic conditions.
This is clearly not the case in the eccentric binary models, thus leading to the observed oscillations
and strong departure from the equilibrium tide configuration displayed particularly by the outer layers.
We note also that a similar comment applies to $e$=0 models in which the product $\nu\rho$ is smaller
than a critical threshold, as is evident in the right panel of Fig.~\ref{om_gradient_compareIND}.

Finally, the appearance of a positive gradient in the $\left<\omega''\right>$ profile of the inner 
layers is perplexing. It seems to counterbalance the  significant negative gradient in the outer layers,
which clearly is a result of the retarding effect of the tidal force, and which would be expected
also for the inner layers.  This curious result may simply be a consequence of the neglect of feedback 
into the orbital motion in the TIDES code. Specifically, if the angular momentum loss of the outer
layers were to be transferred to the orbital motion, this would avoid it being transferred
to the inner layers through the viscous coupling.   However, the timescale over which  the positive 
gradient develops in our calculations ($\leq$100~d) is too fast compared to the tidal timescales 
(assumed to be $\sim$10$^7$ yr),  unless the viscosity is at least $\sim$10$^8$ smaller, if not more.  
Also, if the effect were due to the neglect in TIDES of feedback to the orbital motion, one might 
also expect it to appear in the circular orbit calculations (that are initiated with a uniform rotation
structure) but here it is not evident, as shown in Fig.~\ref{omdp_compare_circ_ecc_IND3}.
On the other hand, the tidal amplitude in all the layers is always larger for the positive velocities
than the negative velocities.  Thus, the average velocity that is computed over all azimuths is
always $>$0, hence feeding a positive angular momentum.  This effect is more pronounced in the
eccentric case than in the analogous circular because of the closer orbital separation at
periastron and because of the larger $\beta$ at apastron.  This issue remains to be further analyzed,
although we note that \citet{1981A&A....99..126H} finds regimes in which a supersynchronous  binary star may 
increase its rotation rate instead of decreasing it, depending on the ratio of angular momenta in the 
orbit and the stellar rotation. 

\begin{figure}
\centering
\includegraphics[width=0.89\columnwidth]{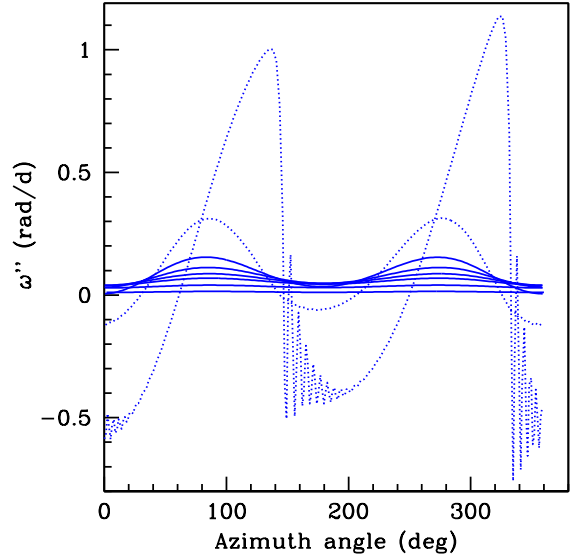}   
\caption{Angular velocity $\left<\omega''\right>$+$\delta\omega''$ at the equator from a TIDES
calculation for the model that is used to evaluate the shear instabilities (Case 41 in Table~\ref{table_models_nominal}).
The curves correspond to the ten layers used in these
calculations, with dashes corresponding to layers closest to the surface, $r/R_\odot$=5.0  and 5.3.
The abscissa is azimuth angle measured in degrees with $\varphi'$=0 corresponding to the sub-binary
longitude.
}
\label{fig_om_and_gradientBEC}
\end{figure}

\section{Discussion}

\subsection{Relevance for mixing: Vertical shear instability}

The physical mechanisms that transport chemical elements from the nuclear-processing region 
toward the surface are driven by a complex interplay between temperature, velocity and
chemical gradients, and their implementation into stellar structure models is generally
parametrized through the use of diffusion coefficients as discussed in \citet{1993SSRv...66..285Z, 
2000ApJ...528..368H, 2000ARA&A..38..143M, 2004A&A...425..243M, 2007A&A...462.1063M, 
2014A&A...566A.110P, 2016ApJ...821...49G, 2020ApJ...901..146G}, and references
therein.   

The basic idea is that shear instabilities can lead to turbulent eddies that then propagate
through the fluid carrying with them the chemical abundances of the region from which they
originated.  The turbulent eddies propagating in the vertical direction (that is, parallel
to the stellar radius), provide transport from the core to the surface.  Those propagating
horizontally (within a shell at a fixed radius) tend to homogenize the chemical abundances
in that shell, but are also found to generate turbulence that propagates vertically.

In this section we illustrate the manner in which the TIDES code calculations can contribute
toward assessing whether the  shear instability may be triggered in deep layers, near the
core, and thus examine the role of the tidal effects in mixing.  We base this analysis on the
results of numerical experiments that are reported in \citet{2016ApJ...821...49G, 2020ApJ...901..146G},
where the instability criteria are provided.

In a fluid in which the temperature and/or chemical composition is stratified, the displacement
of turbulent eddies in the vertical direction is limited by the buoyancy force which tends
to restore them to their equilibrium location, thus rendering the fluid stable against the
shear instabilities unless  $Ri<Ri_{\rm crit}$. Here, $Ri_{\rm crit}$ is a critical value
that is typically thought to be of order unity \citep{1920RSPSA..97..354R,1992A&A...265..115Z},
and $Ri$ is the Richardson number:

\begin{equation}
Ri=\frac{N^2}{S^2}
\end{equation}

\noindent where S= |d{\bf{v}}/dr| is the local shearing rate of the flow field {\bf{v}} and
$N$ is the buoyancy frequency,

\begin{equation}
N^2=\frac{g}{H_p} (\nabla_{ad}-\nabla)
\end{equation}

\noindent where $\nabla_{ad}$=($\partial \ln$T/$\partial \ln$P)$_{ad}$ is the gradient at constant
entropy and composition, and  $\nabla$=$d\ln$T/$d\ln$P. $H_p$ is the pressure scale height and $g$
the gravity.

Typical stars have values $Ri>>$1, suggesting that they are stable against this instability.
This is true also for binaries. Adopting the tidal velocity $\delta\vv''$ as the typical flow 
velocity of the parcels of fluid that are interacting, as listed in Table ~\ref{table_case34_non-uniform},
one finds $Ri\geq$10$^3$ for all layers except perhaps the surface.
However, if a vertically displaced parcel of fluid has time to undergo heat transfer that 
allows it to rapidly come into thermal equilibrium with its new surroundings,  the correct
criterion for shear instability is now the one associated with  {\it diffusive 
shear instabilities}\footnote{Also known as "secular shear instability"} 
\citep{1974IAUS...59..185Z, 1999A&A...348..933L, 2014A&A...566A.110P}:

\begin{equation}
RiPr  < (RiPr)_c
\label{eq_instability}
\end{equation}

\noindent where $(RiPr)_c\sim$ 0.007 \citep{2014A&A...566A.110P, 2016ApJ...821...49G}
and the Prandl number $Pr$=$Pe/Re$=$\nu/\kappa_T$.  Here, $Pe$ is the P\'eclet 
number, $Re$=$UL/\nu$ is the Reynolds number, $\nu$ is the molecular kinematical viscosity,  
$\kappa_T$ is the thermal diffusivity, $U$ is a characteristic velocity scale and $L$ the corresponding
length scale. 

The condition in Equation~\ref{eq_instability} for the onset of the diffusive shear instability 
is valid only if heat transfer occurs rapidly, which corresponds to a low  P\'eclet number (LPN), 
$Pe <$ 1 \citep{1999A&A...348..933L}, a condition that is met only near the stellar surface.  
However, because the definition of the P\'eclet number depends on a length scale and a velocity 
scale that need to be identified,  \citet{1999A&A...348..933L} associated these with the turbulent scales and 
defined a turbulent P\'eclet number,

\begin{eqnarray}
Pe_t&=&u_{rms} \ell_\vv/\kappa_T,             
\label{eq_peclet_turb}
\end{eqnarray}

\noindent where  $u_{rms}$ is the $rms$ flow velocity of turbulent eddies
and $\ell_\vv$ is their vertical scale and suggested that criterion Equation~\ref{eq_instability} is valid
for $Pe_t<<$1, even if $Pe>$1.  The problem then is to obtain the values of of $u_{rms}$ and $\ell_\vv$.
The analysis performed by \citet{2016ApJ...821...49G} leads to the conclusion that, for shearing motions 
that are caused by imposing  a force $F_0$,  one can associate $u_{rms}$  with the typical velocity that 
results.  That is,  $u_{rms}$= $U_F$ with, 

\begin{equation}
U_F=\left(\frac{F_0}{k\rho_0}\right)^{1/2},
\end{equation}

\noindent where $F_0$ is the perturbing force per unit volume, $\rho_0$ is the background density, and the 
length scale is renamed in terms of a unit length scale $k^{-1}$.   The condition for the LPN 
$Pe_t<<$1 is now assessed using  

\begin{eqnarray}
Pe_t&=&\left(\frac{F_0}{\rho_0}\right)^{1/2} k^{-3/2} \kappa_T^{-1}               
\label{eq_peclet_turb2}
\end{eqnarray}

\noindent and if it is satisfied, the instabilities will be triggered if 

\begin{equation}
Ri_FPr_F  < (RiPr)_c
\label{eq_instability2}
\end{equation}

\noindent where $Ri_F$ and $Pr_F$ now refer to the Richardson and Prandtl numbers evaluated using $U_F$,

We use below the convenient expressions for the P\'eclet, Reynolds and Richardson 
numbers provided by \citet{2016ApJ...821...49G} (their Equation. 30 and 31):

\begin{eqnarray}
Pe_t&=&100\left(\frac{F_0/\rho_0}{10^5 cm~s^{-2}}\right)^{1/2} \left(\frac{k^{-1}} {10^9 cm}\right)^{3/2} \left(\frac{\kappa_T}{10^{14} cm^2 s^{-1}}\right)^{-1}   \nonumber  \\
Re_F&=&10^{15}\left(\frac{F_0/\rho_0}{10^5 cm~s^{-2}}\right)^{1/2}\left(\frac{k^{-1}} {10^9 cm}\right)^{3/2}\left(\frac{\nu}{10 cm^2 s^{-1}}\right)^{-1} \nonumber\\
Ri_F&=&0.01\left(\frac{N^2}{10^{-6} s^{-2}}\right) \left(\frac{k^{-1}} {10^9 cm}\right)\left(\frac{F_0/\rho_0}{10^5 cm~s^{-2}}\right)^{-1}
\label{eq_Garaud2016}
\end{eqnarray}

The primary contribution of the TIDES models to the above considerations is that it can directly provide 
values of the acceleration term $F_0/\rho_0$=$dV_{\varphi''}/dt$, where $V_{\varphi''}$=
r $\sin(\theta)\delta\omega''$ is the azimuthal component of the tidal velocity.  
We illustrated this application of the TIDES models with the example that follows. 

We performed a TIDES calculation that is tailored to a stellar structure model that provides internal
structure data needed for the values of $N$. The chosen model has 10~M$_\odot$ at an age of 12~Myrs and an
initial rotation velocity 150 km~s$^{-1}$ and is taken from the rotating massive main-sequence stars grids of 
\citet{2011A&A...530A.115B}.  The age is based on the results of \citet{2016MNRAS.458.1964T}. The density 
structure of this model is close to that of a $n$=3.5 polytrope, and the radius is R$_1$=5.48~R$_\odot$. 
These values were used to compute Case 41 (see Table~\ref{table_models_nominal}). Given this radius and an 
orbital period of 4.01d, the synchronicity parameter of the core is $\beta_0^0$=2.2. 
The resulting angular velocities at the equator are shown in Fig.~\ref{fig_om_and_gradientBEC}.

The stellar structure radial model grid is significantly finer than in the TIDES calculation,
so we averaged the stellar structure parameter values over intervals of r$\pm \Delta$R$_1$/4,
with $\Delta$R$_1$/4=0.08~R$_\odot$. We used $\nu$=10\,cm$^2$s$^{-1}$ for the molecular kinematical viscosity,
and for $\kappa_T$ we visually  interpolated its values from  Fig.\,7 of \citet{2016ApJ...821...49G}.

Given the shape of $\delta\omega''$, it is clear that the acceleration term is not constant but depends
on the $\varphi'$ coordinate. The $\varphi'$ dependence maps into a variation over time at a fixed position 
$\varphi''$ when we apply the transformation which, for circular orbits is 
$\varphi'$=$\varphi''-(\Omega_0 - \omega_0)t$. Given the sinusoidal-like shape of the variation, the 
acceleration term oscillates between a maximum and a minimum value passing through zero. 
Thus we adopt as a measure of the azimuthal acceleration,

\begin{equation}
\frac{dV_{\varphi''}}{dt} \rightarrow r_{mid} \sin(\theta)\frac{\omega''_{max}-\omega''_{min}}{\Delta t},
\label{eq_time_derivative}
\end{equation}

\noindent where  $r_{mid}$ is the midpoint of a stellar layer and $\Delta t$ is the time difference
between maximum and minimum angular velocity.   These maximum and minimum values obtained from Case 41
are listed in Table~\ref{table_TIDES_velocities} as a function of radius and for three latitudes.

Recalling that the general shape of the tidal velocity is similar to that given by Equation~\ref{eq_scharl} 
(except in the outer layers)  and that this equation, when scaled, approximately matches the tidal 
velocity obtained with TIDES (Fig.~\ref{fig_propaz_beta1.8_scaled}), we see that the time dependence of
the acceleration term goes as $\cos^4 (\theta) \sin(2\varphi''-4 \pi \Phi (1.-\beta_0))$,
where $\Phi$=$t$/P, and $\Phi$ is the orbital phase and P the orbital period.  
A plot of this sine function for our P=4d, $\beta_0^0$=2.2 model shows that the time between a 
 maximum and its closest minimum is $\sim$0.2 in orbital phase.  Thus, we set $\Delta t$=0.8d.
For the length scale we adopt a value half the size of the layers that were modeled, $k^{-1}$=$\Delta$R/2.

The results of substituting the velocity values at the equator into Equation.~\ref{eq_Garaud2016} are listed 
in Table~\ref{table_garaud2020_horizontal} (under the section labeled "Vert"),
where it can be seen that the LPN condition  $Pe_F<<$1 is satisfied only very near the 
surface.  Thus, even though the condition given by Equation~\ref{eq_instability2} is satisfied throughout, 
the diffusive  shear instability in the LPN approximation does not appear to be triggered. 
Similar results are obtained for other latitudes.

\begin{table*}
\small
\caption{TIDES angular velocities M=10\,M$_\odot$, R=5.48\,R$_\odot$, V$_{rot}$=150\,kms$^{-1}$, 12Myr model.}
\label{table_TIDES_velocities}
\centering
\begin{tabular}{r l r r r r r r r r r r}
\hline\hline
%
    & $\theta$ &&     & 90$^\circ$   &   &  & 72.5$^\circ$ &       &  & 46$^\circ$   &     \\
$k$ &r/R$_\odot$& &$\left<\omega''\right>$ & $\omega_{min}''$& $\omega_{max}''$& $\left<\omega''\right>$ & $\omega_{min}''$& $\omega_{max}''$ & $\left<\omega''\right>$ & $\omega_{min}''$& $\omega_{max}''$ \\
\hline
        &        &        &        &        &        &        &        &        &        &        \\
  
  1&  2.36&   & 0.013   & 0.011   & 0.015   & 0.012   & 0.010   & 0.014   & 0.010   & 0.009   & 0.011 \\
  2&  2.69&   & 0.024   & 0.021   & 0.028   & 0.023   & 0.020   & 0.026   & 0.018   & 0.016   & 0.020 \\
  3&  3.01&   & 0.035   & 0.029   & 0.040   & 0.033   & 0.027   & 0.038   & 0.025   & 0.022   & 0.028 \\
  4&  3.34&   & 0.045   & 0.036   & 0.054   & 0.042   & 0.033   & 0.050   & 0.031   & 0.026   & 0.035 \\
  5&  3.67&   & 0.055   & 0.040   & 0.069   & 0.051   & 0.037   & 0.063   & 0.036   & 0.029   & 0.042 \\
  6&  4.00&   & 0.065   & 0.040   & 0.087   & 0.059   & 0.036   & 0.078   & 0.040   & 0.029   & 0.050 \\
  7&  4.33&   & 0.076   & 0.032   & 0.112   & 0.067   & 0.029   & 0.099   & 0.043   & 0.025   & 0.060 \\
  8&  4.66&   & 0.086   & 0.005   & 0.155   & 0.073   & 0.005   & 0.133   & 0.045   & 0.013   & 0.076 \\
  9&  4.99&   & 0.088   &-0.124   & 0.313   & 0.075   &-0.094   & 0.257   & 0.046   &-0.021   & 0.114 \\
 10&  5.32&   & -0.020   &-0.758   & 1.137   &-0.012   &-0.491   & 1.044   & 0.037   &-0.154   & 0.272 \\
\hline
\hline
\end{tabular}
\tablefoot{$k$ is the layer number, r is the corresponding midpoint radius given in solar units, 
and $\left<\omega''\right>$,  $\omega_{max}''$, $\omega_{min}''$ are, respectively, 
the average angular velocity, its minimum value and its maximum value, and are given in units of 
rad/day; data for Case 41 model at day 277.}
\end{table*}

\begin{figure}
\centering
\includegraphics[width=0.88\columnwidth]{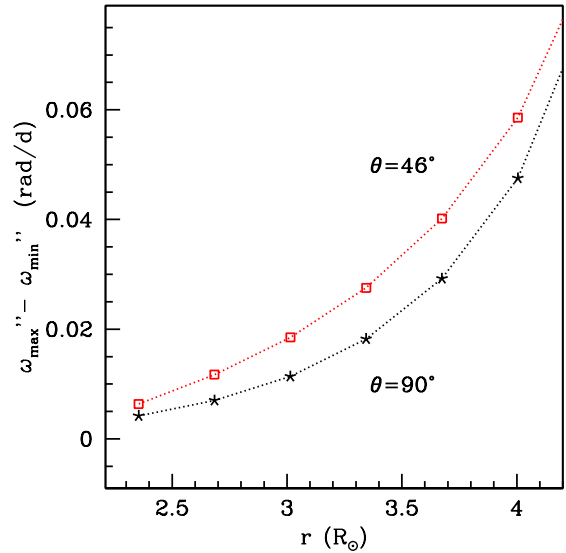}
\caption{Difference of maximum and minimum angular velocity at the equator and at colatitude 46$^\circ$
from the Case 41 model. Only layers at $r<$4.2 R$_\odot$ are illustrated. The abscissa is the midpoint 
radius of the layer. 
}
\label{fig_latitudes}
\end{figure}

The high P\'eclet number regime is only briefly discussed in \citet{2016ApJ...821...49G}.  They report 
that the system transitions from intense mixing  events in which the existing shear is destroyed to 
periods of quiescence during which the flow is nearly laminar and where the shear is gradually amplified by 
the forcing, until intense mixing resumes.  A similar cyclical nature for the instabilities at high P\'eclet 
number has been found in laboratory experiments \citep{2014JFM...753..242M}. Whether such behavior can occur
in the deep stellar layers remains to be assessed.  However, it is interesting to note that the 
$\sin(2\varphi')$ term in the time derivative of Equation~\ref{eq_scharl} implies a time dependence. Large 
amounts of shear may be triggered near the maximum in the flow velocity, after building up during the prior 
times within each flow cycle. 
Thus, it seems like the occurrence of shear instabilities in the high P\'eclet number regime 
cannot be excluded at this time as a mechanism contributing toward mixing.

In addition to the above, it is important to add that the shearing motions are associated with
energy dissipation.  Because the  amplitude of the tidal velocity increases with stellar radius, so does the 
energy dissipation rate. If this were to significantly decrease the temperature stratification, it would 
weaken  the effect of the buoyancy force. 

Another consideration is the potential stellar mass dependance of any mixing induced by the
tidal flows for the following reasons. First, the increase in the importance of
radiation pressure with increasing mass weakens the stability of the stratification.
In addition, due to pronounced peaks in the radiative opacity coefficient as
function of temperature, several convection zones may appear in the subsurface
layers of massive main sequence stars \citep{2009A&A...499..279C}. These
convection zones, which are thought to induce observable turbulent velocity
fields at the stellar surface, will also induce statistical pressure
fluctuations in the layers beneath \citep{2015ApJ...808L..31G}
which may interact with the periodic tidal forcing.

In conclusion, although diffusive vertical shear instabilities in the LPN regime
are excluded as a mixing process in layers close to the convective core,  the same cannot yet
be said about the high P\'eclet limit regime and particularly when the oscillating nature of
the tidal flows is considered, given that such a scenario has not been fully analyzed from the
fluid dynamics perspective.  Furthermore, the tidal shear energy dissipation, strong radiation
pressure and localized regions of sub-photospheric convection may all contribute to reduce
the stratification, triggering the instability.

Finally, we note that the tidal velocity field is also present in low-mass binary stars in
asynchronous rotation.  However, unlike the massive main sequence stars in which altering the
surface chemical abundances requires  mixing from the central convective core throughout the 
entire radiative envelope, the lower-mass stars present boron abundance anomalies associated with
layers closer to the surface.  This is because boron is
destroyed by proton capture,for example, in the inner $\sim 90$\% of the radiative
envelope of a 15\,M$_\odot$ main sequence star \citep{1996A&A...308L..13F},
mixing would need to occur only in the outermost layers in order to
produce a drop in the boron surface abundance. While rotationally induced mixing
is a strong contender to produce such mixing \citep{2016ApJ...824....3P},
the tidal mixing discussed here may have an additional
impact in short-period massive pre-Roche-lobe-overflow binaries. 


\subsection{Relevance for mixing: Horizontal shear instability}

A second type of shearing flows initially discussed by \citet{1992A&A...265..115Z} are those
caused by velocity differences over latitude. In this case, the horizontal shear generates 
secondary turbulent eddies that propagate in the vertical direction, thus potentially contributing
toward mixing in this direction.   The stability of these flows in the high P\'eclet
number regime has recently been analyzed by \citet{2020ApJ...901..146G}, where now the
bifurcation parameter that divides the stable from the unstable regimes is:
 
\begin{equation}
Pe_t=\omega_{rms} \ell_z Pe
\label{eq_peclet_turb_horizontal} 
\end{equation}

\noindent where   

\begin{eqnarray}
\ell_z &\simeq& 2.1 \left(\frac{U^2}{N^2L^2}\right)^{1/3} L           \nonumber \\
w_{rms}& \simeq& 1.3 \left(\frac{U^2}{N^2L^2}\right)^{1/3} U
\end{eqnarray}

\noindent The parameters $U$ and $L$ are the actual characteristic horizontal flow velocity
and length scale, respectively.  We now use \citep{2020ApJ...901..146G}:

\begin{eqnarray}
Re&=&\frac{10^{14}}{2\pi}\left(\frac{U}{10^4cm/s}\right)\left(\frac{L_y}{10^{11}cm} \right)\left(\frac{\nu}{10cm^2/s} \right)^{-1}, \nonumber \\
Pe&=&\frac{10^{8}}{2\pi}\left(\frac{U}{10^4cm/s} \right)\left(\frac{L_y}{10^{11}cm} \right)\left(\frac{\kappa_T}{10^7cm^2/s} \right)^{-1}, \nonumber \\
B &=&\frac{10^{8}}{4\pi^2} \left(\frac{U}{10^4cm/s} \right)^{-2} \left(\frac{L_y}{10^{11}cm}\right)^2 \left(\frac{N}{10^{-3}s^{-1}} \right)^2. \nonumber
\end{eqnarray}

\noindent  where the symbol $B$ is analogous to the Richardson number and denotes  the
ratio of the buoyancy frequency to the horizontal shearing rate. 

As in the previous section, we used the velocities obtained in the Case 41 TIDES model and listed
in Table 6 to quantify the above parameters.  Specifically, we set 
$U$=$r_{mid}[\sin(\theta_1)\left<\omega''(\theta_1)\right>-\sin(\theta_2) \left<\omega''(\theta_2)\right>]$ and 
$L_y \sim$r$_{mid}$($\theta_1-\theta_2$) where $\theta_1$ and $\theta_2$ correspond to two adjacent
colatitudes.  Choosing from Case 41 $\theta_1$=90$^\circ$ and $\theta_2$=72.5$^\circ$ and the
corresponding average velocity values $\left<\omega''\right>$ and substituting into the above relations,
we get values of $\ell_z$ and $w_{rms}$, which when substituted 
into Equation~\ref{eq_peclet_turb_horizontal} yield $Pe_t$.

We list the results in Table~\ref{table_garaud2020_horizontal} under the section labeled Horiz. We 
find that the conditions $Pe_t<<$1 and $BPr<<$0.007 are simultaneously met not only near the surface,
but now also at $r/R_\odot\leq$3. Thus, this suggests that the horizontal shear instability may be
triggered in stellar layers that lie relatively close to the nuclear burning core. Clearly, the same
caveats mentioned in the previous section apply here and a more conclusive answer must await further
developments in the fluid dynamics computations (see, for example \citealt{2021A&A...646A..64P}).

\begin{table*}
\small
\caption{Diffusive shear instability. M=10\,M$_\odot$, R=5.48\,R$_\odot$, 12Myr model.}
\label{table_garaud2020_horizontal}
\centering
\begin{tabular}{c l l l l l l l l l l}
\hline\hline
r       & 2.4   & 2.7   & 3.0    & 3.3   & 3.7   & 4.0   & 4.3   & 4.7   & 5.0   & 5.3    \\
\hline
T       & 7.e+06& 5.e+06& 4.e+06& 3.e+06& 3.e+06& 2.e+06& 1.e+06& 8.e+05& 5.e+05& 2.e+05 \\
g       & 5.e+04& 4.e+04& 3.e+04& 2.e+04& 2.e+04& 2.e+04& 1.e+04& 1.e+04& 1.e+04& 1.e+04 \\
Hp      & 2.e+10& 2.e+10& 2.e+10& 2.e+10& 2.e+10& 2.e+10& 1.e+10& 1.e+10& 9.e+09& 5.e+09 \\
Ny      & 1.e-03& 1.e-03& 1.e-03& 1.e-03& 1.e-03& 1.e-03& 1.e-03& 1.e-03& 1.e-03& 1.e-03 \\
$\kappa_T$&5.e10& 8.e10 & 1.e11 &3.e11  &7.e11  & 3.e12 & 1.e13 & 4.e14 & 5.e15 & 1.e18 \\ 
\hline
Vertical&        &        &        &        &        &        &        &        &        &        \\
$F_0/\rho$& 1.e-01& 2.e-01& 4.e-01& 7.e-01& 1.e+00& 2.e+00& 4.e+00& 8.e+00& 3.e+01& 1.e+02   \\
$Re_F$    & 4.e+13& 6.e+13& 8.e+13& 1.e+14& 1.e+14& 2.e+14& 2.e+14& 3.e+14& 6.e+14& 1.e+15    \\
$Ri_F$    & 2.e+05& 8.e+04& 4.e+04& 2.e+04& 9.e+03& 5.e+03& 3.e+03& 2.e+03& 5.e+02& 2.e+02    \\
$Pr_F$    & 2.e-10& 1.e-10& 1.e-10& 3.e-11& 1.e-11& 3.e-12& 1.e-12& 3.e-14& 2.e-15& 1.e-17    \\
$Pe_t$    & 8.e+03& 7.e+03& 8.e+03& 3.e+03& 2.e+03& 6.e+02& 2.e+02& 9.e+00& 1.e+00& 1.e-02    \\
$RiPr_F$  & 4.e-05& 1.e-05& 4.e-06& 6.e-07& 1.e-07& 2.e-08& 3.e-09& 4.e-11& 1.e-12& 2.e-15    \\
\hline
Horiz   &        &        &        &        &        &        &        &        &        &        \\
U       &  2.e+03& 6.e+03& 9.e+03& 1.e+04& 2.e+04& 3.e+04& 4.e+04& 6.e+04& 7.e+04& 4.e+04  \\
Pe      &  4.e+02& 6.e+02& 1.e+03& 6.e+02& 4.e+02& 1.e+02& 6.e+01& 2.e+00& 2.e-01& 7.e-04  \\
Re      &  2.e+12& 5.e+12& 1.e+13& 2.e+13& 3.e+13& 4.e+13& 6.e+13& 9.e+13& 1.e+14& 7.e+13  \\
B       &  2.e+07& 4.e+06& 2.e+06& 7.e+05& 3.e+05& 2.e+05& 1.e+05& 8.e+04& 8.e+04& 5.e+05  \\
Pr      &  2.e-10& 1.e-10& 1.e-10& 3.e-11& 1.e-11& 3.e-12& 1.e-12& 2.e-14& 2.e-15& 1.e-17  \\
$\ell_z$&  1.e+08& 2.e+08& 3.e+08& 5.e+08& 7.e+08& 9.e+08& 1.e+09& 1.e+09& 2.e+09& 9.e+08  \\
$w_{rms}$& 3.e+00& 1.e+01& 3.e+01& 6.e+01& 1.e+02& 2.e+02& 3.e+02& 5.e+02& 6.e+02& 2.e+02  \\
$Pe_t$  &  7.e-03& 3.e-02& 1.e-01& 1.e-01& 1.e-01& 6.e-02& 4.e-02& 2.e-03& 2.e-04& 2.e-07  \\
$BPr$   &  4.e-03& 5.e-04& 2.e-04& 2.e-05& 5.e-06& 6.e-07& 1.e-07& 2.e-09& 2.e-10& 5.e-12  \\
\hline
\hline
\end{tabular}
\tablefoot{Velocities are from TIDES model Case 41 
All units are cgs except $r$ which is given in units of R$_\odot$. $T$ is in $^\circ$K. Stellar
parameters from the BEC  structure model and $\kappa_T$ from Fig. 7 of \citet{2016ApJ...821...49G}.
}
\end{table*}

\subsection{Implications for binary star observations}

Interactions in binary stars can be classified into two general classes.  The first includes
binary systems where only the observational diagnostics are affected, with no impact on
the stellar structure. These interactions are:  physical eclipses, which cause variations in the 
total measured light; wind eclipses, in which the stellar wind from one star absorbs and  scatters 
light from the companion (see, for example, M\"unch 1950);  Wind-wind collisions, in which the two 
stars possess a stellar wind that collides far from the stellar photosphere, producing excess 
emission at certain wavelengths and distorting the profiles of lines that are formed within the winds.  

The second class includes all stars in which the presence of a companion has an impact on the internal 
structure and subsequent evolution.  The most evident members of this class are systems undergoing Roche 
lobe overflow or wind accretion, and those in which the companion finds itself embedded in the extended
post-main sequence envelope of the primary star and intervenes in its ejection. Less evident members
are stars whose surfaces are shocked by the companion's stellar wind, are subjected to external
irradiation, or undergo tidally induced oscillations. In the first two of these processes, only 
the external layers are likely to be  affected. Tidal perturbations, however, penetrate deeper
layers and, in addition, the associated frequencies may resonante with the normal oscillation modes of
the star. 

Tidally induced oscillations are detected through photometric and photospheric line-profile variability. 
The observed oscillation in the light curves are generally analyzed in the context of  theoretical
frameworks described in \citet{1995ApJ...449..294K}, \citet{2017MNRAS.472.1538F} and
\citet{2020ApJ...888...95G}, and references therein. The theoretical study of analogous effects in the 
spectral lines has been significantly more limited.  Calculations from first principles of the line 
profile and its orbital phase-dependent variability was performed for the eccentric binaries $\epsilon$ Orionis
\citep{Moreno:2005cq} and  $\alpha$ Virginis \citep{2009ApJ...704..813H, 2013A&A...556A..49P,
2013A&A...552A..39P, 2016A&A...590A..54H} and a sample of other massive binary systems known to exhibit 
line-profile variability \citep{2013A&A...552A..39P}.  These studies used the one-layer TIDES model and 
confirmed that the Doppler shifts due to the horizontal component of the tidal perturbation dominate 
over the radial component (see \citealt{2009ApJ...704..813H}, Fig. 13). 

In the case of Spica, for which a large set of spectroscopic observations were available, the
TIDES model adequately predicted the periodic changes in the photospheric absorption-line wings and 
the presence of discrete features that travel from the blue to the red wing of the line. A
detailed fit to the line profiles and variations was not attempted because of uncertainties in
the orbital inclination and stellar properties, which have become available more recently
\citep{2016MNRAS.458.1964T}.  Also, a phase-difference between predicted and observed profiles 
was hypothesized to possibly be associated with the neglect of layers below the stellar surface.

\begin{figure}
\centering
\includegraphics[width=0.88\columnwidth]{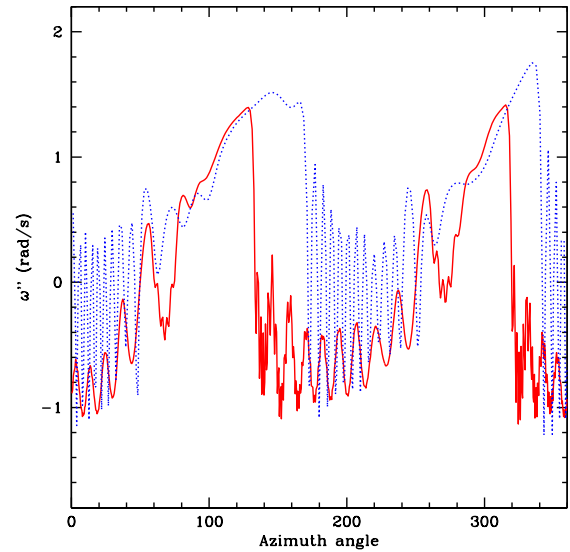}
\caption{Comparison of the azimuthal velocity predicted with the one-layer TIDES model (red) used in
Harrington et al. (2009) and the $n$-layer model discussed in this paper (blue dash, Case 25 in  
Table~\ref{table_models}).  
Both curves correspond to the periastron. 
}
\label{fig_CaseGGD}
\end{figure}

Although revisiting the line-profile variability of Spica is beyond the scope of this paper, 
we used the $n$-layer version of the TIDES code to explore the extent to which the 
azimuthal motions of the surface layer are influenced by the layers below.  In Fig.~\ref{fig_CaseGGD}  
the surface angular velocity from the one-layer TIDES model at periastron obtained with the same input 
parameters as used in Harrington et al (2009) is compared with the $n$-layer TIDES model using the
same input parameters (Case 25 in Table~\ref{table_models}). The general shape of both curves 
is very similar, showing broad maxima around $\varphi'\sim$100$^\circ$ and 300$^\circ$ and 
high-frequency oscillations around $\varphi'\sim$0$^\circ$ and 200$^\circ$. However, the high-frequency
oscillations are significantly different and the maxima are broader in the $n$-layer calculation.
Both of these will have an effect on the detailed structure of the line profiles and their variability,
but not the more general features that are detectable through observations.  Also noteworthy is that
the $n$-layer TIDES model predicts variations in the variability pattern over long timescales, thus
allowing for the possibility of superorbital periodicities, contrary to the one-layer model which 
predicts strictly orbital phase-locked variations over long timescales. 

\begin{figure*}
\centering
\includegraphics[width=0.79\columnwidth]{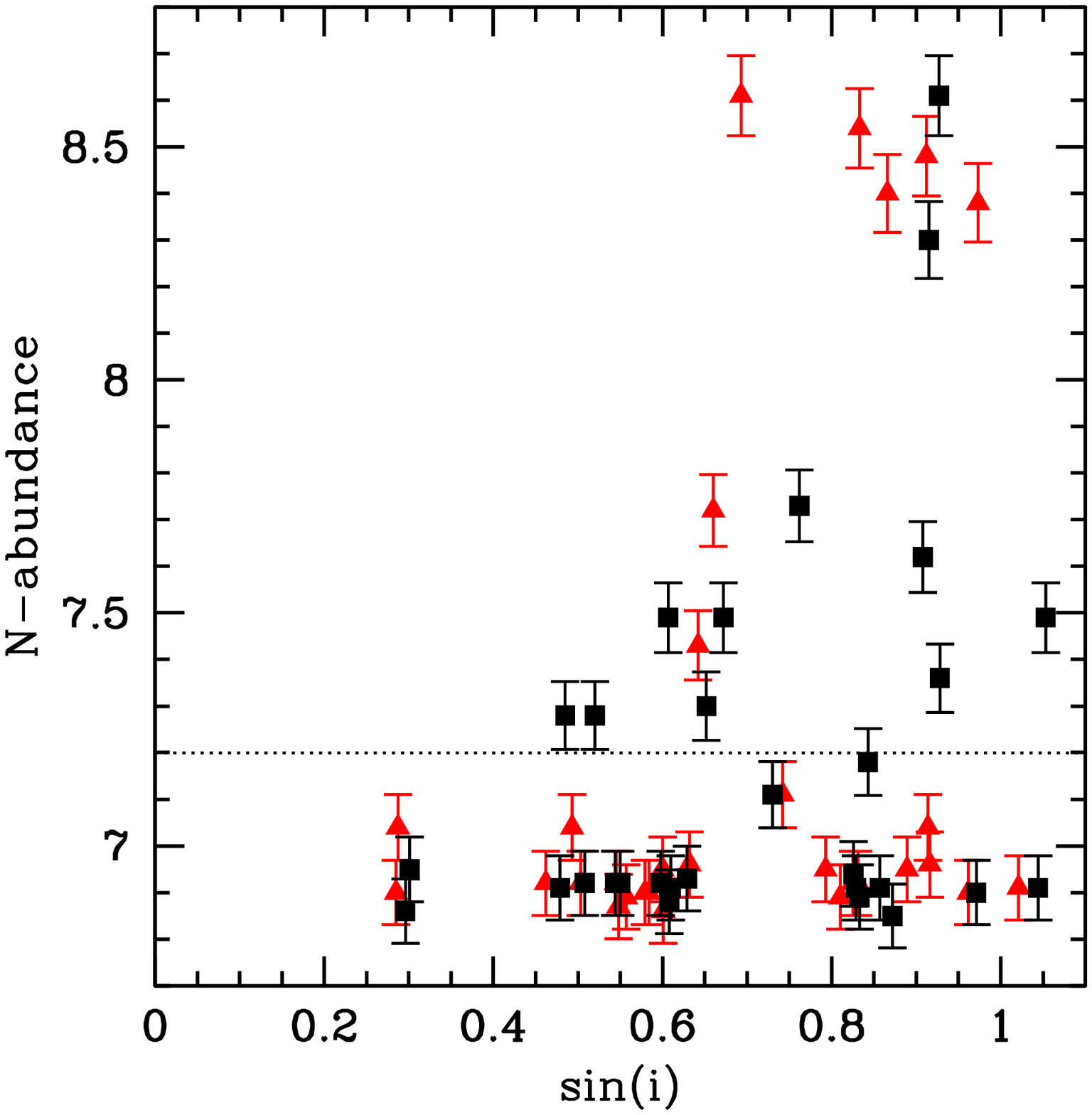}
\includegraphics[width=0.79\columnwidth]{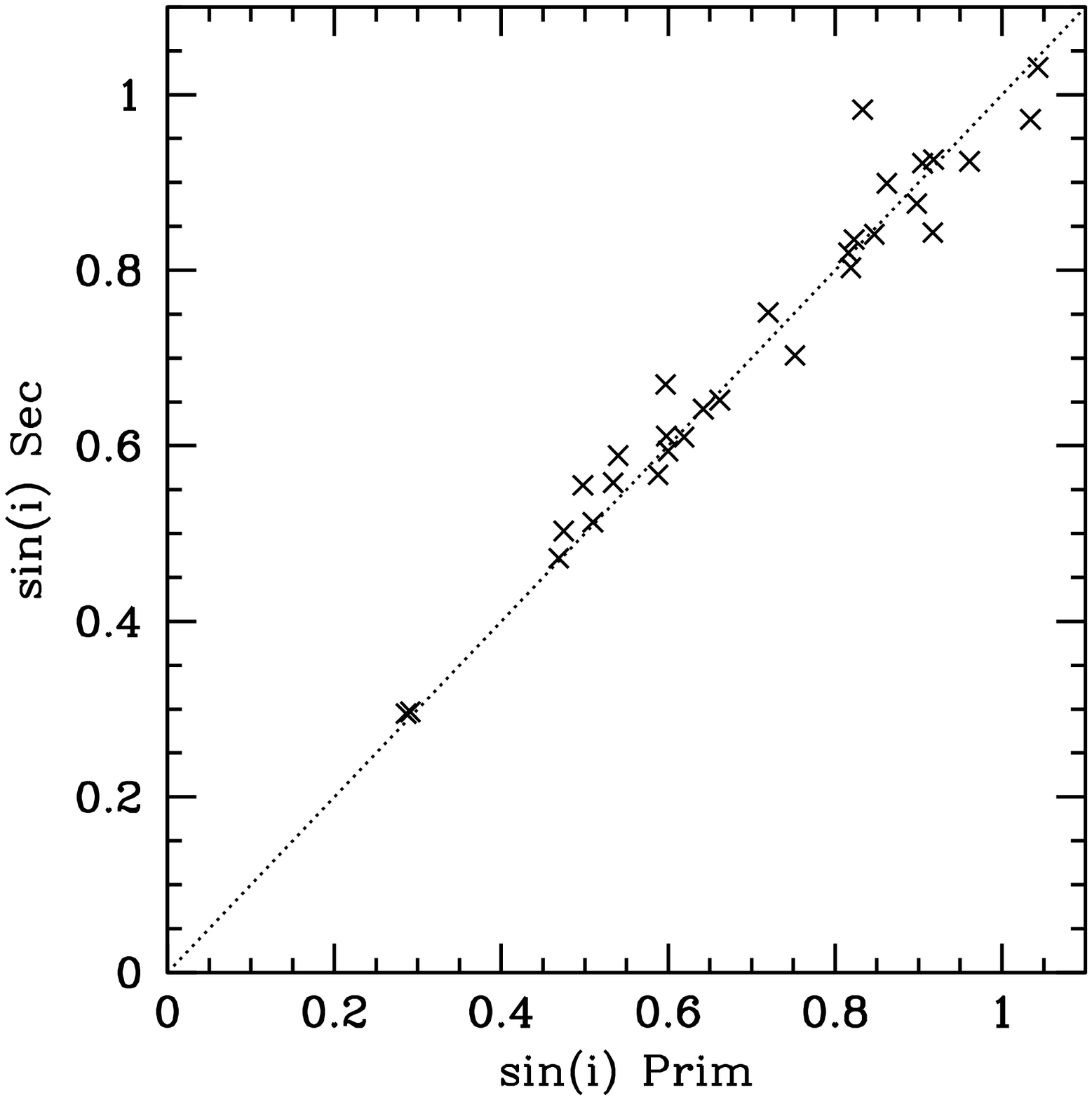}
\caption{Data from \citet{2020A&A...634A.118M} suggesting a possible correlation between nitrogen 
abundance and orbital inclination (left). Black squares correspond to the primary component and 
red triangles to the secondary of each system.
{\it Right:} Orbital inclination deduced from the data in Mahy et al. (see text) for the primary and
secondary components of the binary systems. 
}
\label{fig_Mahy}
\end{figure*}

The amplitude of observed tidal effects depends  on the orbital inclination because
the perturbations  are strongest at and around the equatorial latitude.\footnote{We offer a
reminder that in our model the stellar and orbital angular velocity vectors are parallel.
Other configurations may lead to different implications.}  Curiously, some observational
evidence can be found supporting the idea that tidal effects influence the results obtained
from spectroscopic analyses.   \citet{2020A&A...634A.118M}  obtained chemical abundances for a sample of
massive stars,  members of 32 double-line binary systems.  Using their data for $m \sin{i}^3$
and $M_{spec}$ (their Table 1), one may deduce an estimate for $\sin(i)$, the orbital inclination.
In Fig.~\ref{fig_Mahy} we plot the derived nitrogen abundance (from their Table 2) as a function of
$\sin(i)$ and find a clear trend for larger N-abundance with larger inclination angle.  Whether this
implies a non-homogeneous distribution of nuclear-processed elements or whether the tidal effects
introduce an as yet unaccounted for bias in the data analysis remains to be ascertained.
If most binary stars do not develop strong shear instabilities, the latter would be most likely.
However, further analysis is needed to shed light on this issue.

An additional implication of the results that we present in this paper is the dependence
on the synchronicity parameter $\beta$ of the tidal velocity amplitude:  The further $\beta$ 
departs from synchronicity, the larger the oscillation amplitude. Here, $\beta$ is the
ratio between the stellar rotation angular velocity and the instantaneous orbital angular
velocity, and in a differentially rotating star, there are different values of $\beta$ at
different radii. Furthermore, in eccentric binaries, the orbital angular velocity is smallest at 
apastron and largest at periastron.  Thus, although the orbital separation at apastron is 
larger than at periastron, the larger value of $\beta$ may induce greater surface variability than
at periastron, especially for cases in which $\beta \simeq$1 around the periastron phase.  
If one of the processes driving mass loss is associated with stellar pulsations (see, for example,
\citealt{2007AIPC..948..345T} and  \citealt{2015A&A...581A..75K}), then phenomena that depend on
the mass loss rate (such as wind-wind collision emission and accretion onto a companion) may not 
present themselves as expected from models that only take the orbital separation into account.

Finally, we address the apparently curious result obtained by \citet{2018MNRAS.481.3129P} whose 
analysis of a sample of massive binary stars failed to reveal the expected correlation between 
stellar rotation velocity and surface chemical composition.  In light of the numerical simulations
presented here in Sect. 2.6, this result might not be unexpected.  As already pointed out by 
\citet{2013EAS....64..339K}, binaries that are tidally locked in the strict sense of the term, 
meaning that $\beta$=1 at all times and in all layers, have very stable structures and rapidly
damp out differential rotation and localized flows that would have otherwise contributed to
the mixing  efficiency.  Additional observational evidence supporting the idea that differential 
rotation may be  suppressed in tidally locked low-mass stars has been presented by 
\citet{2006MNRAS.367.1699H} and discussed  by \citet{2007AN....328.1030C}. 
A more controversial question concerns the effect on meridional circulation currents.  
The recent analysis by \citet{2020A&A...641A..86H} leads to the conclusion
that the mixing efficiency due to meridional currents in tidally locked binaries should be 
significantly larger than in their single-star counterparts. The results of \citet{2018MNRAS.481.3129P}
indicate the contrary, thus leading to the speculation that meridional currents may be if not suppressed,
at least diminished, in tidally locked systems.

\section{Summary of results and conclusions}

Binary stars are generally assumed to be tidally locked and thus their internal mixing processes are 
assumed to follow those of single stars in uniform rotation (but see \citealt{2020A&A...641A..86H}).  
However, the picture changes for  eccentric-orbit binaries, for young stars that have 
not had time to circularize and synchronize their orbits, and for stars that are undergoing core 
contraction and envelope expansion as they evolve off the  main sequence. These systems
are all in asynchronous rotation, a condition in  which tidal flows introduce a 3D time-varying 
rotation structure.  Scharlemann (1980) provided an analytical expression that quantifies the amplitude 
of these flows when the departure from equilibrium is very small. In this paper we illustrate a method 
for arbitrary rotation rates and eccentricity.

Our method is based on an upgraded version of the TIDES code, which solves
the equations of motion on a 3D grid of volume elements. The equations of motion include
gravitational, centrifugal, Coriolis, gas pressure and viscous forces and they are solved
simultaneously with the orbital motion of the companion.  The model does not include the 
effects of buoyancy and meridional circulation, nor does it include thermal diffusivity or 
radiative transport, all of which have an impact on the evolution over time of the rotation 
structure \citep{1967ApJ...150..571G, 2016ApJ...821...49G}.  However, as our model is used 
only to compute the  behavior of angular velocity over short timescales (less than a few decades), 
neglecting the effects with long timescales is not likely to significantly impact our results. 

In circular-orbit binaries with moderate departures from synchronicity, the angular velocity at a fixed 
position on the perturbed star, $\omega''$, undergoes sinusoidal-like oscillations that are driven 
by the tidal forcing. The amplitude and frequency of the oscillations grow with growing departures from 
synchronicity. For these systems, the analytical expression of Scharlemann (1981) is shown to reproduce 
the azimuthal component of the tidal velocity within factors of $\sim$2, even though this expression
was derived for much smaller departures ($\sim$1\%) from synchronicity and convective envelopes.  This 
suggests the potential use 
of the analytical expression to obtain a rough estimate of the tidal amplitudes. In the cases of large 
departures from synchronicity and in eccentric-orbit binaries, $\omega''$ is distorted from the
sinusoidal-like shape and it undergoes localized high-frequency oscillations.  Characterizing these
high-frequency oscillations requires  numerical simulations such as those that we present in this paper.
Such oscillations may now be observed through high precision photometry of objects such
as the ``heartbeat'' stars (see, for example, \citealt{2017MNRAS.472.1538F} and \citealt{2020ApJ...888...95G}),

The instantaneous value of $\omega''$ as a function of radius, latitude, and longitude that is provided
by our calculations can be used to inform on the potential impact of the tidal interactions on the mixing 
processes. As an example, we examined whether the velocity amplitudes in the internal layers of our test 
binary are able to trigger the diffusive shear instability in the context of the constraints that are 
given in \citet{2016ApJ...821...49G} and \citet{2020ApJ...901..146G}, and we find that the conditions for 
triggering the horizontal diffusive shear instability may be met in layers that lie relatively close to 
the stellar core. We note, however, that the constraints were not derived for a time-dependent oscillating 
fluid such as the tidal flows, and thus further work is needed to address this issue.

We analyzed the angular velocity averaged over the azimuth angle at a fixed latitude and  its
dependence on radius, which we call the  $\left <\omega''\right>$ profile. We showed that an
initially pronounced differential rotation structure flattens over time and tends toward
uniform rotation  on the viscous timescale.  For the viscosity values considered in this paper
(3$-$160$\times$10$^{13}$ cm$^2$~s$^{-1}$), the  $\left<\omega''\right>$ profile $\rightarrow$ constant
on $\tau<$250\,yr, which is negligible compared to most evolutionary timescales.  

We discussed some of the observational implications that may be drawn from our numerical experiments.
For example, the instantaneous value of $\beta$ in eccentric binaries depends on the orbital phase,
reaching maximum at apastron.  Thus, physical processes that may be affected by the tidally induced
perturbations (such as high-frequency oscillations and mass loss) may be orbital-phase-dependent if
their characteristic timescales are short enough.  Thus, we propose that the synchronicity 
parameter $\beta$ and its orbital-phase dependence be taken into account when analyzing eccentric systems.
For tidally locked systems we find that perturbations are rapidly damped down, providing a possible
explanation for the observational finding that the mixing efficiency in some of these systems appears to be
lower than expected given their rotation rate \citep{2018MNRAS.481.3129P}.

We uncover a curious result in observational data that suggests higher derived nitrogen abundances
in binaries with larger orbital inclinations.  If real, this would suggest an inhomogeneous chemical
distribution over the stellar surface, although a more likely explanation may reside in the tidal
perturbations, which are strongest at the equator, and may distort the spectral lines that are used
for the abundance determinations. 

In our analysis we have neglected the effects of radiative damping, feedback between rotation and
orbital motion and magnetic fields. The latter may be omnipresent in the stellar interior (see for example, 
\citealt{2021A&A...646A..19T}). It has been suggested that internal magnetic fields reduce the 
radial shear in stars \citep{2005A&A...440L...9E, 2005ApJ...626..350H, 2008A&A...481L..87S},
which is also required by observations \citep{2014A&A...564A..27D, 2012A&A...540A.143M}.
On the other hand, the growth time of the tidally induced oscillations is very short, and the associated  
shear motion reverses periodically, such that it may be possible that the shearing rates obtained by
our models persist even when magnetic fields are considered.
Future investigations are needed to address these issues.

\begin{acknowledgements}
We thank an anonymous referee for comments and suggestions that significantly improved this paper.
GK is grateful to Fr\'ed\'eric Masset and Fabian R.N. Schneider who commented and made helpful 
suggestions on an earlier draft of this paper, and to Matteo Cantiello, Francois Leyvraz, Luis Mochan and 
Sergio Cuevas for many helpful discussions.
We acknowledge support from Conacyt project 252499 and UNAM/PAPIIT IN103619
\end{acknowledgements}

\bibliographystyle{aa} 
\bibliography{TIDES} 

\begin{thebibliography}{82}
\expandafter\ifx\csname natexlab\endcsname\relax\def\natexlab#1{#1}\fi

\bibitem[{{Aerts} {et~al.}(2019){Aerts}, {Mathis}, \&
  {Rogers}}]{2019ARA&A..57...35A}
{Aerts}, C., {Mathis}, S., \& {Rogers}, T.~M. 2019, \araa, 57, 35

\bibitem[{{Alexander}(1973)}]{1973Ap&SS..23..459A}
{Alexander}, M.~E. 1973, \apss, 23, 459

\bibitem[{{Brott} {et~al.}(2011){Brott}, {de Mink}, {Cantiello}, {Langer}, {de
  Koter}, {Evans}, {Hunter}, {Trundle}, \& {Vink}}]{2011A&A...530A.115B}
{Brott}, I., {de Mink}, S.~E., {Cantiello}, M., {et~al.} 2011, \aap, 530, A115

\bibitem[{{Cantiello} {et~al.}(2009){Cantiello}, {Langer}, {Brott}, {de Koter},
  {Shore}, {Vink}, {Voegler}, {Lennon}, \& {Yoon}}]{2009A&A...499..279C}
{Cantiello}, M., {Langer}, N., {Brott}, I., {et~al.} 2009, \aap, 499, 279

\bibitem[{{Chen}(2013)}]{2013PhDT.......256C}
{Chen}, E.~M. 2013, PhD thesis, University of California, Santa Cruz

\bibitem[{{Collier Cameron}(2007)}]{2007AN....328.1030C}
{Collier Cameron}, A. 2007, Astronomische Nachrichten, 328, 1030

\bibitem[{{De Marco} \& {Izzard}(2017)}]{2017PASA...34....1D}
{De Marco}, O. \& {Izzard}, R.~G. 2017, \pasa, 34, e001

\bibitem[{{de Mink} {et~al.}(2009){de Mink}, {Cantiello}, {Langer}, {Pols},
  {Brott}, \& {Yoon}}]{2009A&A...497..243D}
{de Mink}, S.~E., {Cantiello}, M., {Langer}, N., {et~al.} 2009, \aap, 497, 243

\bibitem[{{Deheuvels} {et~al.}(2014){Deheuvels}, {Do{\u{g}}an}, {Goupil},
  {Appourchaux}, {Benomar}, {Bruntt}, {Campante}, {Casagrande}, {Ceillier},
  {Davies}, {De Cat}, {Fu}, {Garc{\'\i}a}, {Lobel}, {Mosser}, {Reese},
  {Regulo}, {Schou}, {Stahn}, {Thygesen}, {Yang}, {Chaplin},
  {Christensen-Dalsgaard}, {Eggenberger}, {Gizon}, {Mathis},
  {Molenda-{\.Z}akowicz}, \& {Pinsonneault}}]{2014A&A...564A..27D}
{Deheuvels}, S., {Do{\u{g}}an}, G., {Goupil}, M.~J., {et~al.} 2014, \aap, 564,
  A27

\bibitem[{Dolginov \& Smel'Chakova(1992)}]{Dolginov:1992up}
Dolginov, A.~Z. \& Smel'Chakova, E.~V. 1992, A{\&}A, 257, 783

\bibitem[{{Eggenberger} {et~al.}(2005){Eggenberger}, {Maeder}, \&
  {Meynet}}]{2005A&A...440L...9E}
{Eggenberger}, P., {Maeder}, A., \& {Meynet}, G. 2005, \aap, 440, L9

\bibitem[{{Ekstr{\"o}m} {et~al.}(2012){Ekstr{\"o}m}, {Georgy}, {Eggenberger},
  {Meynet}, {Mowlavi}, {Wyttenbach}, {Granada}, {Decressin}, {Hirschi},
  {Frischknecht}, {Charbonnel}, \& {Maeder}}]{Ekstrom_etal2012}
{Ekstr{\"o}m}, S., {Georgy}, C., {Eggenberger}, P., {et~al.} 2012, \aap, 537,
  A146

\bibitem[{{Fliegner} {et~al.}(1996){Fliegner}, {Langer}, \&
  {Venn}}]{1996A&A...308L..13F}
{Fliegner}, J., {Langer}, N., \& {Venn}, K.~A. 1996, \aap, 308, L13

\bibitem[{{Fuller}(2017)}]{2017MNRAS.472.1538F}
{Fuller}, J. 2017, \mnras, 472, 1538

\bibitem[{{Gagnier} {et~al.}(2019){Gagnier}, {Rieutord}, {Charbonnel},
  {Putigny}, \& {Espinosa Lara}}]{2019A&A...625A..89G}
{Gagnier}, D., {Rieutord}, M., {Charbonnel}, C., {Putigny}, B., \& {Espinosa
  Lara}, F. 2019, \aap, 625, A89

\bibitem[{{Garaud}(2020)}]{2020ApJ...901..146G}
{Garaud}, P. 2020, \apj, 901, 146

\bibitem[{{Garaud} \& {Kulenthirarajah}(2016)}]{2016ApJ...821...49G}
{Garaud}, P. \& {Kulenthirarajah}, L. 2016, \apj, 821, 49

\bibitem[{{Goldreich} \& {Schubert}(1967)}]{1967ApJ...150..571G}
{Goldreich}, P. \& {Schubert}, G. 1967, \apj, 150, 571

\bibitem[{{Grassitelli} {et~al.}(2015){Grassitelli}, {Fossati},
  {Sim{\'o}n-Di{\'a}z}, {Langer}, {Castro}, \& {Sanyal}}]{2015ApJ...808L..31G}
{Grassitelli}, L., {Fossati}, L., {Sim{\'o}n-Di{\'a}z}, S., {et~al.} 2015,
  \apjl, 808, L31

\bibitem[{{Guo} {et~al.}(2020){Guo}, {Shporer}, {Hambleton}, \&
  {Isaacson}}]{2020ApJ...888...95G}
{Guo}, Z., {Shporer}, A., {Hambleton}, K., \& {Isaacson}, H. 2020, \apj, 888,
  95

\bibitem[{{Harrington} {et~al.}(2009){Harrington}, {Koenigsberger}, {Moreno},
  \& {Kuhn}}]{2009ApJ...704..813H}
{Harrington}, D., {Koenigsberger}, G., {Moreno}, E., \& {Kuhn}, J. 2009, \apj,
  704, 813

\bibitem[{{Harrington} {et~al.}(2016){Harrington}, {Koenigsberger},
  {Olgu{\'\i}n}, {Ilyin}, {Berdyugina}, {Lara}, \&
  {Moreno}}]{2016A&A...590A..54H}
{Harrington}, D., {Koenigsberger}, G., {Olgu{\'\i}n}, E., {et~al.} 2016, \aap,
  590, A54

\bibitem[{{Hastings} {et~al.}(2020){Hastings}, {Langer}, \&
  {Koenigsberger}}]{2020A&A...641A..86H}
{Hastings}, B., {Langer}, N., \& {Koenigsberger}, G. 2020, \aap, 641, A86

\bibitem[{{Heger} {et~al.}(2000){Heger}, {Langer}, \&
  {Woosley}}]{2000ApJ...528..368H}
{Heger}, A., {Langer}, N., \& {Woosley}, S.~E. 2000, \apj, 528, 368

\bibitem[{{Heger} {et~al.}(2005){Heger}, {Woosley}, \&
  {Spruit}}]{2005ApJ...626..350H}
{Heger}, A., {Woosley}, S.~E., \& {Spruit}, H.~C. 2005, \apj, 626, 350

\bibitem[{{Hussain} {et~al.}(2006){Hussain}, {Allende Prieto}, {Saar}, \&
  {Still}}]{2006MNRAS.367.1699H}
{Hussain}, G.~A.~J., {Allende Prieto}, C., {Saar}, S.~H., \& {Still}, M. 2006,
  \mnras, 367, 1699

\bibitem[{{Hut}(1981)}]{1981A&A....99..126H}
{Hut}, P. 1981, \aap, 99, 126

\bibitem[{{Koenigsberger} \& {Moreno}(2013)}]{2013EAS....64..339K}
{Koenigsberger}, G. \& {Moreno}, E. 2013, in EAS Publications Series, Vol.~64,
  EAS Publications Series, ed. K.~{Pavlovski}, A.~{Tkachenko}, \& G.~{Torres},
  339--342

\bibitem[{{Koenigsberger} \& {Moreno}(2016)}]{2016RMxAA..52..113K}
{Koenigsberger}, G. \& {Moreno}, E. 2016, \rmxaa, 52, 113

\bibitem[{{Koenigsberger} {et~al.}(2012){Koenigsberger}, {Moreno}, \&
  {Harrington}}]{2012A&A...539A..84K}
{Koenigsberger}, G., {Moreno}, E., \& {Harrington}, D.~M. 2012, \aap, 539, A84

\bibitem[{{Kraus} {et~al.}(2015){Kraus}, {Haucke}, {Cidale}, {Venero},
  {Nickeler}, {N{\'e}meth}, {Niemczura}, {Tomi{\'c}}, {Aret}, {Kub{\'a}t},
  {Kub{\'a}tov{\'a}}, {Oksala}, {Cur{\'e}}, {Kami{\'n}ski}, {Dimitrov},
  {Fagas}, \& {Poli{\'n}ska}}]{2015A&A...581A..75K}
{Kraus}, M., {Haucke}, M., {Cidale}, L.~S., {et~al.} 2015, \aap, 581, A75

\bibitem[{{Kumar} {et~al.}(1995){Kumar}, {Ao}, \&
  {Quataert}}]{1995ApJ...449..294K}
{Kumar}, P., {Ao}, C.~O., \& {Quataert}, E.~J. 1995, \apj, 449, 294

\bibitem[{{Langer}(1991)}]{1991A&A...243..155L}
{Langer}, N. 1991, \aap, 243, 155

\bibitem[{{Langer}(1998)}]{1998A&A...329..551L}
{Langer}, N. 1998, \aap, 329, 551

\bibitem[{{Langer}(2012)}]{2012ARA&A..50..107L}
{Langer}, N. 2012, \araa, 50, 107

\bibitem[{{Langer} {et~al.}(1997){Langer}, {Heger}, \&
  {Fliegner}}]{1997IAUS..189..343L}
{Langer}, N., {Heger}, A., \& {Fliegner}, J. 1997, in IAU Symposium, Vol. 189,
  IAU Symposium, ed. T.~R. {Bedding}, A.~J. {Booth}, \& J.~{Davis}, 343--348

\bibitem[{{Lanza} \& {Mathis}(2016)}]{2016CeMDA.126..249L}
{Lanza}, A.~F. \& {Mathis}, S. 2016, Celestial Mechanics and Dynamical
  Astronomy, 126, 249

\bibitem[{{Ligni{\`e}res}(1999)}]{1999A&A...348..933L}
{Ligni{\`e}res}, F. 1999, \aap, 348, 933

\bibitem[{{Lovekin}(2020)}]{2020FrASS...6...77L}
{Lovekin}, C.~C. 2020, Frontiers in Astronomy and Space Sciences, 6, 77

\bibitem[{{Maeder}(1987)}]{1987A&A...178..159M}
{Maeder}, A. 1987, \aap, 178, 159

\bibitem[{{Maeder} \& {Meynet}(2000)}]{2000ARA&A..38..143M}
{Maeder}, A. \& {Meynet}, G. 2000, \araa, 38, 143

\bibitem[{{Mahy} {et~al.}(2020){Mahy}, {Sana}, {Abdul-Masih}, {Almeida},
  {Langer}, {Shenar}, {de Koter}, {de Mink}, {de Wit}, {Grin}, {Evans},
  {Moffat}, {Schneider}, {Barb{\'a}}, {Clark}, {Crowther}, {Gr{\"a}fener},
  {Lennon}, {Tramper}, \& {Vink}}]{2020A&A...634A.118M}
{Mahy}, L., {Sana}, H., {Abdul-Masih}, M., {et~al.} 2020, \aap, 634, A118

\bibitem[{{Mathis} {et~al.}(2004){Mathis}, {Palacios}, \&
  {Zahn}}]{2004A&A...425..243M}
{Mathis}, S., {Palacios}, A., \& {Zahn}, J.~P. 2004, \aap, 425, 243

\bibitem[{{Mathis} {et~al.}(2007){Mathis}, {Palacios}, \&
  {Zahn}}]{2007A&A...462.1063M}
{Mathis}, S., {Palacios}, A., \& {Zahn}, J.~P. 2007, \aap, 462, 1063

\bibitem[{{Mathis} {et~al.}(2018){Mathis}, {Prat}, {Amard}, {Charbonnel},
  {Palacios}, {Lagarde}, \& {Eggenberger}}]{2018A&A...620A..22M}
{Mathis}, S., {Prat}, V., {Amard}, L., {et~al.} 2018, \aap, 620, A22

\bibitem[{{Meyer} \& {Linden}(2014)}]{2014JFM...753..242M}
{Meyer}, C.~R. \& {Linden}, P.~F. 2014, Journal of Fluid Mechanics, 753, 242

\bibitem[{{Meynet} \& {Maeder}(2000)}]{2000A&A...361..101M}
{Meynet}, G. \& {Maeder}, A. 2000, \aap, 361, 101

\bibitem[{{Moreno} \& {Koenigsberger}(1999)}]{1999RMxAA..35..157M}
{Moreno}, E. \& {Koenigsberger}, G. 1999, \rmxaa, 35, 157

\bibitem[{Moreno {et~al.}(2011)Moreno, Koenigsberger, \&
  Harrington}]{Moreno:2011jq}
Moreno, E., Koenigsberger, G., \& Harrington, D.~M. 2011, A{\&}A, 528, 48

\bibitem[{Moreno {et~al.}(2005)Moreno, Koenigsberger, \&
  Toledano}]{Moreno:2005cq}
Moreno, E., Koenigsberger, G., \& Toledano, O. 2005, A{\&}A, 437, 641

\bibitem[{{Mosser} {et~al.}(2012){Mosser}, {Goupil}, {Belkacem}, {Michel},
  {Stello}, {Marques}, {Elsworth}, {Barban}, {Beck}, {Bedding}, {De Ridder},
  {Garc{\'\i}a}, {Hekker}, {Kallinger}, {Samadi}, {Stumpe}, {Barclay}, \&
  {Burke}}]{2012A&A...540A.143M}
{Mosser}, B., {Goupil}, M.~J., {Belkacem}, K., {et~al.} 2012, \aap, 540, A143

\bibitem[{{Ogilvie}(2014)}]{2014ARA&A..52..171O}
{Ogilvie}, G.~I. 2014, \araa, 52, 171

\bibitem[{{Pablo} {et~al.}(2017){Pablo}, {Richardson}, {Fuller}, {Rowe},
  {Moffat}, {Kuschnig}, {Popowicz}, {Handler}, {Neiner}, {Pigulski}, {Wade},
  {Weiss}, {Buysschaert}, {Ramiaramanantsoa}, {Bratcher}, {Gerhartz}, {Greco},
  {Hardegree-Ullman}, {Lembryk}, \& {Oswald}}]{2017MNRAS.467.2494P}
{Pablo}, H., {Richardson}, N.~D., {Fuller}, J., {et~al.} 2017, \mnras, 467,
  2494

\bibitem[{{Palate} {et~al.}(2013{\natexlab{a}}){Palate}, {Koenigsberger},
  {Rauw}, {Harrington}, \& {Moreno}}]{2013A&A...556A..49P}
{Palate}, M., {Koenigsberger}, G., {Rauw}, G., {Harrington}, D., \& {Moreno},
  E. 2013{\natexlab{a}}, \aap, 556, A49

\bibitem[{{Palate} {et~al.}(2013{\natexlab{b}}){Palate}, {Rauw},
  {Koenigsberger}, \& {Moreno}}]{2013A&A...552A..39P}
{Palate}, M., {Rauw}, G., {Koenigsberger}, G., \& {Moreno}, E.
  2013{\natexlab{b}}, \aap, 552, A39

\bibitem[{{Park} {et~al.}(2021){Park}, {Prat}, {Mathis}, \&
  {Bugnet}}]{2021A&A...646A..64P}
{Park}, J., {Prat}, V., {Mathis}, S., \& {Bugnet}, L. 2021, \aap, 646, A64

\bibitem[{{Pavlovski} {et~al.}(2018){Pavlovski}, {Southworth}, \&
  {Tamajo}}]{2018MNRAS.481.3129P}
{Pavlovski}, K., {Southworth}, J., \& {Tamajo}, E. 2018, \mnras, 481, 3129

\bibitem[{{Penev} {et~al.}(2009){Penev}, {Barranco}, \&
  {Sasselov}}]{2009ApJ...705..285P}
{Penev}, K., {Barranco}, J., \& {Sasselov}, D. 2009, \apj, 705, 285

\bibitem[{{Penev} {et~al.}(2007){Penev}, {Sasselov}, {Robinson}, \&
  {Demarque}}]{2007ApJ...655.1166P}
{Penev}, K., {Sasselov}, D., {Robinson}, F., \& {Demarque}, P. 2007, \apj, 655,
  1166

\bibitem[{{Prat} \& {Ligni{\`e}res}(2014)}]{2014A&A...566A.110P}
{Prat}, V. \& {Ligni{\`e}res}, F. 2014, \aap, 566, A110

\bibitem[{{Press} {et~al.}(1975){Press}, {Smarr}, \&
  {Wiita}}]{1975ApJ...202L.135P}
{Press}, W.~H., {Smarr}, L.~L., \& {Wiita}, P.~J. 1975, \apjl, 202, L135

\bibitem[{{Proffitt} {et~al.}(2016){Proffitt}, {Lennon}, {Langer}, \&
  {Brott}}]{2016ApJ...824....3P}
{Proffitt}, C.~R., {Lennon}, D.~J., {Langer}, N., \& {Brott}, I. 2016, \apj,
  824, 3

\bibitem[{{Richard} \& {Zahn}(1999)}]{1999A&A...347..734R}
{Richard}, D. \& {Zahn}, J.-P. 1999, \aap, 347, 734

\bibitem[{{Richardson}(1920)}]{1920RSPSA..97..354R}
{Richardson}, L.~F. 1920, Proceedings of the Royal Society of London Series A,
  97, 354

\bibitem[{{Scharlemann}(1981)}]{1981ApJ...246..292S}
{Scharlemann}, E.~T. 1981, \apj, 246, 292

\bibitem[{{Song} {et~al.}(2013){Song}, {Maeder}, {Meynet}, {Huang},
  {Ekstr{\"o}m}, \& {Granada}}]{2013A&A...556A.100S}
{Song}, H.~F., {Maeder}, A., {Meynet}, G., {et~al.} 2013, \aap, 556, A100

\bibitem[{{Suijs} {et~al.}(2008){Suijs}, {Langer}, {Poelarends}, {Yoon},
  {Heger}, \& {Herwig}}]{2008A&A...481L..87S}
{Suijs}, M.~P.~L., {Langer}, N., {Poelarends}, A.~J., {et~al.} 2008, \aap, 481,
  L87

\bibitem[{{Takahashi} \& {Langer}(2021)}]{2021A&A...646A..19T}
{Takahashi}, K. \& {Langer}, N. 2021, \aap, 646, A19

\bibitem[{{Talon} {et~al.}(1997){Talon}, {Zahn}, {Maeder}, \&
  {Meynet}}]{1997A&A...322..209T}
{Talon}, S., {Zahn}, J.~P., {Maeder}, A., \& {Meynet}, G. 1997, \aap, 322, 209

\bibitem[{{Tassoul}(1987)}]{1987ApJ...322..856T}
{Tassoul}, J.-L. 1987, \apj, 322, 856

\bibitem[{{Tassoul} \& {Tassoul}(1982)}]{1982ApJ...261..265T}
{Tassoul}, J.~L. \& {Tassoul}, M. 1982, \apj, 261, 265

\bibitem[{{Tkachenko} {et~al.}(2016){Tkachenko}, {Matthews}, {Aerts},
  {Pavlovski}, {P{\'a}pics}, {Zwintz}, {Cameron}, {Walker}, {Kuschnig},
  {Degroote}, {Debosscher}, {Moravveji}, {Kolbas}, {Guenther}, {Moffat},
  {Rowe}, {Rucinski}, {Sasselov}, \& {Weiss}}]{2016MNRAS.458.1964T}
{Tkachenko}, A., {Matthews}, J.~M., {Aerts}, C., {et~al.} 2016, \mnras, 458,
  1964

\bibitem[{{Toledano} {et~al.}(2007{\natexlab{a}}){Toledano}, {Koenigsberger},
  \& {Moreno}}]{2007ASPC..367..437T}
{Toledano}, O., {Koenigsberger}, G., \& {Moreno}, E. 2007{\natexlab{a}}, in
  Astronomical Society of the Pacific Conference Series, Vol. 367, Massive
  Stars in Interactive Binaries, ed. N.~{St. -Louis} \& A.~F.~J. {Moffat}, 437

\bibitem[{{Toledano} {et~al.}(2007{\natexlab{b}}){Toledano}, {Moreno},
  {Koenigsberger}, {Detmers}, \& {Langer}}]{2007A&A...461.1057T}
{Toledano}, O., {Moreno}, E., {Koenigsberger}, G., {Detmers}, R., \& {Langer},
  N. 2007{\natexlab{b}}, \aap, 461, 1057

\bibitem[{{Townsend}(2007)}]{2007AIPC..948..345T}
{Townsend}, R. 2007, in American Institute of Physics Conference Series, Vol.
  948, Unsolved Problems in Stellar Physics: A Conference in Honor of Douglas
  Gough, ed. R.~J. {Stancliffe}, G.~{Houdek}, R.~G. {Martin}, \& C.~A. {Tout},
  345--356

\bibitem[{{Vidal} {et~al.}(2019){Vidal}, {C{\'e}bron}, {ud-Doula}, \&
  {Alecian}}]{2019A&A...629A.142V}
{Vidal}, J., {C{\'e}bron}, D., {ud-Doula}, A., \& {Alecian}, E. 2019, \aap,
  629, A142

\bibitem[{{Welsh} {et~al.}(2011){Welsh}, {Orosz}, {Aerts}, {Brown},
  {Brugamyer}, {Cochran}, {Gilliland}, {Guzik}, {Kurtz}, {Latham}, {Marcy},
  {Quinn}, {Zima}, {Allen}, {Batalha}, {Bryson}, {Buchhave}, {Caldwell},
  {Gautier}, {Howell}, {Kinemuchi}, {Ibrahim}, {Isaacson}, {Jenkins}, {Prsa},
  {Still}, {Street}, {Wohler}, {Koch}, \& {Borucki}}]{2011ApJS..197....4W}
{Welsh}, W.~F., {Orosz}, J.~A., {Aerts}, C., {et~al.} 2011, \apjs, 197, 4

\bibitem[{{Zahn}(1974)}]{1974IAUS...59..185Z}
{Zahn}, J.~P. 1974, in Stellar Instability and Evolution, ed. P.~{Ledoux},
  A.~{Noels}, \& A.~W. {Rodgers}, Vol.~59, 185

\bibitem[{{Zahn}(1975)}]{1975A&A....41..329Z}
{Zahn}, J.~P. 1975, \aap, 41, 329

\bibitem[{{Zahn}(1992)}]{1992A&A...265..115Z}
{Zahn}, J.~P. 1992, \aap, 265, 115

\bibitem[{{Zahn}(1993)}]{1993SSRv...66..285Z}
{Zahn}, J.~P. 1993, \ssr, 66, 285

\bibitem[{{Zahn}(2008)}]{2008EAS....29...67Z}
{Zahn}, J.~P. 2008, in EAS Publications Series, Vol.~29, EAS Publications
  Series, ed. M.~J. {Goupil} \& J.~P. {Zahn}, 67--90

\end{thebibliography}

\clearpage
\vfill\eject

\appendix

\section{Method and examples} \label{appendix_method}

The full set of models that were computed for this paper is listed in Table \ref{table_models}.
Column 1 lists the case number, Col. 2 indicates the value of the core asynchronicity parameter 
$\beta_0^0$, Col. 3 the viscosity value (in units of R$_\odot^2$ d$^{-1}$), Col. 4 the layer 
thickness, and Col. 5 the number of layers. Column 6  indicates the initial rotation condition, where 
uniform rotation is described by the constant $\beta_0^{k>0}$ value while nonuniform rotation is described 
by (a), (b), (c), (f) and the values listed in the footnote to this table.  Column 7 lists the number 
of orbital cycles over which the run was performed. The last column of the table indicates other input 
parameters that were varied from their nominal values, specifically, eccentricity $e$, the number of azimuthal or
latitudinal elements $N_\varphi$, $N_\theta$ and  the secondary mass $m_2$.  The number following
these variable names gives the modified value.

The rotation angular velocities in the inertial reference frame $S$ and the rotating frame $S'$ are
listed for the initial uniform rotation cases in Table~\ref{table2}.  The equatorial rotation velocity
at the surface of the equivalent rigidly rotating stars is 0\,km~s$^{-1}$ ($\beta_0^0$=0), 86\,km~s$^{-1}$
($\beta_0^0$=1) and 155\,km~s$^{-1}$  ($\beta_0^0$=1.8). The analogous list of angular velocities for the
nonuniform initial rotation states (a) and (c) are listed in Table \ref{table_VarBeta}.

\subsection{Transitory, stationary and equilibrium states}
\begin{figure}
\centering
\includegraphics[width=\columnwidth]{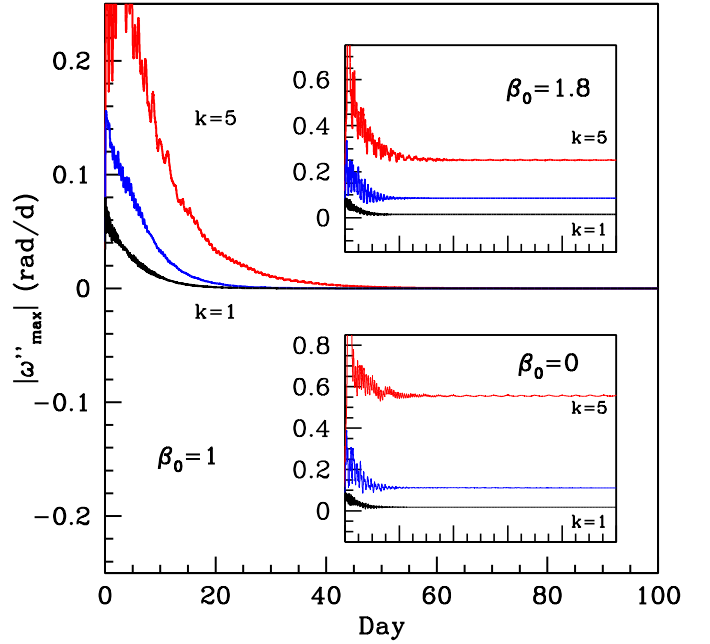}                
\caption{Evolution over time of the maximum angular velocity (absolute value) along the equator
in the rest frame of the rotating primary star ($S''$ coordinate system) at radius 6.6 R$_\odot$ (k=5, red),
6.2 R$_\odot$ (k=4, blue) and 5.0 R$_\odot$ (k=1, black), where k indicates the number assigned
to the layer.  Initial rotation is uniform. Time zero is the start of the calculation, and the
time axis in the three plots is the same.  Data are smoothed using a 100-point moving average.
{\it Main plot:} Synchronous rotation, Case 15 in Table~\ref{table_models}.
{\it Inset, bottom:} Super-synchronous rotation Case 1.
{\it Inset, top:} No rotation, Case 17.
\label{fig_omdpmax}}
\end{figure}

An example of the temporal evolution of $\omega_{max}''$ is illustrated in Fig.~\ref{fig_omdpmax} for cases
with  $\beta_0^0$=0 (nonrotating, Case 17), 1.8 (super-synchronous, Case 1), and 1 (synchronous, Case 15).
The synchronous case converges to uniform rotation while the asynchronous cases converge to one in which 
$\omega''_{max}$ is stratified, with maximum amplitude at the surface.

\begin{figure}
\centering
\includegraphics[width=0.85\columnwidth]{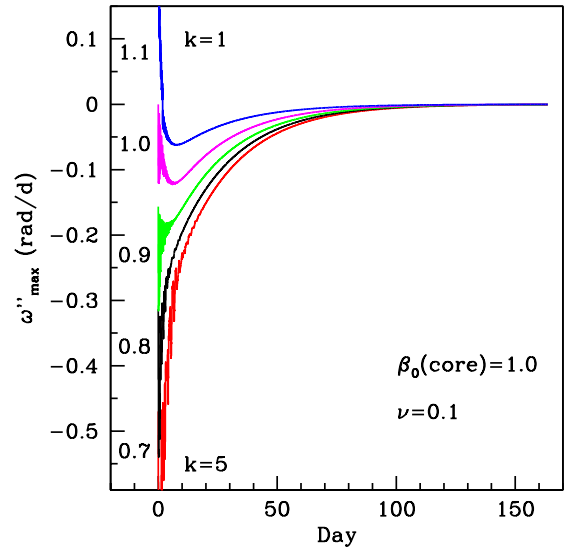}
\caption{Evolution over time of the maximum angular velocity $\omega_{max}''$  at the equator
for a core rotation $\beta^0_0$=1 and an initial envelope differential rotation structure
defined by the $\beta^k_0$ values listed to the left of the curves (Case 16).
}
\label{stability_case8}
\end{figure}

Examples of synchronous rotation are illustrated in Figs. \ref{fig_omdpmax} and \ref{stability_case8}.
These are Cases 15 and 16 in Table \ref{table_models}, and are presented here to illustrate
the stability of the TIDES numerical algorithm. Even when the initial conditions depart significantly
from the equilibrium state, the calculation converges to the equilibrium state.
In Case 15, the star is unperturbed and is in uniform synchronous rotation at the start of the calculation.
When the perturbation due to the companion's gravitational field is applied, the star undergoes large
amplitude oscillations as it adjusts to the presence of the external potential. As is shown in the
main plot of Fig.~\ref{fig_omdpmax}, the perturbed surface attains velocities as high as $\sim$ 0.25 rad\,d$^{-1}$
which, however, rapidly decay and converge to $<5\times10^{-11}$ rad\,d$^{-1}$ (corresponding to
a linear velocity of $\sim$10$^{-4}$ cm\,s$^{-1}$) after 800\,d.
The layers below the surface suffer a similar but smaller amplitude initial perturbation and attain
the stationary state more rapidly.

In Case 16, the calculation starts with three layers rotating slower than synchronous rotation and the
layer closest to the core rotating faster.  These initial velocities are tabulated in Cols. 6-8 of 
Table~\ref{table_VarBeta}. As illustrated in Fig.~\ref{stability_case8}, during the transitory state 
all layers converge toward $\omega''$=0, as expected for the equilibrium state.  In this figure, we 
chose to plot  $\omega_{max}''$ instead of its absolute value as in Fig.~\ref{fig_omdpmax} in order to 
more clearly illustrate the sub-synchronous and super-synchronous motions in the perturbed star's 
rest frame and the manner in which they reach the stationary state.

\begin{table}
\caption{All models and case numbers}
\label{table_models}
\centering
\begin{tabular}{ c c c l l l r l}
\hline\hline
Case &$\beta_0^0$ & $\nu$  & $\Delta$R/$R_1$  & N$_{r}$  & $\beta_0^{k>0}$ &$Cy$  & Var  \\ 
\hline
\hline
    &         &          &$n$=1.5&             &    &      &          \\
\hline
 1   & 1.8    & 0.028     &0.06  &5            &1.8 &   100&      \\
 2  &  1.8    & 0.028     &0.06  &8            &1.8 &   100&      \\
 3  &  1.8    & 0.1       &0.06  &10           &1.8 &    80&      \\ 
 4  &  1.8    & 0.1       &0.06  &10           &(b) &   161&      \\ 
 5  &  1.8    & 0.028     &0.06  &10           &(b) &   200&      \\ 
 6  &  1.8    & 0.028     &0.06  &5            &(c) &   200&      \\ 
 7  &  1.8    & 0.005     &0.06  &5            &(c) &   500&      \\ 
 8  &  1.8    & 0.0005    &0.06  &5            &(c) &   2002&      \\
 9  &  1.8    & 0.1       &0.03  &10           &1.8 &  100&  e1   \\
10  &  1.8    & 0.028     &0.06  & 5           &1.8 &  100&  e1   \\%
11  &  1.8    & 0.1       &0.03  &10           &1.8 &  80 &        \\ 
12  &  1.8    & 0.028     &0.03  &10           &1.8 &  30&N$_\varphi$  \\ 
13  &  1.8    & 0.014     &0.03  &10           &1.8 &  30&N$_\varphi$  \\ 
14  &  1.8    & 0.028     &0.045 &10           &1.8 &  30&         \\ 
15  &  1.0    & 0.028     &0.06  &5            &1.0 &200 &        \\  
16  &  1.0    & 0.028     &0.06  &5            &(a) &120 &       \\   
17  &  0      & 0.028     &0.06  &5            &0.0 &100 &        \\  
18  &  0      & 0.028     &0.06  &5            &0.0 &130&N$_\varphi$  \\
19a &  0      & 0.028     &0.06  &5            &(f) & 100&        \\
19b &  0      & 0.1       &0.06  &5            &(f) & 85 &        \\
21  &  2.1    & 0.1       &0.06  &10           &(b) &   100&      \\ 
22  &  2.1    & 0.028     &0.06  &10           &(b) &   190  &    \\
24  &  2.1    & 0.028     &0.06  &10           &(b) & 100    &    \\ 
25  &  2.07   & 0.018     &0.07  & 5           &2.07&300& e2      \\
26  &  1.1    & 0.028     &0.06  & 5           &1.1 &   40&       \\ 
27  &  1.4    & 0.028     &0.06  & 5           &1.4 &   40&       \\ 
28  &  1.9    & 0.028     &0.06  & 5           &1.9 &   50&       \\ 
\hline
    &         &          &$n$=3  &             &    &      &          \\
\hline
30  & 1.01    & 0.1       &0.06  &5            &1.01&  100 &          \\
31  &  1.8    & 0.1       &0.06  &10           &1.8 &    47&           \\ 
32  &  1.8    & 0.028     &0.06  &10           &1.8 &    40&           \\ 
32b &  1.8    & 0.028     &0.06  & 5           &(c) &    50&           \\
33  &  1.8    & 0.2       &0.06  &10           &(b) &    57&           \\ 
34  &  1.8    & 0.1       &0.06  &10           &(b) &    59&           \\ 
35  &  1.8    & 0.1       &0.06  &10           &1.8 &    89&  e1      \\ 
36  &  0.0    & 0.1       &0.06  &10           &1.8 &    73&           \\ 
\hline
    &         &          &$n$=3.5  &           &    &      &          \\
\hline
41   & 2.2    & 0.1       &0.06  &10           &2.2 &    78&  BEC      \\
\hline
    &         &          &$n$=4.5  &           &    &      &          \\
\hline
51  &  1.8    & 0.028     &0.06  & 5           &(c) &    50&           \\
\hline
\hline
\end{tabular}
\tablefoot{The units of $\nu$ are R$_\odot^2~d^{-1}$. {\it Column 6} indicates whether the
initial rotation was uniform by listing the corresponding value of the initial asynchronicity
parameter $\beta_0^{k>0}$ or it was non-uniform by listing the following letters, which
correspond to the values of $\beta_0^{k>0}$: 
{\bf (a)} $\beta_0^{k>0}$=[1.1, 1.0, 0.9, 0.8, 0.7];
{\bf (b)} $\beta_0^{k>0}$=[2.0, 1.9, 1.8, 1.7, 1.6, 1.4, 1.25, 1.14, 0.92, 0.87];
{\bf (c)} $\beta_0^{k>0}$=[1.4, 1.25, 1.14, 0.92, 0.87]; 
{\bf (f)} $\beta_0^{k>0}$=[0.2, 0.4, 0.6, 0.8, 1.0].
The first value in each bracket corresponds to the layer closest to the core and the last to the surface layer.
{\it Column 8:} A parameter that differs from the nominal case as follows: {\bf e1} $e$=0.1; {\bf N$_\varphi$}=500;
{\bf e2} $e$=0.067; {\bf BEC:} $m_1$=10 M$_\odot$, R$_1$=5.48 R$_\odot$
}
\end{table}

\begin{table}
\caption{Initial uniform rotation velocities}
\label{table2}
\centering
\begin{tabular}{ l l c c }     
\hline\hline
           $\beta_0^0$       &  Description &\multicolumn{2}{c}{Rotation velocity (rad~d$^{-1}$)}   \\
                           &              &  $\omega_0$     & $\omega_0^{\prime}$     \\
\hline
            0.00   &No rotation     &           0         &             $-$1.57        \\
            1.00   &Synchronous     &        $+$1.57      &                0.0          \\
            1.80   &Supersyncronous &        $+$2.82      &             $+$1.25        \\
            1.01   &Scharlemann     &        $+$1.58      &             $+$0.01       \\
\hline
\hline
\end{tabular}
\tablefoot{$\beta_0^0=\omega_0/\Omega_0$, where $\omega_0$ is the rotation angular velocity of the
rigidly rotating core of the primary and $\Omega_0$  is the orbital angular velocity in the circular
orbit cases, both in the inertial reference frame. $\omega_0'$ is the rotation angular velocity of the
rigidly rotating core in the non-inertial reference frame $S'$ rotating with angular velocity $\Omega_0$.
The values in this table correspond to $\Omega_0$=1.57.}
\end{table}

\begin{table}
\caption{Initial non-uniform rotation velocities (a) and (c).}
\label{table_VarBeta}
\centering
\begin{tabular}{l l l l l c l l l}
\hline\hline
         &             & \multicolumn{3}{c}{{\it (c)}}&\multicolumn{1}{c}{}  &\multicolumn{3}{c}{{\it (a)}  }  \\
    $k$  &r$_{mid}$   & $\beta_0^k$ & $\omega_0$ &    $\omega_0''$& |  &$\beta_0^k$ & $\omega_0$ &    $\omega_0''$     \\
\hline
   0& 4.79  & 1.80  &  2.82 & 0.00  &| & 1.0 &1.57& 0.00 \\       
   1& 4.99  & 1.40  &  2.20 & -0.63 &| & 1.1 &1.72& 0.16\\       
   2& 5.40  & 1.25  &  1.96 & -0.86 &| & 1.0 &1.57&-0.00\\
   3& 5.81  & 1.14  &  1.79 & -1.03 &| & 0.9 &1.41&-0.16\\
   4& 6.22  & 0.92  &  1.44 & -1.38 &| & 0.8 &1.25&-0.31\\
   5& 6.63  & 0.87  &  1.37 & -1.45 &| & 0.7 &1.10&-0.47\\
\hline
\hline
\end{tabular}
\tablefoot{Initial angular velocity as a function of the radius used in the models.
 r$_{mid}$ is the radius of the layer midpoint. $\beta_0^k$ is the synchronicity
parameters of the core ($k$=0) and the $k$ layers above it; $\omega_0$ is the initial
angular velocity in the inertial frame $S$.  $\omega_0''$ is the initial angular velocity in
the star's rest frame $S''$. The units of $r_{mid}$ are R$_\odot$ and of
angular velocity are radians~d$^{-1}$.}
\end{table}

\subsection{Comparison with the one-layer model}  \label{app_compare_1D}

\begin{figure}
\centering
\includegraphics[width=0.49\columnwidth]{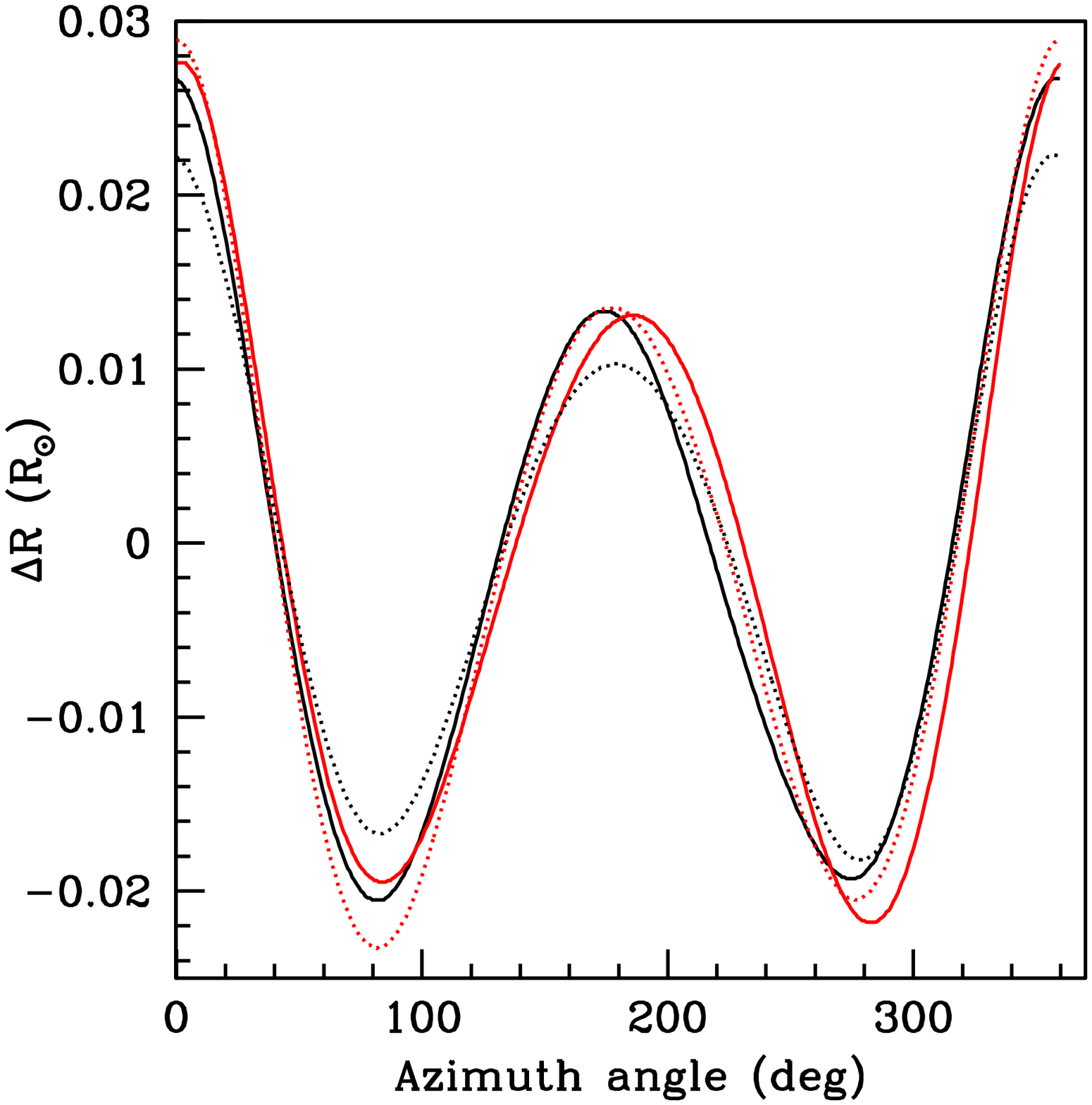}
\includegraphics[width=0.49\columnwidth]{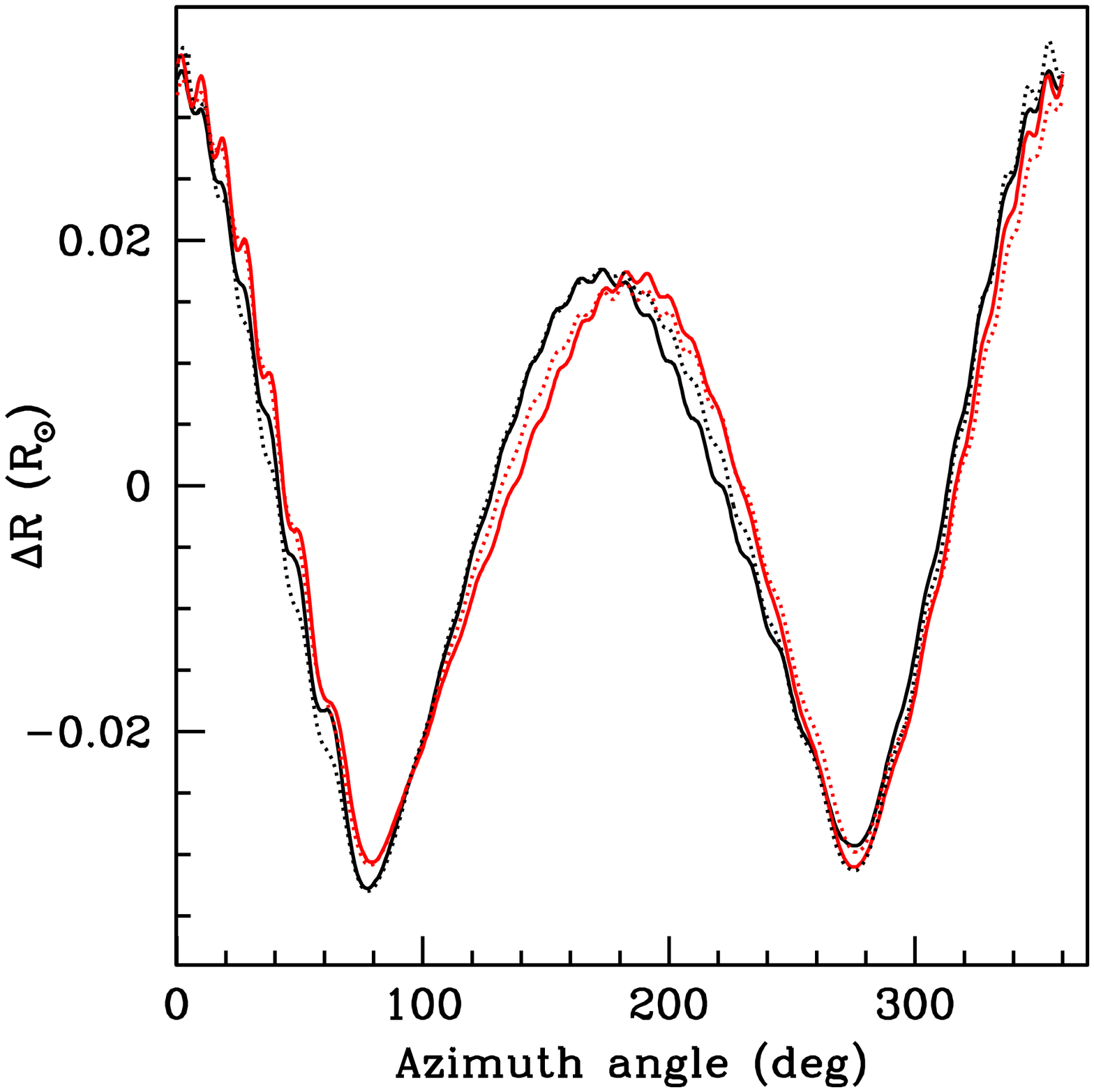}
\includegraphics[width=0.49\columnwidth]{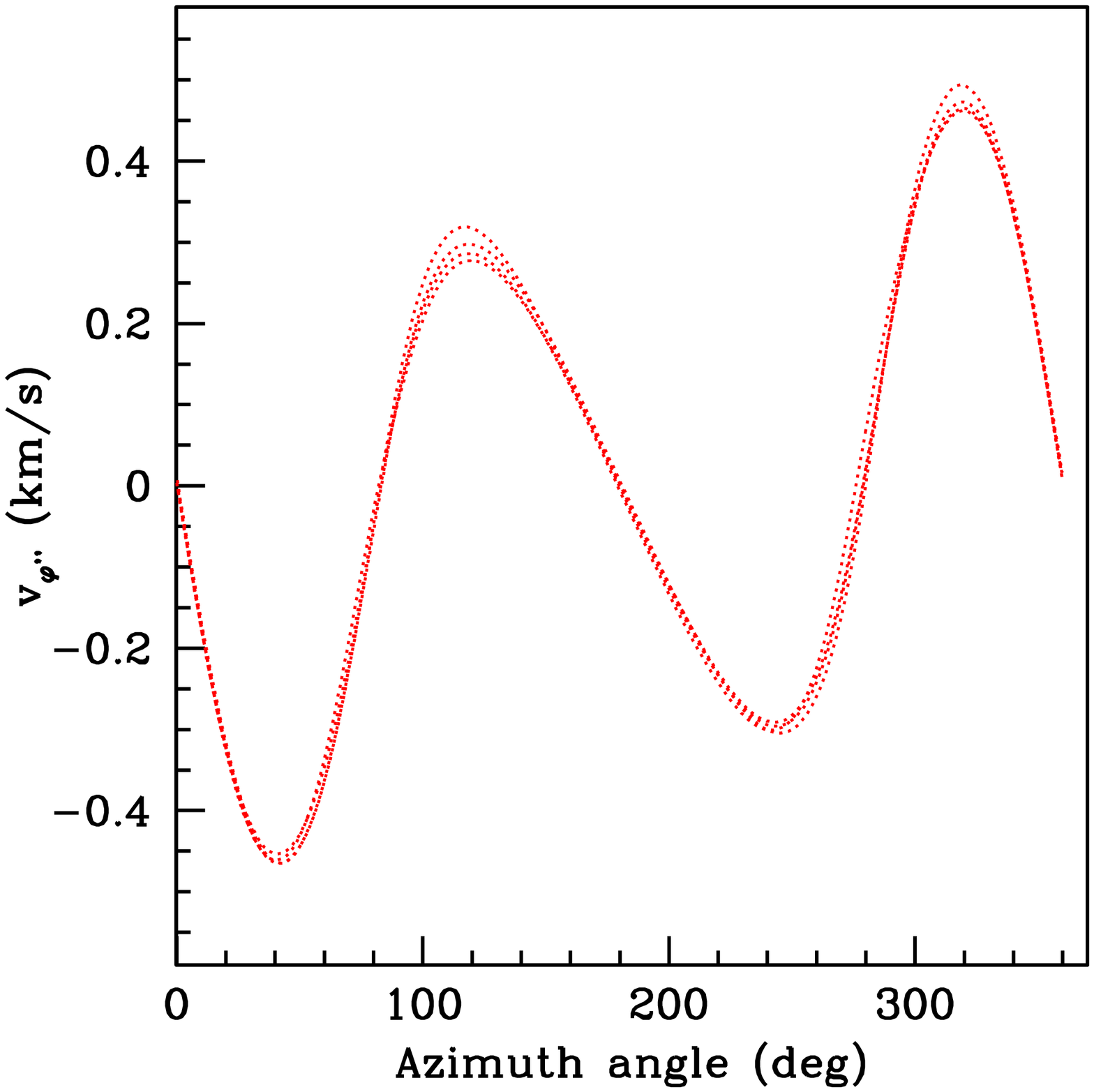}
\includegraphics[width=0.49\columnwidth]{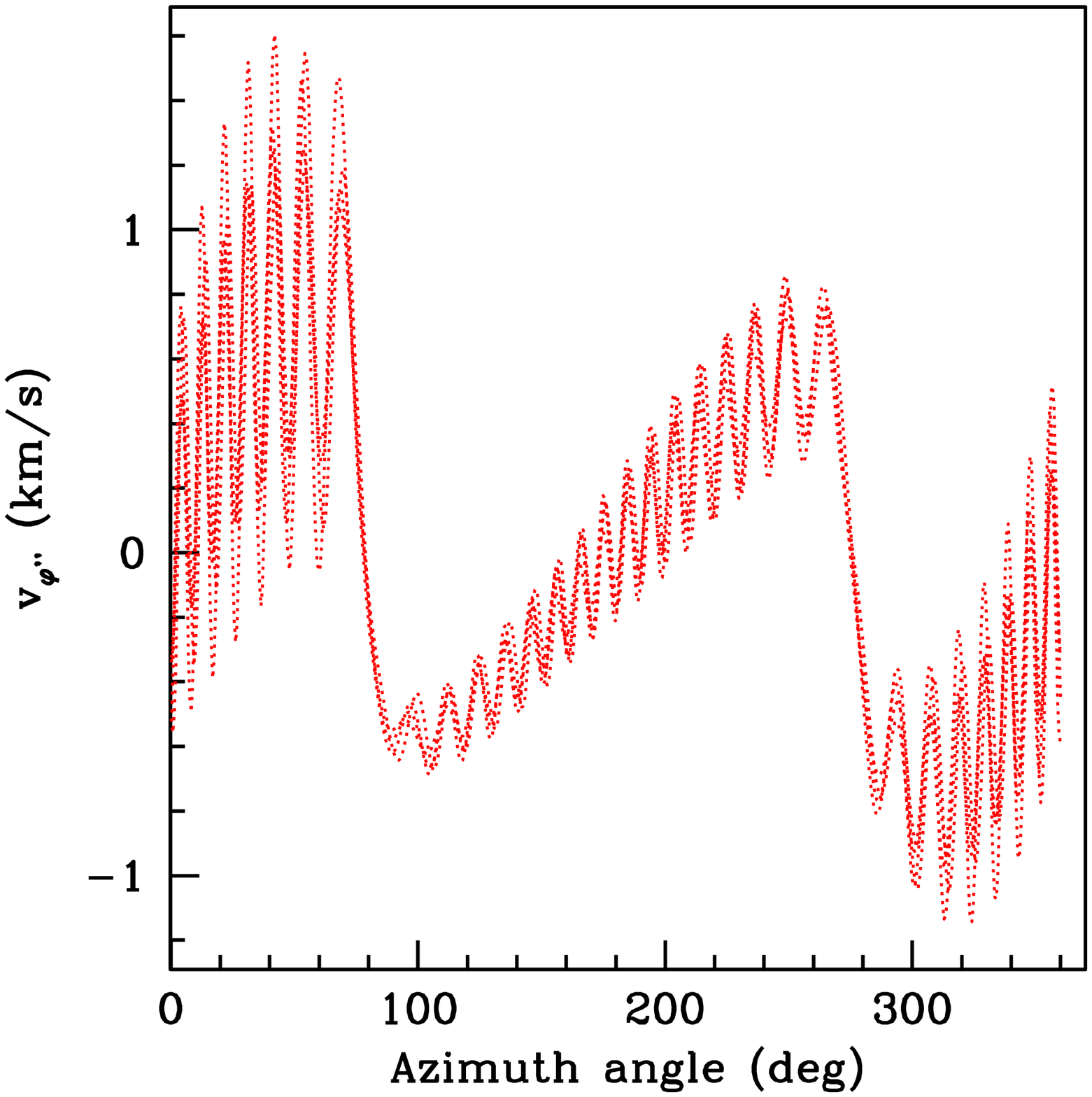}
\caption{Results of calculations with the one-layer TIDES model using input 
parameters corresponding to those listed in Table A.1.  {\it Top:} Surface deformation 
at the equator for the analogous Case 1 model (left, $\beta_0$=1.8) and the analogous Case 17 model 
(right, $\beta_0$=0). Black and red curves indicate different times within 
the same orbital cycle.
{\it Bottom:} Radial component of the tidal velocity for each of the models shown 
in the top panels.
\label{fig_radius1D}}
\end{figure}

The $n$-layer TIDES calculation suppresses motions in the radial and polar
directions.  This implies that there is no tidal deformation.
In this section we examine the implications of neglecting radial
motion and deformation by comparing with results of the original one-layer
calculation (Moreno et al. 2011).  The input parameters are as in the
nominal case (Case 1, Table~\ref{table_models}).  The maximum tidal deformation is
illustrated in Fig.~\ref{fig_radius1D} for orbital phases
distributed equally over the orbital cycle.  Contrary to a
static calculation, the location and amplitude of the tidal bulges undergo
time-dependent variations as a consequence of the dynamical processes.

\begin{figure}
\centering
\includegraphics[width=0.49\columnwidth]{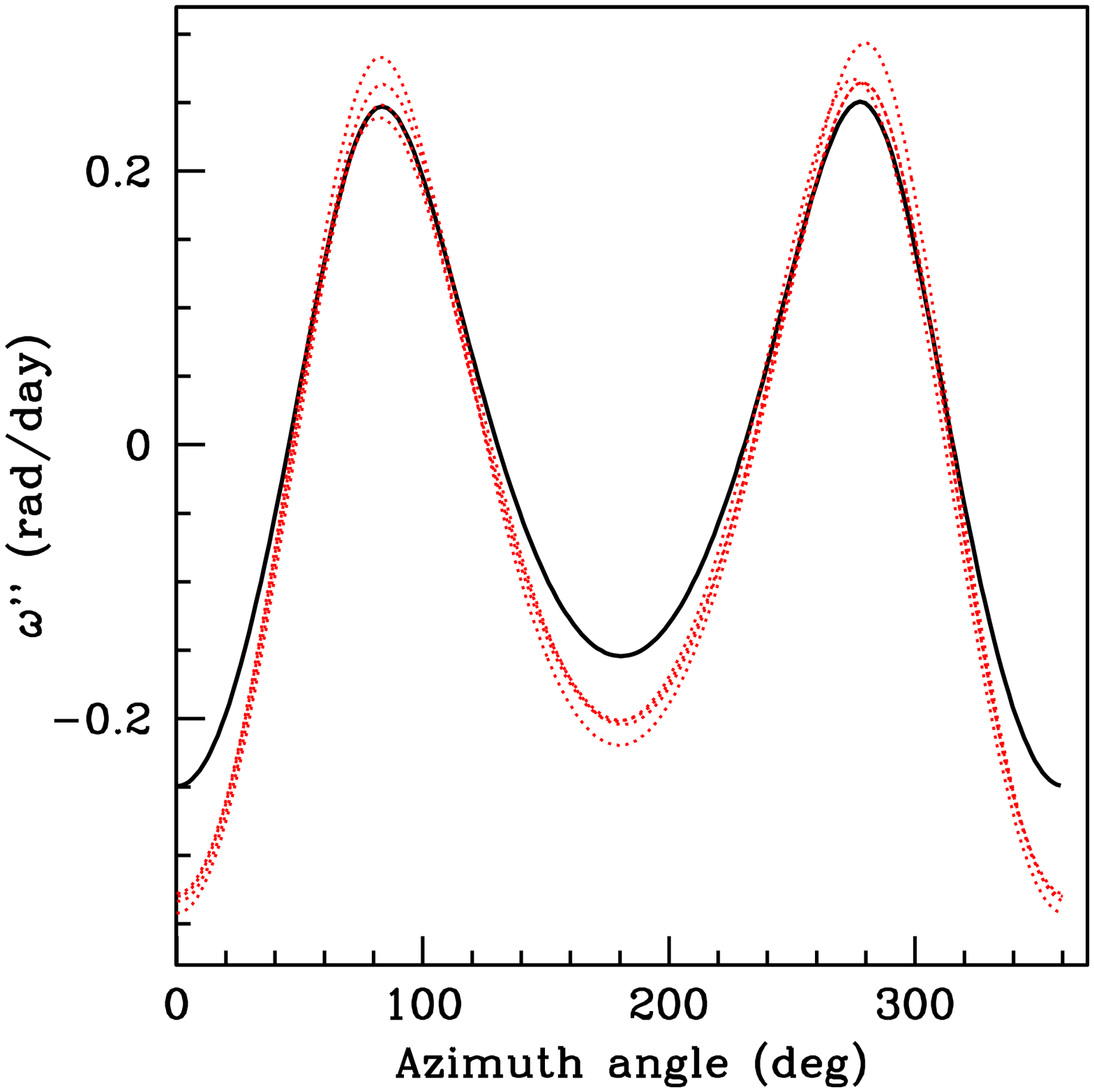}
\includegraphics[width=0.49\columnwidth]{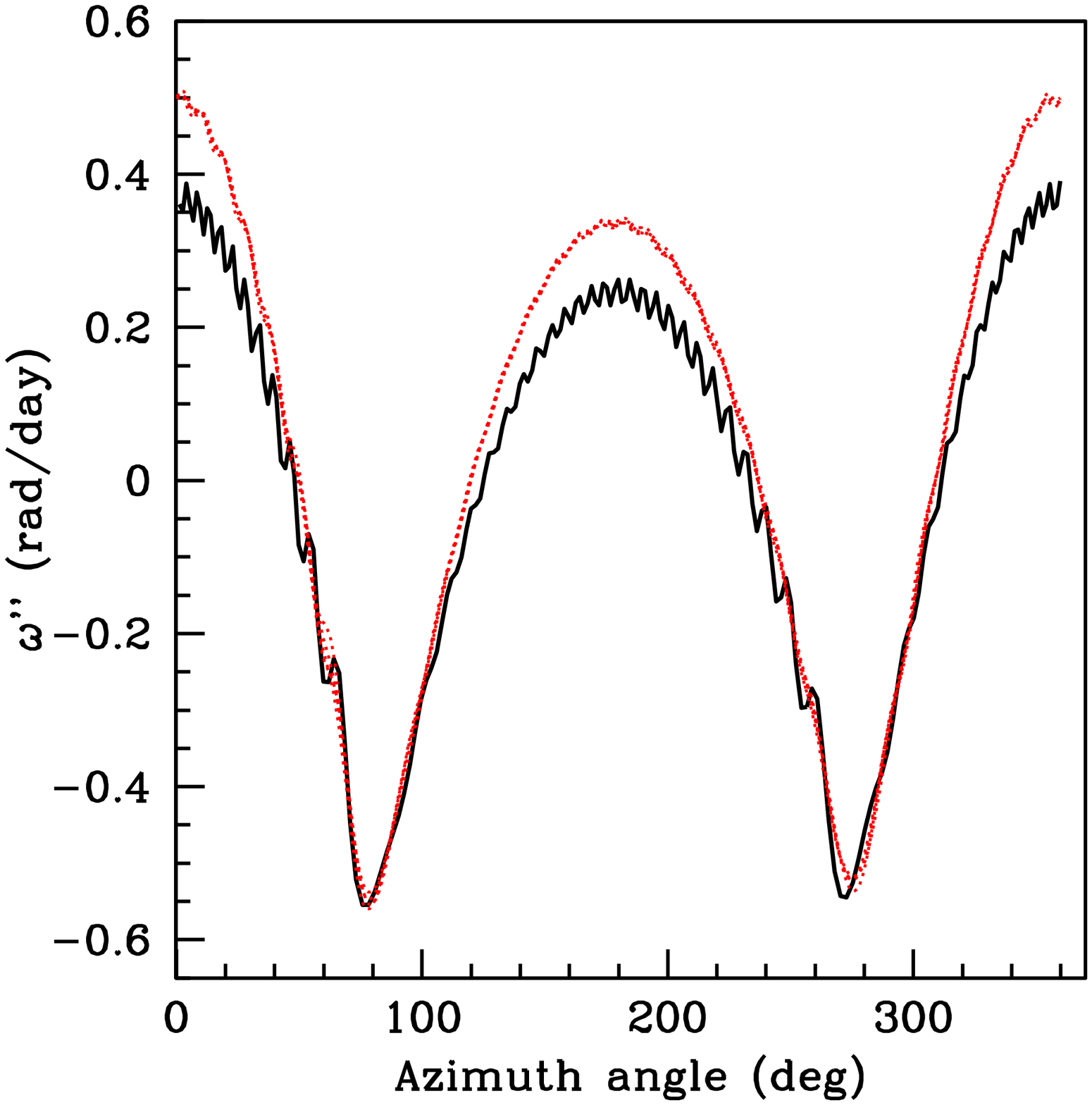}
\caption{Angular velocity at the equator obtained with the one-layer (red) and $n$-layer (black) models
illustrating the differences due to the neglect in the $n$-layer model of the radial deformation 
and radial velocity component.
{\it Left:} Case 1  ($\beta_0$=1.8). {\it Right:} Case 17  ($\beta_0$=0) 
\label{fig_compare_TIDES1and2}}
\end{figure}

Figure~\ref{fig_compare_TIDES1and2} compares the  angular
velocity in the surface layer of the Case 1 model with that obtained 
from the one-layer model. The most important differences are seen at 
azimuth angles corresponding to the tidal bulges (see Fig.~\ref{fig_radius1D}, which
indicates the location of the tidal bulges).  The 
explanation lies in the fact that the bulges are at a larger
radius and are therefore more strongly affected by the tidal
force.  The difference at the sub-binary longitude is $\sim$0.08 rad\, d$^{-1}$,
and somewhat smaller in the opposite hemisphere.  
Hence, neglecting the radial deformation leads to  angular velocities 
that are underestimated at the location of the bulges. Another difference
is that the shape and amplitude of $\omega''$ obtained with the $n$-layer
calculation are constant with orbital phase, contrary to what is obtained 
from the one-layer model.

\begin{table}
\caption{Number of volume elements per colatitude for $N_\theta$=20.}
\label{table_parsnum}
\centering
\begin{tabular}{ r r r r r r }     
\hline\hline
Parl  & $\theta$ & Num &Parl  & $\theta$ & Num \\
\hline
 1 & 7.179 & 80       & 11 &50.769 &155       \\
 2 &11.538 & 80       & 12 &55.128 &164       \\      
 3 &15.897 & 80       & 13 &59.487 &172       \\      
 4 &20.256 & 80       & 14 &63.846 &180       \\      
 5 &24.615 & 83       & 15 &68.205 &186       \\      
 6 &28.974 & 97       & 16 &72.564 &191       \\      
 7 &33.333 &110       & 17 &76.923 &195       \\      
 8 &37.692 &122       & 18 &81.282 &198       \\      
 9 &42.051 &134       & 19 &85.641 &199       \\      
10 &46.410 &145       & 20 &90.000 &200       \\
\hline
\hline
\end{tabular}
\tablefoot{Columns 1 and 4: the identifying number of the parallel, with 20 corresponding to the
equator; Cols. 2 and 5: value of the corresponding colatitude midpoint in degrees; Cols. 3 and 6:
number of elements in the corresponding parallel. Calculation was performed for colatitudes in the
range 5-90$^\circ$.  The total number of volume elements for all layers in a $N_r$=10 run 
is 28510 (one hemisphere) and this is the number of equations of motion that are solved 
simultaneously with the orbital motion.
}
\end{table}

\section{Dependence on input parameters} \label{app_input_params}

\begin{figure}
\centering
\includegraphics[width=0.49\columnwidth]{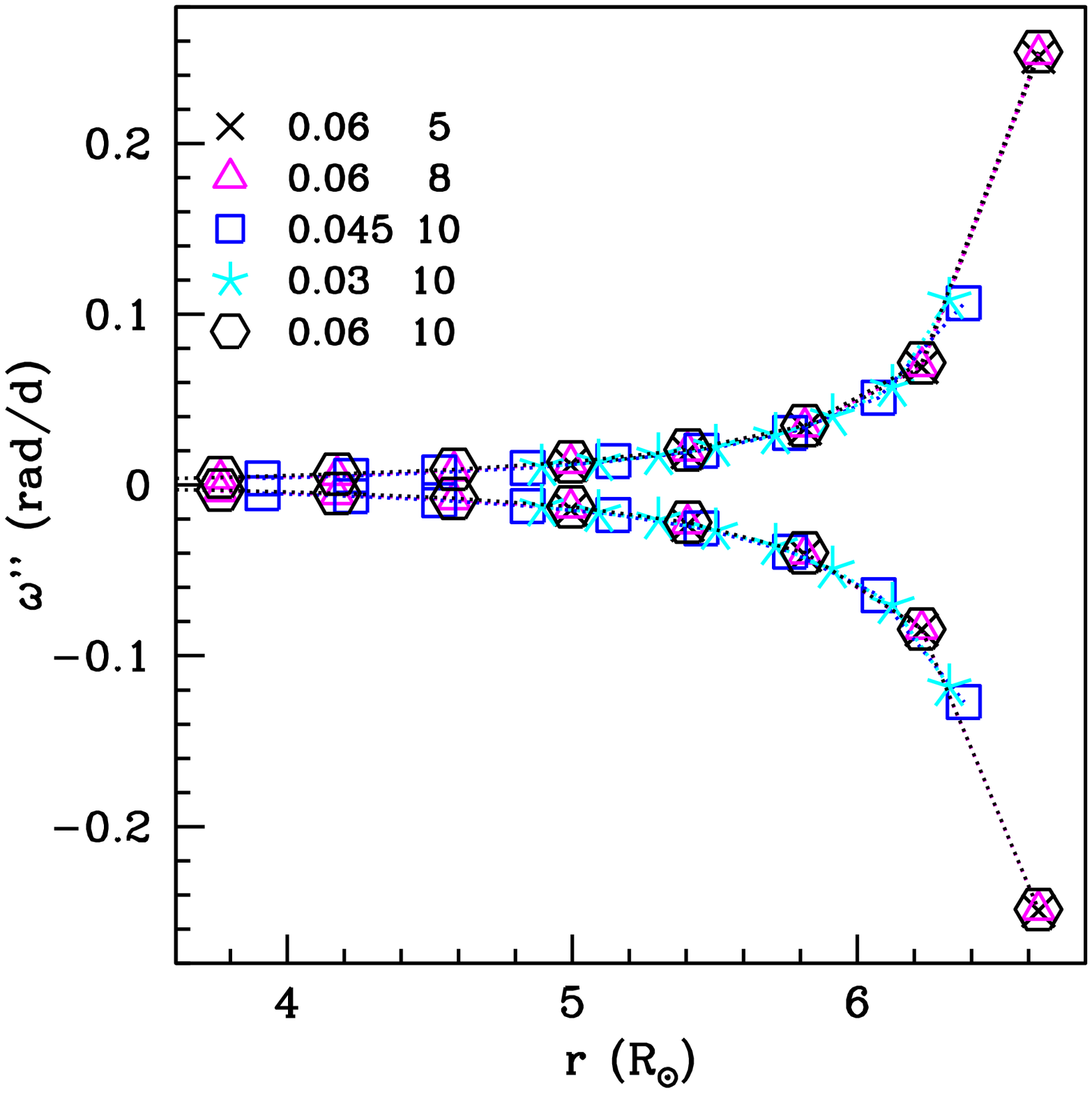}
\includegraphics[width=0.49\columnwidth]{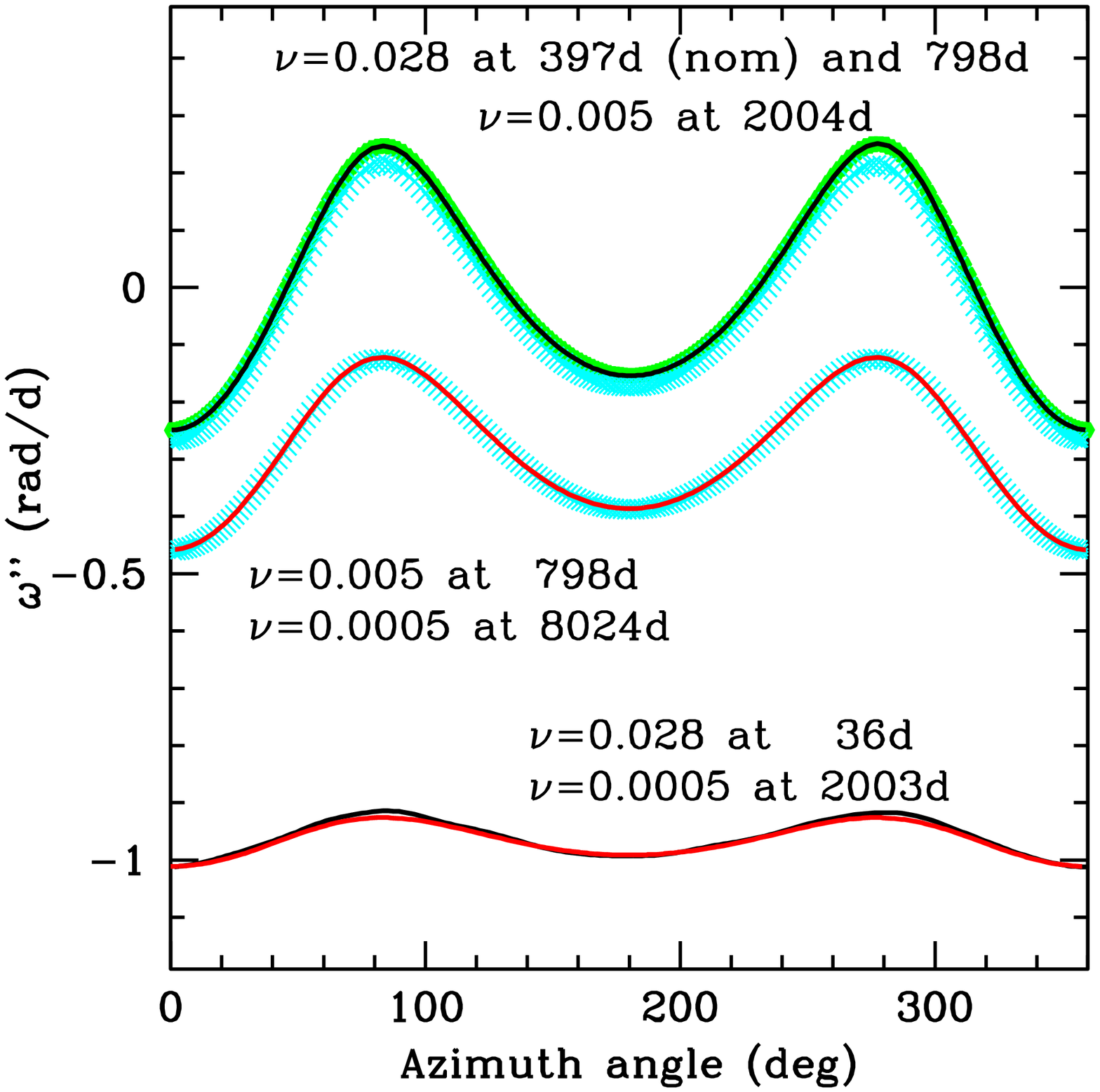}
\caption{Sample results from tests that were performed to assess the dependence on the layer
thickness $\Delta R$/R$_1$, number of layers $N_r$, and viscosity $\nu$. {\it Left:}  $\omega''$ 
profiles for azimuth angles at which the most pronounced velocity gradients occur, computed with 
the combinations of $\Delta R$/R$_1$, $N_r$, and $\nu$ as indicated by the legend which, from
top to bottom, refers to cases to Cases 1, 2, 14, 11 and 3 in Table A.1.
The abscissa is the radial distance from the center of the star to the layer midpoint.
{\it Right:} Surface angular velocity from models holding all input parameters constant
except for $\nu$,  showing that calculations with smaller values of $\nu$ reach the same rotation 
state as those with larger values, but after longer times. The three curves show:
{\it Top:}  Case 1 at Day 397 (green), Case 6 at Day 798 (black), Case 7 at Day 2004 (cyan). 
{\it Middle:} Case 7 at Day 798 (cyan), Case 8 at Day 8024 (red). {\it Bottom:}  Case 6 at Day 36 (black), 
Case 8 at Day 2003 (red).}
\label{fig_minmax_Ncap}
\end{figure}

We tested the dependence of the $\omega''$ profile on a variety of input 
parameter combinations, specifically $N_r$, $\Delta R/R_1$, $\nu$, and polytropic index.   
We illustrate the $\omega''$ profiles with the most pronounced velocity gradients 
in Fig.~\ref{fig_minmax_Ncap} (left) for runs performed 
with different combinations of the first three of the above mentioned input parameters.  
The  $\omega''$ profiles are the same in all cases.  However, models run with smaller 
viscosity values require a longer time to attain the same state as models that were
run with greater viscosities. This also holds true for the azimuthal dependence of 
$\omega''$, as illustrated in Fig.~\ref{fig_minmax_Ncap} (right).  This figure also shows 
that the same result is obtained regardless of whether the calculation is initiated in 
uniform rotation or with a pronounced differential rotation structure.  

\begin{figure}
\centering
\includegraphics[width=0.49\columnwidth]{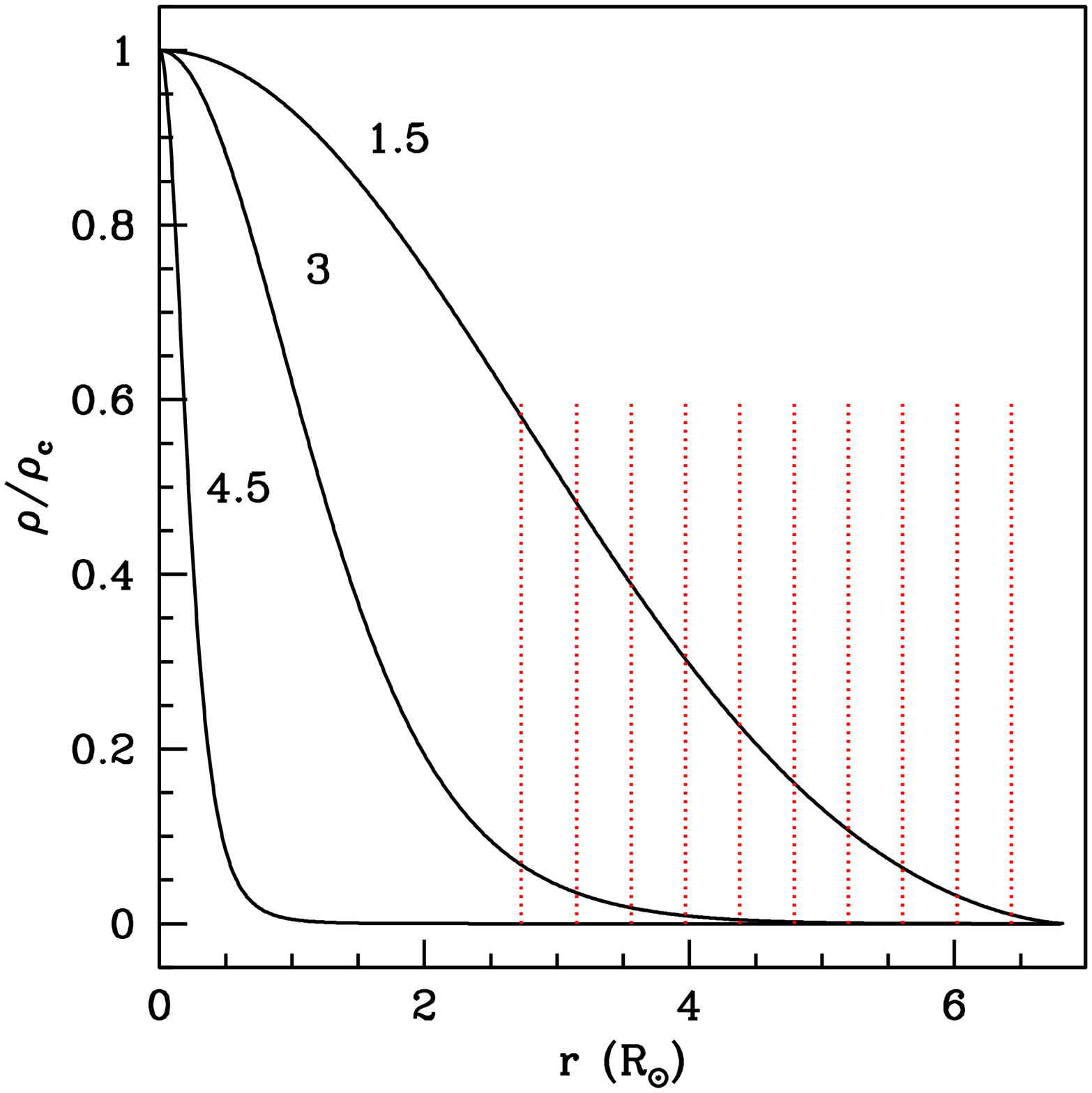}
\includegraphics[width=0.49\columnwidth]{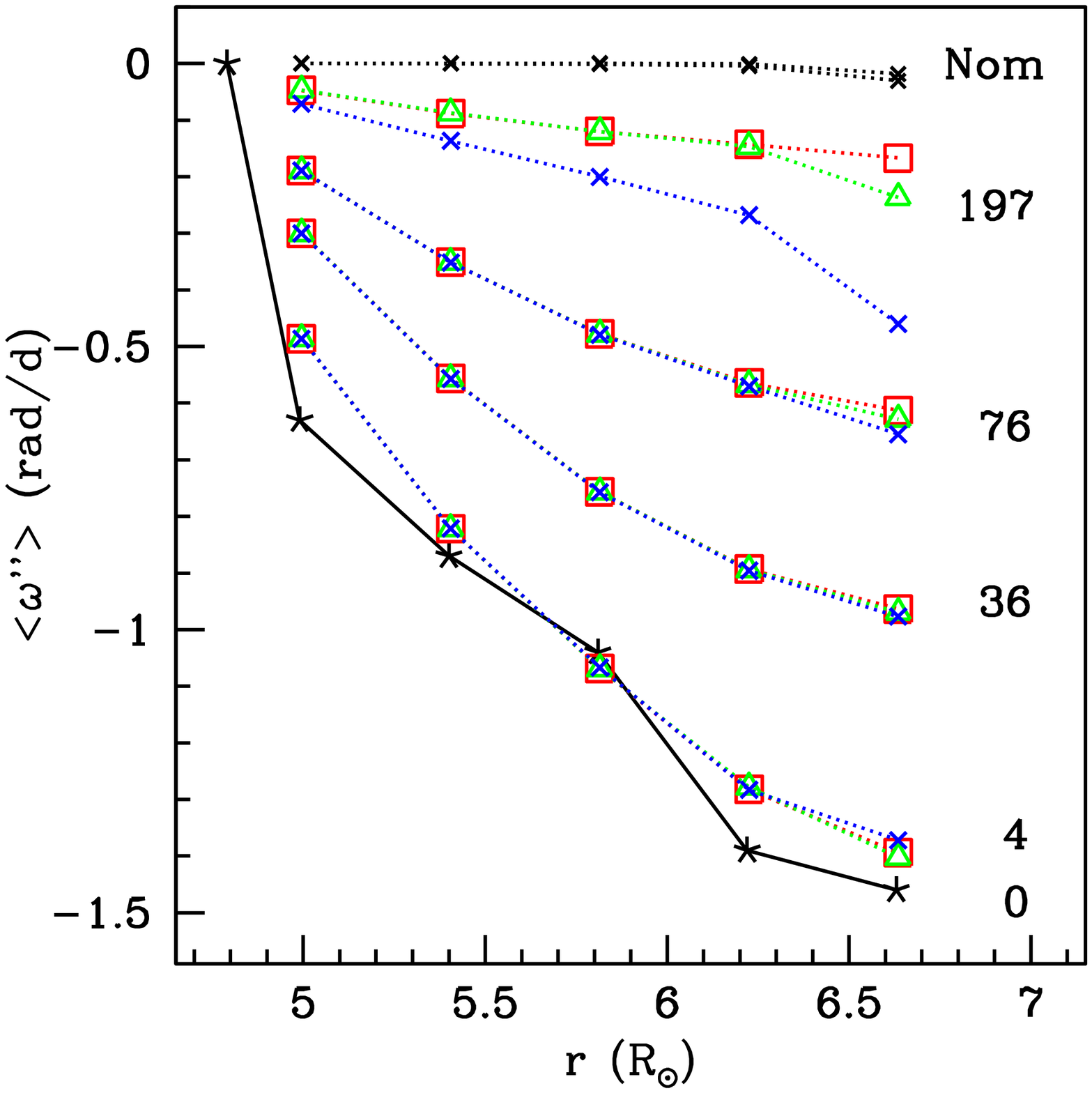}
\caption{Tests performed for different polytropic indices $n$, where $\gamma$=1+1/$n$, the
exponent of the polytropic equation of state. {\it Left:} Polytropic density structures as a 
function of radius for the 6.84~R$_\odot$ models. Density is given in units of $\rho_c$, the 
central density. Each curve is labeled with its corresponding value of $n$.  
The vertical lines indicate the base of layers for layer thickness $\Delta$R/$R_1$=0.06.
{\it Right:} Azimuthally averaged angular velocity $\left<\omega''\right>$ for 
$n$=1.5 (square, Case 1), 3.0 (triangle, Case 32b) and 4.5 (large cross, Case 51).  
Illustrated are the values for times $t/day$=4, 36, 76, and 197 after the start of the 
calculation, showing that all cases converge toward average uniform rotation.
The initial $\omega''$ profile is shown with the asterisks joined by a continuous line. 
}
\label{fig_poly}
\end{figure}

Results of tests with different polytropic indices ($n$=1.5, 3 and 4.5) are illustrated 
in Fig.~\ref{fig_poly} (right). The corresponding density structures are shown in 
Fig.~\ref{fig_poly} (left).  The most significant difference in the
results occurs for $n$=4.5 because uniform average rotation is approached much more 
slowly than the other cases, particularly in the outer layers.  This is explained by 
the fact that such a large polytropic index corresponds to a star with a more pronounced 
density gradient and thus the transport of angular momentum in these layers is less 
efficient than in the case of smaller values of $n$.

Case 4 ($\beta_0^0$=1.8) and Case 21 ($\beta_0^0$=2.1) were used to explore the effect of different 
degrees of synchronicity.   In both cases, the average rotation tends toward uniform, but with a faster
final rotation speed in Case 21 as expected.  Also, the tidal amplitude is larger for Case 21 than Case 4,
as can be verified by comparing the data in Tables~\ref{table_case4_non-uniform} and 
\ref{table_case21_non-uniform_beta2.1_Nu0.1}.  Tests with $\beta_0^0$=1.1, 1.4, and 1.9 confirm that
the tidal amplitude increases with increasing departures from synchronicity.  It is important to
note, however, that in order to attain average uniform rotation in a super-synchronously rotating
star, the outer layers may approach the critical speed at which surface material can become detached.
The equatorial latitude would be the first to experience this instability, thus potentially leading
to equatorial outflows.

Case 17 is one of various tests performed for a nonrotating ($\beta_0$=0) star and which
serves to illustrate the case of sub-synchronous rotation.  The surface angular velocity is
shown in Fig.~\ref{fig_compare_TIDES1and2}. It is noteworthy that although the core is nonrotating,
the tidal flows at the surface on the equator attain speeds of  $\sim$0.4~rad\,d$^{-1}$,  which from
an observational standpoint is relevant for the interpretation of the line broadening in observed spectra.
That is, depending on the orientation of the orbital axis with respect to an observer, the detected 
spectroscopic line profiles could simulate a $\vv \sin i$  as large as 20 km\,s$^{-1}$ ($i$ being the 
inclination of the orbital plane with respect to the observer), despite the fact that the star is 
nonrotating.

\begin{table}
\caption{Azimuthally averaged angular velocity and tidal amplitude
Case 4: $\beta_0^0$=1.8, $\nu$=0.1 ${\rm R_\odot^2\, d^{-1}}$}
\label{table_case4_non-uniform}
\centering
\begin{tabular}{ c c c l l r}
\hline\hline
$k$ &$r_{mid}$ &$\omega''_0$& $<\omega''>$ & $\delta\omega''$ & $\delta \vv''$ \\
    & $R_\odot$&\multicolumn{3}{c}{|---------- rad\,d$^{-1}$ --------|}              & km\,s$^{-1}$  \\
\hline
    &          &Day 397   &          &          &         \\
  1 & 2.941  &   0.313  &  -0.006  &   0.003  &   0.079  \\
  2 & 3.352  &   0.157  &  -0.011  &   0.005  &   0.135  \\
  3 & 3.762  &   0.000  &  -0.015  &   0.007  &   0.225  \\
  4 & 4.172  &  -0.157  &  -0.018  &   0.011  &   0.371  \\
  5 & 4.583  &  -0.313  &  -0.021  &   0.017  &   0.614  \\
  6 & 4.993  &  -0.626  &  -0.024  &   0.026  &   1.033  \\
  7 & 5.404  &  -0.861  &  -0.025  &   0.041  &   1.803  \\
  8 & 5.814  &  -1.033  &  -0.027  &   0.072  &   3.387  \\
  9 & 6.224  &  -1.377  &  -0.029  &   0.151  &   7.574  \\
 10 & 6.635  &  -1.456  &  -0.046  &   0.480  &  25.634  \\
\hline
\hline
\end{tabular}
\tablefoot{Case 4 results at Day 397.
$\omega''_0$ is the initial angular velocity of each layer in the $S''$ reference frame and corresponds
to the $\beta_0^k$ values listed for this case in Table~\ref{table_models}.
$\left<\omega''\right>$  is the average angular velocity over azimuth angle at the equator. $\delta\omega''$ is the
peak to peak amplitude of variation over azimuth angle of the angular velocity, and $\delta\vv''$
is the corresponding linear velocity. The radius listed in column 2 corresponds
to the midpoint  of each layer and is given is $R_\odot$.}
\end{table}

\begin{table}
\caption{Azimuthally-averaged angular velocity and tidal amplitude
 Case 21: $\beta_0^0$=2.1, $\nu$=0.1\,${\rm R_\odot^2\, d^{-1}}$. }
\label{table_case21_non-uniform_beta2.1_Nu0.1}
\centering
\begin{tabular}{ c c c l l r}
\hline\hline
$k$ &$r_{mid}$ &$\omega''_0$& $<\omega''>$ & $\Delta\omega''$ & $\Delta \vv''$ \\
    & $R_\odot$&\multicolumn{3}{c}{|---------- rad\,d$^{-1}$ --------|}              & km\,s$^{-1}$  \\
\hline
     &          &          &          &          &         \\
     &          & Day 196  &          &          &         \\
  1  &   2.941  &  -0.157  &  -0.055  &   0.005  &   0.107  \\
  2  &   3.352  &  -0.313  &  -0.102  &   0.007  &   0.177  \\
  3  &   3.762  &  -0.470  &  -0.143  &   0.010  &   0.288  \\
  4  &   4.172  &  -0.626  &  -0.177  &   0.014  &   0.467  \\
  5  &   4.583  &  -0.783  &  -0.205  &   0.021  &   0.762  \\
  6  &   4.993  &  -1.096  &  -0.228  &   0.032  &   1.268  \\
  7  &   5.404  &  -1.330  &  -0.246  &   0.051  &   2.197  \\
  8  &   5.814  &  -1.503  &  -0.259  &   0.088  &   4.137  \\
  9  &   6.224  &  -1.847  &  -0.269  &   0.190  &   9.493  \\
 10  &   6.635  &  -1.925  &  -0.301  &   0.684  &  36.487  \\
     &          &          &          &          &         \\
     &          &  Day 397 &          &          &         \\
  1  &   2.941  &  -0.157  &  -0.008  &   0.005  &   0.110  \\
  2  &   3.352  &  -0.313  &  -0.014  &   0.007  &   0.187  \\
  3  &   3.762  &  -0.470  &  -0.020  &   0.010  &   0.311  \\
  4  &   4.172  &  -0.626  &  -0.024  &   0.015  &   0.515  \\
  5  &   4.583  &  -0.783  &  -0.027  &   0.023  &   0.855  \\
  6  &   4.993  &  -1.096  &  -0.029  &   0.036  &   1.448  \\
  7  &   5.404  &  -1.330  &  -0.031  &   0.059  &   2.554  \\
  8  &   5.814  &  -1.503  &  -0.031  &   0.105  &   4.888  \\
  9  &   6.224  &  -1.847  &  -0.032  &   0.245  &  12.243  \\
 10  &   6.635  &  -1.925  &  -0.095  &   1.316  &  70.243  \\

\hline
\hline
\end{tabular}
\tablefoot{Case 21 results at Days 196 and 397. Columns are the same as in
Table~\ref{table_case4_non-uniform}
}
\end{table}

\section{Temporal evolution of the $\left<\omega''\right>$ profile \label{sect_diffrotation}}

We find that, excluding angular momentum redistribution processes other than the viscous interaction
between layers, an initially steep differential rotation
structure tends to flatten over time, first rapidly, and then progressively slower
as the star approaches the average uniform rotation state.  We take a closer
look at this process using the results of Cases 6, 7 and 8 from Table~\ref{table_models}.
These models have a smaller radial grid size than some of the others discussed in this paper
because the layers that lie closest to the surface are the slowest to
approach uniform rotation so they determine the timescale over which the average
uniform rotation state is attained.  Hence, it is computationally more efficient to focus on
these layers.   The initial differential rotation considered here is the same as that of the 5
outer layers of the larger radial grid models (see Table~\ref{table_case34_non-uniform}).

\begin{figure}
\centering
\includegraphics[width=0.49\columnwidth]{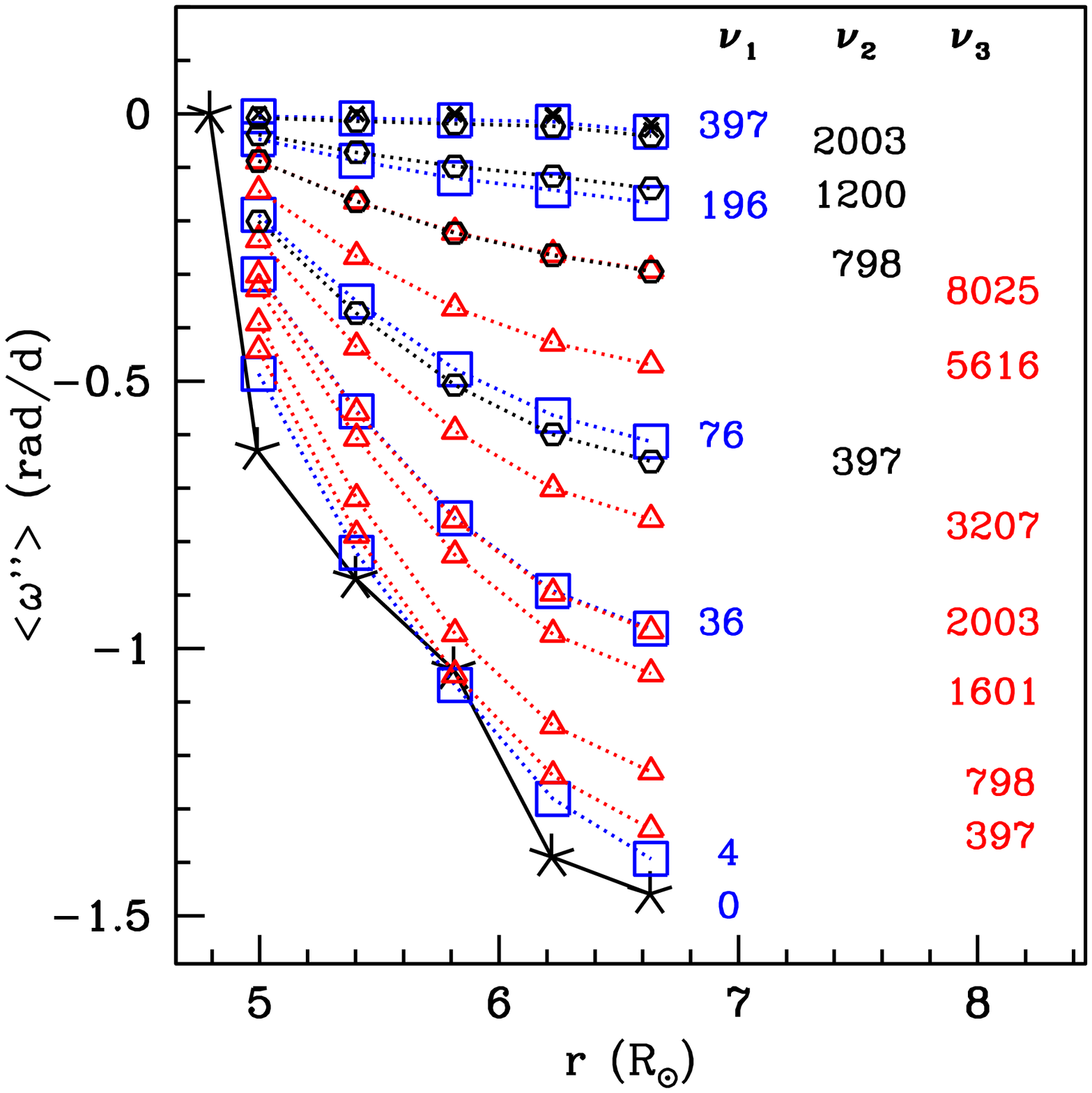}
\includegraphics[width=0.49\columnwidth]{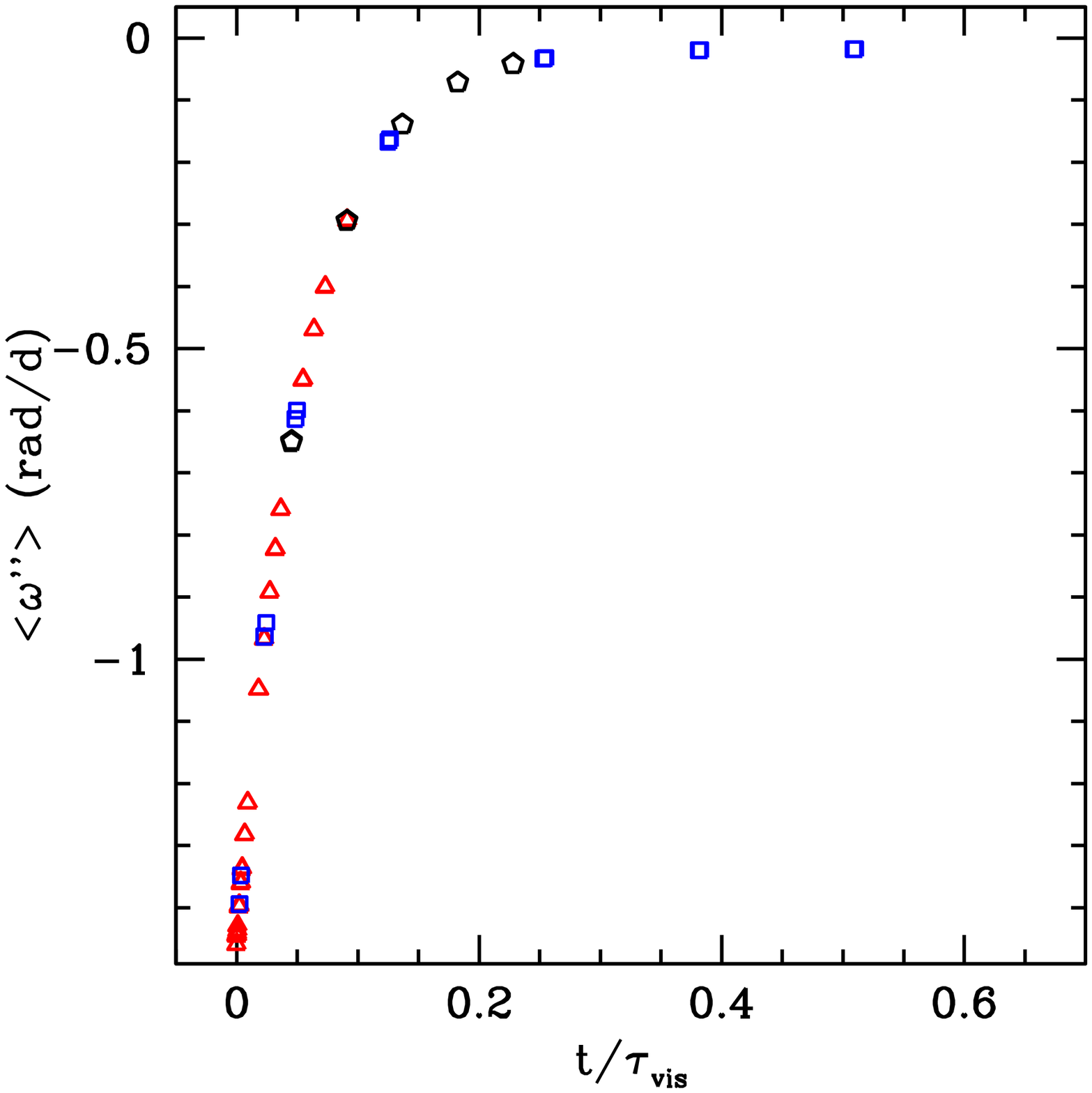}
\caption{Evolution from differential rotation to average uniform rotation for three
different viscosity values, $\nu_1/(R_\odot^2/d)$=0.028 (square), $\nu_2/(R_\odot^2/d)$=0.005 
(hexagon), and $\nu_3/(R_\odot^2/d)$=0.0005 (triangle).
{\it Left:} Azimuthally averaged angular velocity ($\left<\omega''\right>$ at times (in days) 
as listed to the right of each curve and in a column corresponding to the $\nu$ value. The initial 
$\left<\omega''\right>$ profile is indicated with asterisks joined by the continuous line. 
The time in days for each curve is listed under its corresponding viscosity. The small "x" 
(top curve) corresponds to Case 1, in which the calculation was started in uniform rotation.  
{\it Right:} Plot showing that all models tend toward average uniform rotation on the
viscous timescale,  $\tau_{\vv is}=r^2/\nu$.  The ordinate is the $\left<\omega''\right>$ value 
of the surface layer, $r$=6.6R$_\odot$.  The ordinate is $t/\tau_{\vv is}$  where $t$ is the time
listed next to each curve in the left hand panel of this figure. 
Symbols are the same as in the left panel.
}
\label{fig_viscosity}
\end{figure}

The procedure followed is to calculate the values of $\left<\omega''\right>$ at several
times after the start of the calculation and for calculations performed for several
viscosity values.  The result is
illustrated in Fig.~\ref{fig_viscosity} (left). The three viscosity values yield the same
result: the $\left<\omega''\right>$ profile flattens over time but the timescale increases
with decreasing viscosity value.  In addition, we find that the same $\left<\omega''\right>$ profile
is obtained regardless of the viscosity value, given enough time for the flattening to occur.
For example, it takes 2003\,d for the $\left<\omega''\right>$ profile in the model computed with
$\nu$=5$\times$10$^{-4}$ R$_\odot^2$d$^{-1}$ to equal the profile at Day 36 in the model with
$\nu_1$=0.028R$_\odot^2$d$^{-1}$.

Not only does the $\left<\omega''\right>$ profile behave in this manner but also the 
$\omega''$ dependence on azimuth angle (as shown in Fig. B.1), at least within the range of 
viscosities that we used in our simulations.  This result is useful for, among other things, 
allowing us to employ a relatively large viscosity value so as to accelerate the approach to the 
stationary state in the computations. Now, defining a viscous timescale 
$\tau_{\vv is}$=$r^2/\nu$ and using $r$=6.63~R$_\odot$, the surface radius, 
Fig.~\ref{fig_viscosity} (right) confirms that the time that it takes for the surface to 
(almost) corotate with the core is $\tau_{\vv is}$,  for the range of viscosities that we used.

The fact  that $\left<\omega''\right>\simeq$0 is approached asymptotically on the viscous timescale
$\tau_{\vv is}$=$r^2/\nu$ brings to light the importance of constraining the viscosity value, since
for turbulent viscosity 10$^{10}<\nu <$ 10$^{14}$ cm$^2$~s$^{-1}$ \citep{2007ApJ...655.1166P}, the
surface of our Case 6 binary star attains average uniform rotation on timescales ranging from one yr 
to 600 thousand years. Although this timescale corresponds to only a small fraction of the
$\sim$12\,Myr lifetime of a star like Spica's primary, viscosities on the low end of the
range would require consideration of wind mass loss, meridional currents and orbital evolution, 
which are here neglected.  However, for the larger values of $\nu$ in the above range, the timescale 
for reaching average uniform rotation is significantly shorter than that of these other processes, 
which suggests that stars with very large $\nu$ values may effectively be considered to be in
average uniform rotation.

\end{document}